\documentclass[11pt]{article}
\usepackage[titletoc,title]{appendix}
\usepackage{amsmath,amsfonts,amssymb,bm,amsthm}
\usepackage[colorlinks,citecolor=blue,linkcolor=blue,pagebackref]{hyperref}
\usepackage{tikz}
\usetikzlibrary{shapes,arrows}
\usepackage[all,cmtip]{xy}
\usetikzlibrary{intersections,shapes.arrows}
\usetikzlibrary{decorations.pathreplacing}
\usetikzlibrary{arrows.meta}
\usepackage{xcolor}
\definecolor{bostonred}{rgb}{0.8, 0.0, 0.0}
\usepackage{hyperref}
\hypersetup{
    colorlinks=true,
    linkcolor=bostonred,
    urlcolor=bostonred,
    citecolor=bostonred
}

\numberwithin{equation}{section}
\newtheorem{theorem}{Theorem}[section]
\newtheorem{lemma}[theorem]{Lemma}
\newtheorem{proposition}[theorem]{Proposition}
\newtheorem{corollary}[theorem]{Corollary}

\theoremstyle{definition}
\newtheorem{example}[theorem]{Example}
\newtheorem{definition}[theorem]{Definition}
\newtheorem{assumption}[theorem]{Assumption}
\theoremstyle{remark}
\newtheorem{remark}[theorem]{Remark}

\DeclareFontFamily{U}{mathx}{}
\DeclareFontShape{U}{mathx}{m}{n}{<-> mathx10}{}
\DeclareSymbolFont{mathx}{U}{mathx}{m}{n}
\DeclareMathAccent{\widecheck}{0}{mathx}{"71}

\usepackage{multicol}
\usepackage{longtable}
\usepackage{slashed}
\usepackage{accents}
\usepackage{tikz-cd}
\usepackage[T1]{fontenc}
\usepackage{pifont}
\usepackage{mathrsfs}
\usepackage[top=1in, bottom=1in, left=1in, right=1in]{geometry}
\usepackage{fancyhdr}
\newcommand{\loc}{\mathrm{loc}}

\definecolor{amber}{rgb}{0.2, 0.2, 0.6}

\def\mcM{{\mathcal{M}}}

\usepackage{enumerate}
\usepackage{xcolor,cancel}
\usepackage{relsize}

\renewcommand{\tilde}{\widetilde}
\usepackage{orcidlink}

\allowdisplaybreaks

\usepackage{pifont}
\renewcommand*{\backrefalt}[4]{%
\ifcase #1 %
No citations%
\or
\ding{43}~p.~#2%
\else
\ding{43}~pp.~#2%
\fi}
\usepackage{soul}
\begin{document}

\title{Projection of massive and interacting conformal theories to the future null infinity}
\author{
Beatrice Costeri\,\orcidlink{0009-0004-6594-5926}\thanks{INFN, Sezione di Pavia \&
Dipartimento di Fisica ``Alessandro Volta'',
Universit\`a degli Studi di Pavia \& INdAM, Sezione di Pavia, 
Via Bassi 6, 
I-27100 Pavia, 
Italia;
beatrice.costeri01@universitadipavia.it, \url{https://sites.google.com/view/beatricecosteri/}
}
\and
Claudio Dappiaggi\,\orcidlink{0000-0002-3315-1273}\thanks{Dipartimento di Fisica ``Alessandro Volta'',
Universit\`a degli Studi di Pavia \& INFN, Sezione di Pavia \& INdAM, Sezione di Pavia, 
Via Bassi 6, 
I-27100 Pavia, 
Italia;
claudio.dappiaggi@unipv.it,
\url{https://claudiodappiaggi.com}} 
\and 
Alberto Ferrari 
\thanks{Dipartimento di Fisica ``Alessandro Volta'',
Universit\`a degli Studi di Pavia, 
Via Bassi 6, 
I-27100 Pavia, 
Italia;
alberto.ferrari06@universitadipavia.it}
}


\date{}

\maketitle

\vspace{-.6cm}

\begin{abstract}
On a four-dimensional, globally hyperbolic and asymptotically flat spacetime, we consider a real, linear, massive scalar field. Within the framework of perturbative algebraic quantum field theory and inspired by the principle of perturbative agreement, we regard the mass term in the underlying Lagrangian as an interaction potential. This allows us to construct the algebra of observables of the massive theory and, using a M\o ller map, to represent each of its elements as a formal power series in the squared mass, whose coefficients take values in the algebra of observables of a massless, conformally coupled real scalar field. Following a procedure first outlined in \cite{DMP_2006}, we formulate a bulk-to-boundary correspondence which identifies each element of the algebra of observables of the massless, conformally coupled theory with a counterpart intrinsically defined at future null infinity $\mathscr{I}^+$. We extend this construction to the massive theory by projecting each observable onto $\mathscr{I}^+$ as a formal power series, whose convergence we prove when the underlying spacetime is Minkowski and an infrared cut-off is introduced. The latter is necessary for the construction and should eventually be removed by taking an adiabatic limit. Still focusing on four-dimensional Minkowski spacetime, we show that such a limit exists in the relevant topology but it is trivial. This obstruction can be circumvented by exploiting the time-slice axiom, which allows us to restrict our attention to observables localized in a causally convex neighbourhood containing a Cauchy hypersurface of the underlying spacetime. Consequently, we can remove the cut-off only locally, obtaining a non-trivial limit and hence a bulk-to-boundary correspondence for the massive theory. Secondly, we extend our procedure to a self-interacting real scalar field, representing each observable of the theory as a formal power series in the coupling constant, with coefficients taking values in the algebra of observables of a massless, conformally coupled scalar field. Each coefficient can then be projected to future null infinity using the same procedure outlined in \cite{DMP_2006}. As a concrete example, we apply this method to the stress-energy tensor of a massless scalar field with a quartic self-interaction, computing its projection up to first order in perturbation theory. 

\


\

{\bf 2020 MSC classes:} 81T05, 81T20  

\end{abstract}

\tableofcontents

\allowdisplaybreaks

\section{Introduction}\label{Sec: Introduction} 

The aim of this work is to investigate the extent to which the bulk-to-boundary correspondence for a massless, conformally coupled scalar field can be generalized to massive and self-interacting theories. More precisely, we consider a four-dimensional, globally hyperbolic and asymptotically flat spacetime $(M,g)$ with future null infinity $\mathscr I^+$, using the formulation of \cite{Friedrich_1986a,Friedrich_1986b}. Starting from the algebraic description of quantum fields on $\mathscr I^+$, see, {\it e.g.}, \cite{DMP_2017}, we address two related questions. We investigate, on the one hand, whether observables of a massive Klein--Gordon field admit a non-trivial realization at the null, conformal boundary, while, on the other hand, we discuss whether the same procedure can be implemented for observables of a perturbatively interacting theory. The two problems require different analytic tools and they lead to distinct notions of convergence.

Future null infinity plays a distinguished role in the study of asymptotically flat spacetimes. Within Penrose's conformal framework, $\mathscr I^+$ is the endpoint of future-directed null geodesics and it provides the natural, covariant, geometric setting for describing outgoing radiation data. Most notably, its intrinsic geometric structure is universal. It does not depend on the details of the bulk metric, up to a conformal freedom. Its group of asymptotic symmetries is the infinite dimensional Bondi--Metzner--Sachs (BMS) group \cite{Bondi_1962,Sachs_1962}. These properties make $\mathscr I^+$ a natural candidate for encoding field-theoretic information associated with the bulk, an idea which has been investigated within the algebraic approach to quantum field theory in \cite{DMP_2006,Moretti_2006,DMP_2017}.

The relevance of such boundary description is not confined to quantum field theory on curved spacetimes. In gravitational theories, asymptotic symmetries and the associated charges enter the infrared structure of scattering, including their relation with soft theorems and memory effects, see, \textit{e.g.}, \cite{Str18}. In addition, the degenerate geometry induced on null hypersurfaces has a natural interpretation in terms of Carrollian geometry, which has become a prominent ingredient in Carrollian formulations of asymptotically flat gravity and in approaches to flat-space holography \cite{Ruz26}. We mention these developments to emphasize the broader interest in field theories at null infinity. The correspondence studied in the present work is, however, an algebraic construction for scalar quantum fields on a prescribed background. It is not our goal neither to assume nor to establish a duality for gravitational degrees of freedom.

For a massless, conformally coupled, real scalar field, the construction is well-understood. Under suitable hypotheses on the conformal completion, the rescaled solutions of the Klein--Gordon equation admit a restriction to $\mathscr I^+$, yielding an injective symplectic map from the space of bulk observables into a boundary counterpart. Upon quantization, this induces an injective $*$-homomorphism between the corresponding field algebras \cite{DMP_2006,DMP_2017}. On the boundary, the universal BMS structure allows one to single out a quasifree state, which is unique in its folium. Its pullback to the bulk satisfies the Hadamard condition, thereby providing a geometrically motivated selection of a physically admissible state even in the absence of a preferred bulk vacuum \cite{Moretti_2006,DMP_2017}. The usefulness of this perspective is not limited to the construction of Hadamard states since it has also been exploited in the analysis of modular Hamiltonians, relative entropy and energy inequalities on suitable asymptotically flat backgrounds \cite{DMMR25}.

The situation changes considerably when a non-vanishing mass is introduced. The mass term breaks conformal covariance and the rescaled massive Klein--Gordon equation develops a singular coefficient at the conformal boundary. In particular, ordinary radiation data on $\mathscr I^+$ cannot be expected to retain the information carried by a massive field, whose asymptotic propagation is associated instead with timelike infinity. This obstruction does not exclude other procedures for representing massive data at null infinity. In \cite{Dappiaggi_2008}, for example, harmonic analysis on the mass hyperboloid and global estimates for the wave equation were used to associate boundary data with norm-finite massive solutions on Minkowski spacetime. A different, recent development is \cite{Mol26}, where microlocal scattering methods yield distinguished incoming and outgoing Hadamard states for massive Klein--Gordon fields on non-trapping asymptotically flat spacetimes. The latter approach concerns asymptotic positive- and negative-frequency data and scattering, with particular attention to timelike infinity. Neither construction amounts to restricting the massive field itself to the ordinary massless radiation space at $\mathscr I^+$.

In this paper we follow a different route, based on the framework of perturbative algebraic quantum field theory (pAQFT) \cite{Rejzner_2016}. The starting point is the Principle of Perturbative Agreement \cite{Hollands_2005,Drago_2016}, which allows us to regard the mass term as a quadratic interaction of the massless conformally coupled theory. At the level of linear observables, the corresponding retarded Møller map generates an expansion in powers of $m^2$. Every coefficient is an observable of the massless theory and, therefore, it can be projected to $\mathscr I^+$ using the established bulk-to-boundary map. This gives a formal boundary expansion on general backgrounds satisfying our standing hypotheses. To address its convergence and the removal of the infrared cut-off, we then specialize to four-dimensional Minkowski spacetime $\mathbb{M}^4$.

A first result is that, for a smooth compactly supported mass cut-off, the projected series converges in the topology induced by a suitable energy norm on the conformal boundary. Moreover, the resulting map is injective on the on-shell observables of the theory with cut-off mass and it preserves their symplectic form. Nevertheless, convergence in the presence of a cut-off does not entail necessarily the existence of a non-trivial global adiabatic limit. As the cut-off approaches the constant function $1$, we prove that the projected observables converge to zero in the distributional topology on $\mathscr I^+$. This is consistent with the absence of ordinary massless radiation data for a genuinely massive field.

The vanishing of the global limit suggests that a different use of the algebraic structure is needed. To this end, we exploit the time-slice axiom. Let $\mathcal N\subset\mathbb M^4$ be a globally hyperbolic, causally convex neighbourhood containing a Cauchy hypersurface, and let $\mathsf{g}$ be a smooth, past-compact switching function such that $\mathsf{g}|_{\mathcal N}=1$. The massive field algebra on $\mathcal N$ is isomorphic, by the time-slice property, to the algebra on the whole spacetime. We construct a map $\Gamma_{\mathsf g,\mathcal N}$ which assigns to every massive observable localized in $\mathcal N$ a boundary observable at $\mathscr I^+$. We prove that this map is injective and symplectic and that its image coincides with that of the ordinary massless bulk-to-boundary map. In this precise sense, the cut-off can be removed locally without trivializing the correspondence. We stress, however, that the identification of individual massive observables with their boundary representatives can still depend on the switching function.

The same construction provides a way of selecting states for the massive theory. Composing $\Gamma_{\mathsf g,\mathcal N}$ with the two-point function of the distinguished BMS-invariant boundary state yields a quasifree state for the massive Klein--Gordon field on $\mathcal N$. The Hadamard property follows by comparison with the massless Minkowski vacuum in a Cauchy region preceding the switching and by propagation of singularities, in agreement with the extended Møller construction of \cite{Dappiaggi:2015xba}. The time-slice axiom induces a global quasifree Hadamard state for the massive theory on $\mathbb M^4$. In general, this state depends on the cut-off and it does not coincide with the massive Poincaré-invariant ground state. A suitable adiabatic limit at the level of states, as investigated in \cite{Dappiaggi:2015xba}, is a separate problem from the limiting behaviour of the projected observables considered here.

Subsequently, we investigate the projection of observables in a self-interacting massless, scalar field theory. In this case, the Bogoliubov map expresses interacting observables as formal power series in the coupling constant, whose coefficients belong to the free theory. Hence, these can be analyzed using the massless bulk-to-boundary correspondence. Unlike the quadratic perturbation considered above, no convergence of this expansion is claimed. Furthermore, the coefficients involve time-ordered products and hence they require a renormalization prescription \cite{Hollands_2001,HollandsWald2002,Rejzner_2016}. We discuss how the corresponding finite renormalization freedom is inherited by the boundary image, without identifying it with an independent local renormalization procedure intrinsic to $\mathscr I^+$. As a concrete example, we consider a quartic interaction and we compute the boundary projection of the improved stress-energy tensor through first order in the coupling, separating contributions which survive in the classical limit from the genuinely quantum ones. The resulting observables are in general non-local at null infinity, even when their bulk counterparts are local.

We stress that we focused on real scalar fields only for the sake of convenience, but we expect that our construction can be generalized to higher spin, massive fields. As a matter of fact, the two key ingredients of our framework are the Principle of Perturbative Agreement and the existence of a bulk-to-boundary correspondence in the massless case, which identifies in turn a distinguished state on $\mathscr{I}^+$. While the first one is valid for all covariant theories, the second one has been investigated in the past years for spin $\frac12$ \cite{Hack:2010iw}, $1$ \cite{Dappiaggi:2011cj} and $2$ \cite{Benini:2014rya}. 

\paragraph*{Synopsis --} The paper is organized as follows. In Section 2 we review the geometric and algebraic framework underlying the bulk-to-boundary correspondence for massless, conformally coupled scalar fields on asymptotically flat spacetimes, together with the main ingredients of perturbative algebraic quantum field theory.

In Section 3 we study the projection of massive scalar field observables to future null infinity by treating the mass term as a quadratic perturbation. Specializing to four-dimensional Minkowski spacetime, we establish convergence of the projected series in the presence of an infrared cut-off and show that its global adiabatic limit vanishes. Subsequently we exploit the time-slice axiom to construct a non-trivial local bulk-to-boundary correspondence, proving its symplectic properties and establishing the existence of induced quasifree Hadamard states for the massive theory.

Eventually, in Section 4, we extend the bulk-to-boundary construction to perturbatively interacting scalar fields, discussing the associated renormalization freedom. Focusing on a quartic self-interaction, we compute the first-order correction to the projected stress-energy tensor, identifying both classical and genuinely quantum contributions, discussing in addition the non-local character of the resulting boundary observables.

\section{Geometric and algebraic preliminaries}\label{Sec: Geometric and algebraic preliminaries}

In this section we fix the notation and conventions used throughout the paper and establishes some preliminary notions needed for the ensuing discussion. For a more comprehensive account of the underlying geometric and algebraic structures, we refer the reader to \cite{Baer_2008, DMP_2017, Rejzner_2016} and \cite{Wald_1984}.

\subsection{Asymptotically Flat Spacetimes at Null Infinity} 
\label{Sec: Asymptotically Flat Spacetimes}
Henceforth, let $(\mcM,g)$ be a four-dimensional, oriented and time-oriented globally hyperbolic spacetime with empty boundary, in the sense of \cite[Def.~1.3.8]{Baer_2008}. In particular, throughout this work, $g$ denotes a Lorentzian metric on $\mcM$ with signature $(-,+,+,+)$. As it is customary in the literature, we shall refer to $(\mcM, g)$ as the \emph{physical} or \emph{bulk spacetime}. For any open subset $\mathcal{O}\subset\mcM$, we denote by $J^\pm(\mathcal{O}; \mcM)$ its causal future/past, defined as the union of the causal futures/pasts of all points in $\mathcal{O}$ with respect to the causal structure of $\mcM$. Moreover, let $\Sigma\subset\mcM$ be a Cauchy hypersurface, namely a closed, achronal subset such that its domain of dependence $D(\Sigma)=\mcM$. In this work we shall consider asymptotically flat spacetimes, which, heuristically speaking, are Lorentzian manifolds admitting an asymptotic region along null directions in which the metric approaches the Minkowski one with a prescribed fall-off. In the following we shall give a precise definition following \cite{Friedrich_1986b}, requiring in addition global hyperbolicity. Observe that our definition slightly differs from others present in the literature, see for instance, \cite[Ch. 2]{DMP_2017} and \cite[Ch. 1--8]{Wald_1984}. The reason lies in the necessity, for technical reasons, of requiring the existence of future timelike infinity as a distinguished point of the conformal diagram of the underlying background, in contrast to other approaches who require instead the existence of spatial infinity. 

\begin{definition}
\label{Def: asymptotically flat spacetimes}
A physical spacetime $(\mcM, g)$ is called {\bf asymptotically flat (at future null infinity with future timelike infinity)} if there exist
\begin{itemize}
    \item[\ding{104}] a four-dimensional spacetime $(\tilde{\mcM}, \tilde{g})$, called \emph{unphysical spacetime}, and a preferred point therein $i^+ \in \tilde{\mcM}$ dubbed \emph{future timelike infinity}.
    \item[\ding{104}] a smooth embedding $\iota: \mcM \hookrightarrow \iota(\mcM) \subseteq \tilde{\mcM}$, such that $\iota(\mcM)$ is an open subset of $\tilde{\mcM}$. 
    \item[\ding{104}] a smooth, strictly positive function $\Omega \in C^\infty(\iota(\mcM); (0, \infty))$ such that $\iota^* \tilde{g} = (\Omega^2 \iota)^2 g$. Such $\Omega$ is called \emph{conformal factor}.
\end{itemize}
Furthermore, the following conditions must hold true: 
\begin{enumerate}
    \item $J^-(i^+; \tilde{\mcM})$ is a closed set and $\iota(\mcM) \doteq J^-(i^+; \tilde{\mcM}) \setminus \partial J^-(i^+; \tilde{\mcM}) \subseteq \tilde{\mcM}$ is a manifold with boundary. Furthermore, $\partial \iota(\mcM) = \mathscr{I}^+ \cup \{i^+\}$, where $\mathscr{I}^+ \doteq \partial J^-(i^+; \tilde{\mcM}) \setminus \{i^+\}$ is a three-dimensional submanifold of $\tilde{\mcM}$ satisfying $$\mathscr{I}^+ \cap J^-(\iota(\mcM, \tilde{\mcM}) = \emptyset.$$ We call $\mathscr{I}^+$ {\bf future null infinity} of $(\mcM, g)$. 
    \item $(\tilde{\mcM}, \tilde{g})$ is strongly causal in an open neighbourhood of $\mathscr{I}^+$, \textit{cf.}, \cite[Def. 1.3.7]{Baer_2008}.
    \item $\Omega$ admits a non-unique extension, still denoted by the same symbol, to a smooth function defined on the whole $\tilde{\mcM}$ which vanishes on $\mathscr{I}^+$ and such that $d\Omega \vert_{\mathscr{I}^+} \ne 0$ and $d \Omega (i^+)= 0$. Moreover, it holds that $\tilde{\nabla}_\mu \tilde{\nabla}_\nu \Omega(i^+) = - 2 \tilde{g}_{\mu \nu}(i^+)$.
    \item There exists a smooth function $\omega \in C^\infty(\tilde{\mcM})$ that is strictly positive on $\mcM \cup \mathscr{I}^+$ such that, denoting by $n \in T\tilde{\mcM}$ the vector field which reads in local components $n^\mu = \tilde{g}^{\mu \nu} \partial_\nu \Omega$, $\tilde{\nabla}_\mu (\omega^4 n^\mu) \vert_{\mathscr{I}^+} = 0$ and the integral curves of $\omega^{-1} n$ are complete on $\mathscr{I}^+$.
    \item The vacuum Einstein equations $R_{\mu \nu} = 0$ are fulfilled for $(\mcM, g)$ in a neighbourhood of $\mathscr{I}^+$.
\end{enumerate}
\end{definition}
\noindent We observe that Item $4.$ implies that $\mathscr{I}^+$ is diffeomorphic to $\mathbb{R} \times \mathbb{S}^2$, with the second factor corresponding to the range of the parameter of the integral curves of $\omega^{-1} n$. Furthermore, With a slight abuse of notation, we shall henceforth write $\mcM$ in place of $\iota(\mcM)$, leaving the embedding understood. 

\begin{remark}
For completeness, we note that one may similarly define a spacetime which is {\bf asymptotically flat at past null infinity with past time infinity} by replacing $\mathscr{I}^+$ and $i^+$ in the above definition by \emph{past null infinity} $\mathscr{I}^-$ and \emph{past timelike infinity} $i^-$, respectively. In this case, the embedded physical spacetime lies in the causal future of $i^-$. Furthermore, one may define a spacetime which is {\bf asymptotically flat at null and spatial infinity} by adjoining to $\tilde{\mcM}$ a point $i_0$, referred to as the \emph{spatial infinity of $\mcM$}. We shall not discuss this extension further, since the geometric structure near $i_0$ involves additional subtleties that lie beyond the scope of the present work. We refer to \cite{Wald_1984} for a more detailed analysis.
\end{remark}

Note that the conformal completion of an asymptotically flat spacetime as per Definition \ref{Def: asymptotically flat spacetimes} is non-unique. Indeed, in a neighbourhood of $\mathscr{I}^+$, one may rescale the conformal factor by $\Omega \mapsto \omega \Omega$, where $\omega$ is smooth and strictly positive. If $\widetilde h$ denotes the degenerate metric induced on $\mathscr{I}^+$ by $\widetilde g$ and $n^\mu=\widetilde g^{\mu \nu}\widetilde\nabla_\nu\Omega$, then the corresponding boundary data transform as $\widetilde h\mapsto \omega^2\widetilde h$, $n\mapsto \omega^{-1}n.$ Therefore, the intrinsic geometry of $\mathscr{I}^+$ is naturally described by the equivalence class $[(\mathscr{I}^+,\widetilde h,n)]$ under these gauge transformations. A key property of this structure is its universality. More precisely, the future null infinities associated with any pair of asymptotically flat
spacetimes are diffeomorphic in such a way that suitable representatives of their boundary data are identified. This follows from the existence of a \emph{Bondi gauge}. Fixing $\omega$ as per Item 4 of Definition \ref{Def: asymptotically flat spacetimes}, and denoting the
gauge-transformed quantities again by $\Omega$, $\widetilde g$ and $n$, one can introduce coordinates $(u,\Omega,\theta,\phi)$ in a neighbourhood of $\mathscr{I}^+$ such that
\begin{equation*}
    \left.\widetilde g\right|_{\mathscr{I}^+} =
    -\bigl(d\Omega\otimes du+du\otimes d\Omega\bigr) +d\mathbb{S}^2(\theta,\phi),
\end{equation*}
where $u\in\mathbb{R}$ is an affine parameter along the complete null
generators and $n=\partial_u$. Thus
\begin{equation*}
    (\mathscr{I}^+,\widetilde h,n) \simeq
    (\mathbb{R}\times\mathbb{S}^2,d\mathbb{S}^2,\partial_u).
\end{equation*}
The coordinates $(u,\theta,\phi)$ define a \emph{Bondi frame}, see, \textit{e.g.}, \cite[Sec.~11.1]{Wald_1984}. Most notably, the universal structure of $\mathscr{I}^+$ admits a distinguished symmetry group, the {\bf Bondi--Metzner--Sachs (BMS) group} $G_{\mathrm{BMS}}$ \cite{Bondi_1962, Sachs_1962}. It consists of those diffeomorphisms
$\gamma:\mathscr{I}^+\to\mathscr{I}^+$ which preserve the boundary data up to the above gauge transformations, namely $\gamma^*\widetilde h=\omega^2\widetilde h,$ and $\gamma^*n=\omega^{-1}n$, for some smooth, strictly positive $\omega$. For later use, we briefly describe an explicit realization of $G_{\mathrm{BMS}}$, following \cite[Sec.~2.2.3]{DMP_2017}. 

\begin{remark}
\label{Rmk: action of BMS group}
    Let us consider a fixed Bondi coordinate system $(u, \theta, \phi)$, where $u$ is the affine parameter along the null integral curves of $\mathscr{I}^+$, while $(\theta, \phi) \in \mathbb{S}^2$. Via a stereographic projection, we introduce the complex coordinates $(\zeta, \bar{\zeta}) \in \mathbb{C}^2$ such that $\zeta = e^{i\phi} \cot (\frac{\theta}{2})$. Observe that, in such frame, the set $G_{\mathrm{BMS}}$ coincides with $SO(3, 1)^\uparrow \rtimes C^\infty(\mathbb{S}^2)$, where $SO(3,1)^\uparrow$ is the proper orthocronous Lorentz group in four dimensions. We may thus specify the action of a generic element $(\Lambda, f) \in SO(3, 1)^\uparrow \times C^\infty(\mathbb{S}^2)$ on $(u, \zeta, \bar{\zeta})$ as 
    \begin{flalign*}
        u &\mapsto u' \doteq K_\Lambda (\zeta, \bar{\zeta})(u+ f(\zeta, \bar{\zeta})), \\
        \zeta &\mapsto \zeta' \doteq \Lambda \zeta \doteq \frac{a_\Lambda \zeta + b_\Lambda}{c_\Lambda \zeta + d_\Lambda}, \quad \bar{\zeta} \mapsto \bar{\zeta}' \doteq \Lambda \bar{\zeta} \doteq \frac{\overline{a_\Lambda} \bar{\zeta} + \overline{b_\Lambda}}{\overline{c_\Lambda} \bar{\zeta} + \overline{d_\Lambda}}, 
    \end{flalign*}
    where \begin{equation*}
        K_\Lambda(\zeta, \bar{\zeta}) \doteq \frac{(1+\zeta \bar{\zeta})}{(a_\Lambda \zeta + b_\Lambda) (\overline{a_\Lambda} \bar{\zeta} + \overline{b_\Lambda}) + (c_\Lambda \zeta + d_\Lambda) (\overline{c_\Lambda} \bar{\zeta} + \overline{d_\Lambda})}, \qquad \Pi^{-1} (\Lambda) = \begin{pmatrix}
            a_\Lambda & b_\Lambda \\
            c_\Lambda & d_\Lambda
        \end{pmatrix},
    \end{equation*}
    with $a_\Lambda, b_\Lambda, c_\Lambda, d_\Lambda \in \mathbb{C}$ and $a_\Lambda d_\Lambda - b_\Lambda c_\Lambda = 1$, $\Pi$ being the surjective covering homomorphism from $SL(2, \mathbb{C})$ to $ SO(3,1)^\uparrow$.  
\end{remark}

A direct consequence of the above remark is that, in a Bondi frame,
\begin{equation}
\label{Eq: GBMS}
    G_{\mathrm{BMS}} \simeq SO(3,1)^\uparrow\ltimes C^\infty(\mathbb{S}^2). 
\end{equation}
Given two different Bondi frames $(u, \zeta, \bar{\zeta})$ and $(u', \zeta', \bar{\zeta}')$ on $\mathscr{I}^+$ and denoting by $(\Lambda, f)$ and $(\Lambda', f')$ respectively the explicit realizations of an element $\gamma\in G_{\mathrm{BMS}}$ in each frame as in Remark \ref{Rmk: action of BMS group}, one has
\begin{equation*}
    (\Lambda', f') = (R,g)^{-1} \odot (\Lambda, f) \odot (R,g), 
\end{equation*}
where $g \in C^\infty(\mathbb{S}^2)$, $R \in SO(3) \hookrightarrow SO(3,1)^\uparrow$, while $\odot$ denotes the product in $G_{\mathrm{BMS}}$, see \cite[Prop. 2.22]{DMP_2017}. As recalled in Section \ref{Sec: The bulk to boundary correspondence}, the induced action of $G_{\mathrm{BMS}}$ on the boundary algebra will be used to select a distinguished quasifree state on $\mathscr{I}^+$.

\begin{example}
\label{Ex: Minkowski and ESU}
Four-dimensional Minkowski spacetime $(\mathbb{M}^4, \eta)$ is conformally embedded into the Einstein static universe, which plays here the r\^ole of the unphysical spacetime $\tilde{\mcM}$ of Definition \ref{Def: asymptotically flat spacetimes}. Indeed, under the following coordinate transformations, 
\begin{equation*}
(t,r) \longmapsto (u \doteq t-r, v \doteq t+r) \longmapsto (U \doteq \arctan u, V \doteq \arctan v), 
\end{equation*}
where the corresponding domains have been omitted for the sake of brevity, and choosing the conformal factor $\Omega=2\cos U\cos V$, the rescaled metric $\widetilde{\eta}=\Omega^2\eta$ becomes, in the
coordinates $T \doteq U+V$ and $R \doteq V-U$,
\begin{equation*}
    \widetilde{\eta} = -dT^2+dR^2+\sin^2R\,d\mathbb{S}^2(\theta, \phi).
\end{equation*}
This is precisely the metric of $\mathbb{R}\times\mathbb{S}^3$, see, \textit{e.g.}, \cite{Wald_1984} for the full construction. Observe that in these coordinates, the embedding $\iota: \mathbb{M}^4 \to \mathbb{R} \times \mathbb{S}^3$ maps Minkowski spacetime into the open set:
\begin{equation}
    \label{Eq: embedding of Minkowski}
    \iota(\mathbb{M}^4) = \{(T, R, \theta, \phi) \in \mathbb{R} \times \mathbb{S}^3 \, : \, -\pi < T \pm R < \pi, R> 0\} \subset \mathbb{R} \times \mathbb{S}^3.
\end{equation}
In particular, the boundary of $\overline{\iota(\mathbb{M}^4)}$ decomposes as 
\begin{equation}
    \label{Eq: conformal boundary of Einstein static universe}
    \mathscr{C} \doteq  \partial \overline{\iota(\mathbb{M}^4)} = \mathscr{I}^+ \cup \mathscr{I}^- \cup i^+ \cup i^- \cup i_0, 
\end{equation}
where 
 \begin{equation*}
        \mathscr{I}^\pm =\{(T, R, \theta, \phi) \in \mathbb{R} \times \mathbb{S}^3 \, : \, T\pm R=\pm\pi, \, 0<R<\pi\}, 
    \end{equation*}
denote future and past null infinity, 
    \begin{equation*}
        i^\pm =\{(T, R, \theta, \phi) \in \mathbb{R} \times \mathbb{S}^3 \, : \, R= 0, T = \pm \pi\}, 
    \end{equation*}
are, respectively, future and past timelike infinity, while 
    \begin{equation*}
        i_0 = \{(T, R, \theta, \phi) \in \mathbb{R} \times \mathbb{S}^3 \, : \, R= \pi, T = 0\}. 
    \end{equation*}
denotes spatial infinity. 
\end{example}

To conclude the section, in the same spirit of \cite[Thm.~4.1]{Moretti_2006}, we make a further assumption which will play a distinguished role in our construction.

\begin{assumption}\label{Ass: V}
   There exists an open subset $\mathscr V\subseteq\widetilde{\mcM}$ containing $\iota(\mcM)$ such that $(\mathscr V,\widetilde g|_{\mathscr V})$ is globally hyperbolic, where $\iota: \mcM \hookrightarrow \tilde{\mcM}$ denotes the conformal embedding as per Definition \ref{Def: asymptotically flat spacetimes}. 
\end{assumption}
\noindent This hypothesis shall guarantee that, upon a conformal rescaling of the metric of a physical spacetime, the massless, conformally coupled Klein--Gordon equation, which we will introduce in Equation \eqref{Eq: conformally transformed KG}, admits a counterpart on $\mathscr{V}$. In addition herein solutions can be constructed out of globally defined advanced and retarded Green operators, providing the geometric framework for the bulk-to-boundary correspondence, the main goal of this paper.

\begin{remark}
Assumption~\ref{Ass: V} is stronger than the corresponding hypothesis in \cite[Prop.~2.3]{DMP_2006}. Therein, it is required the existence of an open subset $\mathscr{V}\subseteq\widetilde{\mcM}$ such that $$\overline{J^-(\mathscr{I}^+; \tilde{\mcM}) \cap \iota(\mcM)}^{\tilde{\mcM}} \subset \mathscr{V}, \qquad (\mathscr{V}, \tilde{g} \vert_\mathscr{V}) \; \text{is globally hyperbolic}.$$ Notice that this condition does not necessarily imply $\iota(\mcM) \subset \mathscr{V}$. The advantage of the above formulation is hence two-fold. On the one hand, it guarantees that every compactly supported test function on $\mcM$, including those generated by the M\o ller expansion, defines a test function on the globally hyperbolic unphysical region $\mathscr V$. On the other hand, it ensures well-definiteness of the bulk-to-boundary map on solutions with compactly supported Cauchy data, see \cite[Lem. 4.2]{Moretti_2006}. 
\end{remark}

\subsection{The perturbative algebraic framework}\label{Sec: The perturbative algebraic framework}

\paragraph{Free theory --} On top of an asymptotically flat, physical spacetime $(\mcM, g)$ abiding by Assumption \ref{Ass: V}, we consider a massless, real scalar field $\varphi: \mcM \to \mathbb{R}$, whose dynamics is ruled by the conformally coupled Klein--Gordon operator
\begin{equation}
\label{Eq: massless KG op}
    P_0 \doteq \Box_g - \frac{R(x)}{6},  
\end{equation}
where $\Box_g \doteq g^{\mu \nu} \nabla_\mu \nabla_\nu$ is the wave operator built out of $g$, while $R(x)$ is the scalar curvature of the underlying manifold. Henceforth, we shall denote the space of field configurations by $\mathcal{E}(\mcM) \doteq C^\infty(\mcM; \mathbb{R})$ equipped with the standard Fréchet topology, \textit{cf.}, \cite[Def. 3.2]{Rejzner_2016} and by $\mathcal{D}(\mcM) \doteq C^\infty_0(\mcM)$ the locally convex space of compactly supported smooth test functions on $\mcM$, \textit{cf.}, \cite[Def. 3.3]{Rejzner_2016}. In the algebraic approach to quantum field theory, physical observables are characterized as suitable functionals on the space of configurations. We shall denote by:
\begin{itemize}
    \item[\ding{104}] $\mathcal{F}(\mcM)$ the space of all (Bastiani) smooth functionals $F:\mathcal{E}(\mcM) \to \mathbb{C}$ in the sense of \cite[Def. 3.4 \& Prop. 3.1]{Rejzner_2016}, and by $F^{(k)}: \mathcal{E}(\mcM) \times \mathcal{E}(\mcM)^k \to \mathbb{C}$ its $k$-th order Fréchet derivative.
    \item[\ding{104}] $\mathcal{P}(\mcM) \subset \mathcal{F}(\mcM)$ the space of smooth \emph{polynomial} functionals, namely, the collection of those functionals with only a finite number of non-vanishing Fréchet derivatives. 
\end{itemize}
Given $F \in \mathcal{F}(\mcM)$, its \emph{spacetime support} is defined as 
\begin{flalign}
    \label{Eq: spacetime support}
    \notag
    \operatorname{supp}(F) \doteq \{p \in \mcM \, &: \, \forall U \, \text{open neighbourhood of} \, p, \exists \psi, \eta \in \mathcal{E}(\mcM), \\& \operatorname{supp}(\psi) \subset U, \text{s.t.} \, F(\eta + \psi) \ne F(\eta)\}. 
\end{flalign}
Relevant subclasses of functionals can be singled out by imposing suitable conditions on the wavefront sets of their functional derivatives, ensuring that their contractions with the relevant distributional kernels are well-defined by H\"ormander's criterion. More precisely, we call $F \in \mathcal{F}(\mcM)$
\begin{itemize}
    \item[\ding{104}] a \emph{microcausal functional} if, for all $\varphi \in \mathcal{E}(\mcM)$ and for any $k \in \mathbb{N}$, 
    \begin{equation}\label{Eq: Microcausal Functionals}
        \operatorname{WF}(F^{(k)}(\varphi)) \subset T^*(\mcM^k) \setminus \left( \bigcup_{x \in \mcM} (\overline{V}^+_x)^k \cup \bigcup_{x \in \mcM} (\overline{V}^-_x)^k \right), 
    \end{equation}
    where $\overline{V}^\pm_x$ are the closed future $(+)$ and past $(-)$ lightcones in $T^*_x \mcM$. We denote the space of microcausal functionals by $\mathcal{F}_{\mu c}(\mcM)$ and that of microcausal polynomial functionals by $\mathcal{P}_{\mu c} (\mcM)$. 
    \item[\ding{104}] a \emph{regular functional} if, for all $\varphi \in \mathcal{E}(\mcM)$ and $k \in \mathbb{N}$, $F^{(k)}(\varphi) \in \mathcal{D}(\mcM^k) \hookrightarrow \mathcal{E}'(\mcM^k)$. The space of regular functionals and that of regular polynomial functionals will henceforth be denoted by $\mathcal{F}_\mathrm{reg}(\mcM)$ and by $\mathcal{P}_\mathrm{reg}(\mcM)$.  
    \item[\ding{104}] a \emph{local functional} if, for all $\varphi \in \mathcal{E}(\mcM)$, $\operatorname{WF}(F^{(1)}(\varphi))= \emptyset$ and
    $$\operatorname{supp}(F^{(k)} (\varphi)) \subset \operatorname{Diag}_k(\mcM) \doteq \{(x, \ldots, x) \in \mcM^k\}, \qquad \forall \varphi \in \mathcal{E}(\mcM), \, \forall k \in \mathbb{N}.$$ We denote the space of local functionals by $\mathcal{F}_{\mathrm{loc}}(\mcM)$ and that of local polynomial ones by $\mathcal{P}_\mathrm{loc}(\mcM)$. 
\end{itemize}

\begin{remark}
\label{Rem: functionals and integral kernels}
    For future convenience, let us remark that, given any microcausal polynomial functional $F \in \mathcal{P}_{\mu c}(\mcM)$, its action on a field configuration can be fully characterized in terms of a (finite) collection of linear functionals $\{F_k\}_{k=1}^N$ by writing
    \begin{equation*}
        F[\varphi] = \sum_{k=0}^N \frac1{k!} \langle F_k, \varphi^{\otimes k}\rangle_k, \qquad \varphi \in \mathcal{E}(\mcM), \; N \in \mathbb{N}, 
    \end{equation*}
    where $F_k \in \mathcal{E}'(\mcM^k)$ is the compactly supported distribution given by $\langle F_k, h^{\otimes k} \rangle_k = F^{(k)} (0) (h^{\otimes k})$ for any $h \in \mathcal{E}(\mcM)$, while $\langle \cdot, \cdot \rangle_k$ is the standard pairing in $\mcM^k$ between smooth and smooth compactly supported functions.
\end{remark}

With these definitions at hand, we call {\bf classical microcausal algebra} the commutative, unital, $\ast$-algebra $\mathcal{A}_{\mathrm{cl}}(\mcM) = (\mathcal{P}_{\mu c}(\mcM), \cdot, \ast)$ of microcausal polynomial functionals over $\mathcal{E}(\mcM)$, where $\cdot$ denotes the \emph{pointwise product} of functionals defined in agreement with \cite[Eq. (1.7)]{Goi_2026}, while $\ast$ is the involution defined via the relation $F^*(\varphi) = \overline{F(\varphi)}$. The dynamical content of the theory is encoded in the following action
\begin{equation}
    \label{Eq: free action on M}
    S_0[f](\varphi) \doteq -\frac12 \int_\mcM d\mu_x \, \left(g^{\mu \nu}(x) \nabla_\mu \varphi(x) \nabla^\nu \varphi(x) + \frac{R(x)}{6} \varphi^2(x) \right) \, f(x), \quad \varphi \in \mathcal{E}(\mcM), f \in \mathcal{D}(\mcM). 
\end{equation}
The corresponding equation of motion \cite[Def. 4.4]{Rejzner_2016} is precisely the conformally coupled Klein-Gordon equation. Most notably, since $P_0$ is normally hyperbolic, it admits unique {\bf advanced ($\mathrm A$)} and {\bf retarded ($\mathrm R$) Green operators} or {\bf propagators}, namely, continuous maps 
\begin{equation}
    \label{Eq: adv and ret Green operators}
    \Delta_{\mathrm A/ \mathrm R}: \mathcal{D}(\mcM) \to \mathcal{E}(\mcM), 
\end{equation}
satisfying 
\begin{equation*}
    \begin{cases}
        P_0 \circ \Delta_{\mathrm A/ \mathrm R} = \Delta_{\mathrm A/ \mathrm R} \circ P_0 \vert_{\mathcal{D}(\mcM)} = \operatorname{id} \vert_{\mathcal{D}(\mcM)} \\
        \operatorname{supp} (\Delta_{\mathrm A/ \mathrm R}(f)) \subseteq J^{\mp} (\operatorname{supp} (f)), \qquad \forall f \in \mathcal{D}(\mcM), 
    \end{cases}
\end{equation*}
see \cite[Thm. 3.3.1 and Cor. 3.4.3]{Baer_2008}. We stress that it is also possible to extend the action of the advanced and retarded Green operators respectively to the space of future and past compact test functions $C^\infty_{\mathrm{fc}/ \mathrm{pc}}(\mcM)$. With a slight abuse of notation, we shall denote such extensions by the same symbol. Notice, in addition, that the formal adjoint of the retarded Green operator $\Delta_\mathrm{R}$ with respect to the standard pairing between $\mathcal{E}(\mcM)$ and $\mathcal{D}(\mcM)$ is the advanced Green operator $\Delta_{\mathrm{A}}$, and viceversa. The existence of the advanced and retarded Green operators allows us to define the {\bf causal (Pauli-Jordan) propagator} as 
\begin{equation}
    \label{Eq: causal propagator}
    \Delta_\mathrm{S}: \mathcal{D}(\mcM) \to \mathcal{E}(\mcM), \qquad \Delta_\mathrm{S} \doteq \Delta_\mathrm{R}- \Delta_\mathrm{A}, 
\end{equation}
which is formally skew adjoint with respect to the pairing $\langle \cdot, \cdot \rangle: \mathcal{E}(\mcM) \times \mathcal{D}(\mcM) \to \mathbb{C}$. Furthermore, for any $f \in \mathcal{D}(\mcM)$, it holds that $P_0 \Delta_\mathrm{S}(f) = 0$. 

\begin{remark}
The Schwartz kernel theorem applied to the advanced and retarded Green operators yield two distinct bidistributions, denoted with a mild abuse of notation by the same symbol $\Delta_{\mathrm{A}/\mathrm R} \in \mathcal{D}'(\mcM \times \mcM)$, known in the literature as {\bf advanced ($\mathrm A$) and retarded ($\mathrm R$) fundamental solutions}. Similarly, we denote by $\Delta_\mathrm{S} \in \mathcal{D}'(\mcM \times \mcM)$ the bidistribution corresponding to the causal propagator introduced above. We shall henceforth suppress the distinction between Green operators and propagators, specifying the notation only whenever ambiguity may arise.   
\end{remark}

\begin{remark}
\label{Rem: conformal invariance of P0}
The reason for considering a specific coupling with the scalar curvature in Equation \eqref{Eq: massless KG op} lies in the fact that, in this way, the resulting operator and, hence, the equation of motion is conformally invariant. More precisely, given $\Omega \in C^\infty(\mcM, (0, \infty))$ so that Assumption \ref{Ass: V}  holds true, if we denote by $\tilde{P}_0$ the Klein--Gordon operator associated to the conformally rescaled metric tensor $\tilde{g} = \Omega^2 g$, by $\tilde{\Delta}_{\mathrm{A}/\mathrm R}$ the advanced and retarded Green operators and by $\tilde{\Delta}^\mathscr{V}_\mathrm{S}$ the corresponding causal propagator of $\tilde{P}_0$ on $\mathscr{V}$, then it holds that \cite{Pinamonti:2009zqj}
\begin{equation}
    \label{Eq: conformally transformed KG}
\tilde{P}_0 = \Omega^{-3} P_0 \Omega, \qquad
        \tilde{\Delta}_{\mathrm{A}/\mathrm R}|_{\iota(\mcM)} = \Omega^{-1}{\Delta}_{\mathrm{A}/\mathrm R} \Omega^3, \qquad \tilde{\Delta}^\mathscr{V}_\mathrm{S}|_{\iota(\mcM)} = \Omega^{-1} {\Delta}_\mathrm{S} \Omega^3. 
\end{equation}
A notable consequence of this fact is that, if $\varphi \in \mathcal{E}(\mcM)$ is a solution of Equation \eqref{Eq: KG equation}, then $\phi \doteq \Omega^{-1} \varphi$ is a solution of the conformally rescaled Klein--Gordon equation $\tilde{P}_0 \phi = 0$.
\end{remark}

Via the causal propagator we can also give an explicit
characterisation of the solution space for the (conformally coupled) Klein--Gordon equation: Consider the following initial-value Cauchy problem for $\varphi \in \mathcal{E}(\mcM)$
\begin{equation}
    \label{Eq: KG equation}
\begin{cases}
     P_0 \varphi(x) = \left(\Box_g - \frac{R(x)}{6}\right) \varphi(x) = 0 \\
     \varphi \vert_{\Sigma} = \varphi_0, \qquad \nabla_n \varphi \vert_{\Sigma} \equiv n^\mu \partial_\mu \varphi \vert_{\Sigma} = \varphi_1,
\end{cases}
\end{equation}
where $\varphi_0, \varphi_1 \in \mathcal{D}(\Sigma)$ are smooth and compactly supported initial data assigned on a Cauchy hypersurface $\Sigma$, while $n \in \Gamma(T \mcM \vert_{\Sigma})$ is the future-directed, unit normal vector field to $\Sigma$. If we denote by $\mathscr{S}_{\mathrm{sc}}(\mcM) \doteq \{\varphi \in C^\infty_{\mathrm{sc}}(\mcM) \, : \, P_0 \varphi = 0\}$, there exists a vector space isomorphism 
\begin{equation*}
 \frac{C^\infty_0(\mcM)}{P_0[C^\infty_0(\mcM)]}  \ni [f] \mapsto \Delta_\mathrm{S}(f) \in  \mathscr{S}_{\mathrm{sc}}(\mcM),  
\end{equation*}
see \cite[Prop. 3.1.2]{DMP_2017}. Furthermore, denoting by 
\begin{equation}
    \label{Eq: space of observables}
    \mathcal{E}^{\mathrm{obs}}(\mcM) \doteq  \frac{C^\infty_0(\mcM)}{P_0[C^\infty_0(\mcM)]} \equiv \mathscr{S}_{\mathrm{sc}}(\mcM),
\end{equation}
the following facts hold true \cite[Prop. 3.1.3]{DMP_2017}: 
\begin{itemize}
    \item[\ding{104}] The map $\sigma_{\mcM}:  \mathcal{E}^{\mathrm{obs}}(\mcM) \times  \mathcal{E}^{\mathrm{obs}}(\mcM) \to \mathbb{R}$ defined by 
\begin{equation}
    \label{Eq: symplectic form on M}
    \sigma_{\mcM}([f], [f']) \doteq \Delta_\mathrm{S}(f,f') \doteq \int_\mcM f(x) \Delta_\mathrm{S}(f')(x) \, d\mu_g(x), \qquad \forall [f], [f'] \in \mathcal{E}^{\mathrm{obs}}(\mcM), 
\end{equation}
is a well-defined symplectic form which is \emph{weakly non-degenerate}, that is, if $\sigma_{\mcM}([f], [f']) = 0$ for all $[f] \in  \mathcal{E}^{\mathrm{obs}}(\mcM)$, then $f' \in [0]$. 
\item[\ding{104}] Setting $\varphi_f \doteq \Delta_\mathrm{S}(f)$ and $\varphi_{f'} \doteq \Delta_\mathrm{S}(f')$, if $\Sigma$ is a Cauchy surface of $\mcM$ with future-directed unit vector field $n$, 
\begin{equation*}
    \sigma_\mcM([f],[f']) = \int_\Sigma (\varphi_f \nabla_n \varphi_{f'} - \varphi_{f'} \nabla_n \varphi_f) \, d\Sigma_g.
\end{equation*}
where $d \Sigma_g$ is the metric induced measure. 
\end{itemize}

The dynamical information of Equation \eqref{Eq: KG equation} can be encoded algebraically by considering equivalence classes of observables where the defining relation is defined in terms of the equations of motion, as the following definition clarifies.
\begin{definition}
We call {\bf classical on-shell microcausal algebra} (of polynomial functionals) the quotient 
    \begin{equation}
        \label{Eq: classical on-shell algebra}
        \mathcal{A}^{\mathrm{on}}_{\mathrm{cl}}(\mcM) \doteq \frac{\mathcal{A}_{\mathrm{cl}}(\mcM)}{\mathcal{I}}, 
    \end{equation}
where $\mathcal{A}_{\mathrm{cl}}(\mcM)$ is the classical microcausal algebra introduced above, while 
\begin{equation*}
    \mathcal{I} (\mcM) = \{F \in \mathcal{P}_{\mu c} (\mcM) \, : \, F(\varphi) = 0, \forall \varphi \in \mathscr{S}(\mcM)\}, \qquad \mathscr{S}(\mcM) \doteq \{\varphi \in \mathcal{E}(\mcM) \, : \, P_0 \varphi = 0\}, 
\end{equation*}
is the two-sided ideal of the equation of motion.
\end{definition}

\begin{remark}
\label{Rem: relation with Aobs}
The on-shell construction above is closely related to the algebra of observables used in \cite[Sec. 3.1]{DMP_2017}. In that framework one first introduces the real symplectic space $(\mathcal{E}^{\mathrm{obs}}(\mcM), \sigma_\mcM)$, where the symplectic form is as per Equation \eqref{Eq: symplectic form on M}. The algebra $\mathcal{A}^{\mathrm{obs}}(\mcM)$ is then characterized as the complex unital associative $\ast$-algebra 
\begin{equation}
    \mathcal{A}^{\mathrm{obs}}(\mcM) \doteq \frac{\mathcal{T}^{\mathrm{obs}}(\mcM)}{\mathcal{I}^{\mathrm{obs}}(\mcM)},
\end{equation}
where $\mathcal{T}^{\mathrm{obs}}(\mcM)$ is the universal tensor algebra over $\mathcal{E}^{\mathrm{obs}}(\mcM)$ while $\mathcal{I}^{\mathrm{obs}}(\mcM)$ is the two-sided $\ast$-ideal implementing the canonical commutation relations. This algebra is finitely generated by all elements of the form $$\varphi \otimes \varphi' - \varphi' \otimes \varphi - i \sigma_\mcM(\varphi, \varphi')\mathbb{I}, \qquad \forall \varphi, \varphi' \in \mathcal{E}^{\mathrm{obs}}(\mcM),$$ \textit{cf.}, \cite[Def. 3.1.2]{DMP_2017}. In the functional formulation adopted here, $\mathcal{E}^{\mathrm{obs}}(\mcM)$ is naturally identified with the linear sector of $\mathcal{A}_{\mathrm{cl}}^{\mathrm{on}}(\mcM)$ via the correspondence
\begin{equation*}
    \mathcal{E}^{\mathrm{obs}}(\mcM) \ni [f] \longmapsto [\Phi_f] \in \mathcal{A}^{\mathrm{on}}_{\mathrm{cl}}(\mcM),
        \qquad
        \Phi_f(\varphi)
        \doteq
        \int_{\mcM} d\mu_x\, f(x)\varphi(x),
    \end{equation*}
Correspondingly, after deformation quantization, $\mathcal{A}^{\mathrm{obs}}(\mcM)$ can be identified with the on-shell $\ast$-subalgebra generated by the linear observables $[\Phi_f]$ with respect to the product $\star_{\Delta_+}$, see the discussion below. In this sense, the construction of $\mathcal{A}^{\mathrm{obs}}(\mcM)$ is the algebraic counterpart of the on-shell quantization of the regular polynomial sector within the perturbative AQFT framework. Notice, however, that $\mathcal{A}^{\mathrm{obs}}(\mcM)$ does not yet contain generic local non-linear observables, such as Wick polynomials. Their inclusion requires the extended algebra of microcausal functionals, \textit{cf.}, also Remark \ref{Rem: relation with Aobs quantum}.
\end{remark}

We report here a notable property of the on-shell microcausal algebra which will be of paramount relevance in discussing the local adiabatic limit in Section \ref{Sec: Projection of the massive scalar field}, see \cite[Prop. 3.1.5]{DMP_2017}. 

\begin{proposition}[Time-slice axiom]
\label{Prop: time slice}
    Let $\mathcal N \subseteq \mathcal{M}$ be a causally convex open subset containing a Cauchy hypersurface of $\mcM$. Then the inclusion $\iota_\mathcal{N}: \mathcal{N} \hookrightarrow \mcM$ induces an isomorphism of unital $\ast$-algebras $$ \mathcal A^{\mathrm{on}}_{\mathrm{cl}}(\mathcal N; \mathcal{M}) \simeq \mathcal A^{\mathrm{on}}_{\mathrm{cl}}(\mathcal M),$$ where $\mathcal{A}^{\mathrm{on}}_{\mathrm{cl}}(\mathcal N; \mcM)$ is the unital sub-$\ast$-algebra of elements in $\mathcal{A}^{\mathrm{on}}_{\mathrm{cl}}(\mcM)$ which admit a representative supported in $\mathcal N$. 
\end{proposition}

To quantize the theory, we employ the framework of \emph{deformation quantization}, whereby the pointwise product of the classical algebra of microcausal functionals is deformed so as to reproduce the canonical commutation relations (CCR). The key ingredient in this construction is a bidistribution implementing the desired deformation. A particularly convenient choice is the two-point function of a Hadamard state $\Delta_+ \in \mathcal{D}'(\mcM \times \mcM)$, namely, a positive bi-solution of the Klein--Gordon equation satisfying the microlocal spectrum condition, see \cite[Def. 4.1]{Brunetti_1996}, and whose antisymmetric part coincides with $\frac{i}{2}\Delta_{\mathrm{S}}$, \textit{cf.}, \cite[Def. 1.19]{Goi_2026}. The latter condition, in particular, ensures that the resulting deformed product automatically implements the canonical commutation relations. Thus, for any $F,G \in \mathcal{P}_{\mu c} (\mcM)$, we define their deformed product by setting
\begin{equation}
    \label{Eq: deformed product}
    F \star_{\hbar \Delta_+} G = \mathrm{i}^* \circ e^{D_{\hbar \Delta_+}}[F \otimes G], \qquad D_{\hbar \Delta_+} = \hbar \langle \Delta_+, \frac{\delta}{\delta \varphi} \otimes \frac{\delta}{\delta \varphi'} \rangle, 
\end{equation}
where $\mathrm{i}: \mathcal{E}(\mcM) \to \mathcal{E}(\mcM) \times \mathcal{E}(\mcM)$, $\mathrm{i}(\varphi) =(\varphi, \varphi)$ is the diagonal map on the space of field configurations, while the exponential has to be understood as a formal power series in $\hbar$. This allows us to introduce the {\bf quantum microcausal algebra} of polynomial functionals characterized as the unital $\ast$-algebra
\begin{equation}
    \label{Eq: quantum microcausal algebra}
        \mathcal{A}_{\Delta_+} (\mcM) = (\mathcal{P}_{\mu c}(\mcM)[[\hbar]], \star_{\hbar \Delta_+}, \ast).
\end{equation}
Here $\mathcal{P}_{\mu c}(\mcM)[[\hbar]]$ denotes the space of formal power series in $\hbar$ with coefficients in $\mathcal{P}_{\mu c}(\mcM)$. Since at this level we are confining our attention to polynomial microcausal functionals for a free scalar theory, the series is actually a sum of finitely many terms. Notice that, while the antisymmetric part of $\Delta_+$ is fixed by the CCR, its symmetric part depends on the choice of Hadamard state. To disentangle the algebraic construction from this state dependence, we may replace $\Delta_+$ by a Hadamard bidistribution $H$ having the same singular structure and the same antisymmetric part, so that $\Delta_+-H\in C^\infty(\mcM\times\mcM)$. Locally, $H$ coincides, up to smooth contributions, with the \emph{Hadamard parametrix}, whose singular part is determined entirely by the spacetime geometry and by the Klein--Gordon operator. For the construction of the Hadamard parametrix and the associated transport equations for its coefficients, see, \textit{e.g.}, \cite[App.~A]{Moretti_2003} and \cite{Decanini_2008}, while for existence results in globally hyperbolic settings we refer the reader to \cite{Fulling_1981}. This discussion motivates the following definition.

\begin{definition}
\label{Def: algebra microcausal polynomial functionals}
    Let $(\mcM,g)$ be a globally hyperbolic spacetime with empty boundary and let $H\in\mathcal D'(\mcM\times\mcM)$ be a Hadamard bidistribution such that $\Delta_+ - H \in C^\infty(\mcM \times \mcM)$. We call
    {\bf extended algebra of polynomial microcausal functionals} the
    unital $\ast$-algebra
    \begin{equation*}
        \mathcal{A}_e(\mcM)  \doteq
        \bigl(\mathcal P_{\mu c}(\mcM)[[\hbar]],\star_{\hbar H},\ast \bigr).
    \end{equation*}
    The $\ast$-subalgebra generated by the local polynomial functionals  will be denoted by
    \begin{equation*}
        \mathcal A_{\mathrm{loc}}(\mcM)
        \doteq
        \operatorname{Alg}_{\star_{\hbar H}}
        \bigl(\mathcal P_{\mathrm{loc}}(\mcM)\bigr)
        \subseteq
        \mathcal A_e(\mcM), 
    \end{equation*}
where $\operatorname{Alg}_{\star_{\hbar H}}\bigl(\mathcal P_{\mathrm{loc}}(\mcM)\bigr)$ is the algebraic closure of $\mathcal{P}_{\mathrm{loc}}(\mcM)$ with respect to the product $\star_{\hbar H}$. We shall refer to $\mathcal{A}_{\mathrm{loc}}(\mcM)$ as the {\bf locally covariant algebra of microcausal polynomial functionals}. 
\end{definition}

\begin{remark}\label{Rem: relation with Aobs quantum}
This functional realization is ultimately equivalent to the extended algebra of observables $\mathcal A_e^{\mathrm{obs}}(\mcM)$ introduced in \cite[Sec.~5.1]{DMP_2017}. More precisely, their extended algebra is obtained by equipping the (on-shell) algebra of compactly supported microcausal functionals with the Hadamard-deformed product $\star_H$. The only difference relevant here is that the construction in \cite{DMP_2017} is formulated for observables encompassing the Klein--Gordon equation, whereas we shall work off shell whenever convenient switching to the corresponding on-shell equivalence classes when applying the bulk-to-boundary map.
\end{remark}

\begin{remark}
Different admissible choices of $H$ differ only by smooth kernels and, hence, they yield canonically $\ast$-isomorphic deformed algebras. The $\ast$-isomorphism is explicitly given by the map
\begin{equation*}
    \alpha_{H_2 - H_1}: \mathcal{A}^1_\mathrm{loc}(\mcM) \to \mathcal{A}_{\mathrm{loc}}^2(\mcM), \qquad F \mapsto \alpha_{H_2 - H_1} F = e^{\frac12 \langle \hbar (H_2 - H_1), \frac{\delta^2}{\delta \varphi^2}\rangle} F, 
\end{equation*} 
for any two distinct choices of deforming bidistributions $H_1$ and $H_2$. 
\end{remark}

Wick polynomials can be realized within $\mathcal{A}_{\mathrm{loc}}(\mcM)$ by introducing the abstract Wick-ordering prescription
\begin{equation*}
    :F:_{H_{0^+}} \doteq \alpha_{-{H_{0^+}}}(F) = e^{-\frac12 \langle \hbar H_{0^+}, \frac{\delta^2}{\delta \varphi^2} \rangle} F
\end{equation*}
for $F\in\mathcal{A}_{\mathrm{loc}}(\mcM)$, where $H_{0^+}$ denotes the universal singular part of $H$, see, {\it e.g.}, \cite{Rejzner_2016}. A concrete realization is then obtained through the $\ast$-isomorphism
\begin{equation*}
    \alpha_H: \mathcal{A}_\mathrm{loc} (\mcM) \to \mathcal{A}_H(\mcM), 
\end{equation*}
usually referred to as the \emph{operator-ordering map}. Thus, $\alpha_H$ identifies abstract Wick-ordered observables with their representatives in the $H$-dependent realization of the extended algebra.

\paragraph{Interacting theory --} The aim of this section is to incorporate interactions into the algebraic framework developed so far. In the presence of a non-linear interaction, explicit solutions of the equations of motion are generally unavailable, and one is naturally led to a perturbative treatment in which observables are expanded as formal power series in the coupling parameter. This is the setting of \emph{perturbative algebraic quantum field theory}. We refer to \cite{Rejzner_2016} for a detailed account of the perturbative formalism which we now succinctly outline. The starting point is an action functional decomposed into a free part and an interaction term. We adopt the following notation:
\begin{equation}
    \label{Eq: split of the total action}
    S_{\lambda V}(\varphi) = S_{0}(\varphi) + \lambda V(\varphi), \qquad V \in \mathcal{F}_{\mathrm{loc}}(\mcM), 
\end{equation}
where $\lambda \in \mathbb{R}$ is the coupling constant, which will play the r\^ole of perturbative parameter. The corresponding equation of motion hence reads $S^{(1)}_0(\phi) = - \lambda V^{(1)}(\phi)$ and the associated normally hyperbolic operator is
\begin{equation}
\label{Eq: interacting KG operator}
    P_{\lambda V} = P_0 + \lambda V^{(2)}, \qquad \lambda \in \mathbb{R}. 
\end{equation}
We shall denote the set of field configurations that satisfy the free and the interacting equation of motions respectively by $\mathcal{E}_{S_0}(\mcM)$ and $\mathcal{E}_{S}(\mcM)$. Observe that, if the interacting potential is at most quadratic in the field configurations as in Section \ref{Sec: Projection of the massive scalar field}, then the interacting equation of motion is linear and $\mathcal{E}_S(\mcM)$ is an affine subspace of $\mathcal{E}(\mcM)$, \textit{cf.}, \cite[Ch. 4]{Rejzner_2016}. 

\begin{remark}
Most notably, we can relate the advanced and retarded Green operators for the free and the interacting theories. Indeed, denoting by $\Delta_{\mathrm{A}/\mathrm R}^{S_{\lambda V}}$ the advanced/retarded Green operators of $P_{\lambda V}$ in Equation \eqref{Eq: interacting KG operator} and by $\Delta_{\mathrm{A}/\mathrm R}^{S_0}$ those of $P_0$, a direct computation entails the following relation
\begin{equation}
    \label{Eq: advanced propagator for interacting theory}
    \Delta_{\mathrm A/\mathrm R}^{S_{\lambda V}} (\varphi) (\psi) =  \Delta_{\mathrm A/\mathrm R}^{S_0} (\psi) + \sum_{n=1}^\infty (-\lambda)^n \left( \Delta_{\mathrm A/\mathrm R}^{S_0} \circ V^{(2)} (\varphi) \right)^n  \Delta_{\mathrm A/\mathrm R}^{S_0} (\psi),
\end{equation}
for any $\psi \in \mathcal{D}(\mcM)$, where the series is understood as a formal power series in $\lambda$.
\end{remark}

At a classical level, the free and interacting theories can be connected by means of the \emph{(classical) retarded} and \emph{advanced M\o ller maps} acting on field configurations, namely, the maps $r_{\lambda V}: \mathcal{E}(\mcM) \to \mathcal{E}(\mcM)$ and $a_{\lambda V}: \mathcal{E}(\mcM) \to \mathcal{E}(\mcM)$ such that 
\begin{equation*}
\begin{split}
     S^{(1)}_{\lambda V} \circ  r_{\lambda V} = S^{(1)}_0, \qquad r_{\lambda V} (\varphi) \vert_{\mcM \setminus J^+(\operatorname{supp}(V))} = \varphi \vert_{\mcM \setminus J^+(\operatorname{supp}(V))}, \\
      S^{(1)}_{\lambda V} \circ  a_{\lambda V} = S^{(1)}_0, \qquad a_{\lambda V} (\varphi) \vert_{\mcM \setminus J^-(\operatorname{supp}(V))} = \varphi \vert_{\mcM \setminus J^-(\operatorname{supp}(V))}.
\end{split}  
\end{equation*}
Most notably, if $\varphi \in \mathcal{E}_{S_0}(\mcM)$, then $r_{\lambda V} (\varphi) \in \mathcal{E}_S(\mcM)$ and, similarly, for $a_{\lambda V}(\varphi)$. If we assume the free action to be quadratic in the field as it is the case for the conformally coupled Klein--Gordon field considered in this work, then the retarded M\o ller map enjoys some notable properties which we report here for the reader's convenience, \textit{cf.}, \cite[Prop. 3.1]{Drago_2016}. Let $S_{\lambda V}(\varphi) = S_0(\varphi) + \lambda V(\varphi)$ with
\begin{equation*}
    S_{0}(\varphi) = -\frac12 \int_{\mcM} d\mu_x \, f(x) \left( g^{\mu \nu}(x) \nabla_\mu \varphi(x) \nabla_\nu \varphi(x) + \frac{R(x)}{6} \varphi^2(x) \right), \; \; \lambda V(\varphi) = -\frac{\lambda}2 \int_{\mcM} d\mu_x \, h(x) \varphi^2(x)
\end{equation*}
where $\lambda \in \mathbb{R}$, $h \in \mathcal{D}(\mcM)$ is an infrared cut-off, while $f \in \mathcal{D}(\mcM)$ is such that $f \equiv 1$ on $\operatorname{supp}(h)$. The classical retarded M\o ller map takes the form 
\begin{equation}
\label{Eq: classical retarded Moller map}
    r_{\lambda V} = \operatorname{id} - \lambda \Delta_{\mathrm{R}}^{S_{\lambda V}} \circ V^{(1)}, 
\end{equation}
where $\Delta_{\mathrm{R}}^{S_{\lambda V}}$ is the retarded Green operator associated to the normally hyperbolic operator $P_{\lambda V}$ induced by $S_{\lambda V}$, while
$$V^{(1)}: \mathcal{E}(\mcM) \to \mathcal{D}(\mcM),\quad\varphi\mapsto V^{(1)} (\varphi) = - h \varphi.$$
Furthermore, it holds that 
\begin{equation*}
    r_{\lambda V} \circ ( \operatorname{id} + \lambda \Delta_{\mathrm{R}}^{S_0} \circ V^{(1)}) = ( \operatorname{id} + \lambda \Delta_{\mathrm{R}}^{S_0} \circ V^{(1)}) \circ  r_{\lambda V} = \operatorname{id}, 
\end{equation*}
$\Delta_{\mathrm{R}}^{S_0} \equiv \Delta_{\mathrm{R}}$ being the retarded propagator corresponding to the free action $S_0$. 

\begin{remark}
    \label{Rem: on the adiabatic limit h}
    The infrared cut-off $h$ restricts the interaction to a compact subset of spacetime and it is introduced to control infrared divergences in the perturbative expansion. To recover the physical theory, one is ultimately interested in the adiabatic limit $h\to 1$. In Section \ref{Sec: Local algebraic limit}, we shall explain how this limit can be implemented in an appropriate local sense by exploiting the time-slice axiom.
\end{remark}

Equation \eqref{Eq: classical retarded Moller map} is particularly useful since it allows us to represent the classical retarded M\o ller map in terms of the retarded propagator of the free theory and of the Fréchet derivative of the potential, without the need to compute explicitly the retarded Green operator for the interacting theory. We can further manipulate Equation \eqref{Eq: classical retarded Moller map} and express $r_{\lambda V}$ as a formal power series, known as \emph{Neumann series}: 
\begin{equation}
    \label{Eq: retarded Moller map as Neumann series}
    r_{\lambda V} = \sum_{n = 0}^\infty (-\lambda)^n (\Delta_\mathrm{R}^{S_0} \circ V^{(1)})^n. 
\end{equation}
If the underlying background is the Minkowski spacetime, Equation \eqref{Eq: retarded Moller map as Neumann series} can be shown to converge uniformly on any compact set for interactions at most quadratic in the configuration, see \cite[Lem. B.1]{Drago_2016}. In general, within the perturbative framework, $\lambda$ is a formal parameter and the series is understood, unless otherwise stated, as a formal power series in $\lambda$. 

The action of the classical M\o ller maps can be conveniently extended to elements of the algebra of observables as clarified by the following definition. 

\begin{definition}
    \label{Def: Moller maps on functionals}
Assuming Equation \eqref{Eq: split of the total action}, we define the {\bf retarded M\o ller map} on the space of microcausal functionals as $\mathscr{R}_{\lambda V}: \mathcal{F}_{\mu c}(\mcM) \to \mathcal{F}_{\mu c}(\mcM)[[\lambda]],$
\begin{equation}
    \label{Eq: retarded Moller map on the space of functional}
   (\mathscr{R}_{\lambda V} F) (\varphi) \doteq F(r_{\lambda V} (\varphi)), \qquad \forall \varphi \in \mathcal{E}(\mcM), F \in \mathcal{F}_{\mu c}(\mcM). 
\end{equation}
Here $\mathcal{F}_{\mu c}(\mcM)[[\lambda]]$ denotes the space of microcausal functionals regarded as formal power series in $\lambda$. Similarly, we define the {\bf advanced M\o ller map} $\mathscr{A}_{\lambda V}: \mathcal{F}_{\mu c}(\mcM) \to \mathcal{F}_{\mu c}(\mcM)$, with the advanced and retarded Green operators interchanged. 
\end{definition}

\noindent We emphasize that the classical Møller maps introduced in Definition \ref{Def: Moller maps on functionals} intertwine the Poisson structures of the free and interacting theories. More precisely, they relate the Peierls brackets associated with $S_0$ and $S_{\lambda V}$, as shown in \cite[Prop.~2]{DF_2003}.

We conclude the analysis of the classical interacting theory by reporting Proposition 4.10 in \cite{Rejzner_2016}. It yields a perturbative expansion of the M\o ller maps, thereby providing a convenient recursive formula for computing interacting classical observables. 

\begin{proposition}
\label{Prop: expansion of the Moller map}
    Let $(\mcM, g)$ be a globally hyperbolic spacetime with empty boundary and let $\mathcal{F}_{\mathrm{loc}}(\mcM)$ denote the space of local functionals thereon. Let $S_{\lambda V}$ be the action of an interacting theory as per Equation \eqref{Eq: split of the total action}. Then, the retarded M\o ller map $\mathscr{R}_{\lambda V}$ as per Definition \ref{Def: Moller maps on functionals} satisfies 
\begin{equation}
\label{Eq: expansion of the moller map}
    \mathscr{R}_{\lambda V} F = \sum_{n=0}^\infty \frac{\lambda^n}{n!} \mathscr{R}_n (V^{\otimes n}; F), \qquad \forall F \in \mathcal{F}_{\mu c}(\mcM). 
\end{equation}
For every $\varphi \in \mathcal{E}(\mcM)$, the coefficients $\mathscr{R}_n (V^{\otimes k}; F)$ abide by the recursion relations 
\begin{flalign*}
    \mathscr{R}_{n+1} (V^{\otimes (n+1)}; F)(\varphi) = - \sum_{k=0}^n \binom{n}{k} \mathscr{R}_k \left(V^{\otimes k}; \int_{\mcM^2} d\mu_x d\mu_y \; \frac{\delta V}{\delta \varphi}(x) (\Delta_{\mathrm{A}}^{S_0})^{(n-k)} (\varphi) (x,y) \frac{\delta F}{\delta \varphi} (y)\right)(\varphi), 
\end{flalign*}
where $(\Delta_{\mathrm{A}}^{S_0})^{(l)} (\varphi) (x,y) \doteq \frac{d^l}{d\lambda^l}\Delta_{\mathrm{A}}^{S_{\lambda V}}(\varphi) (x,y) \vert_{\lambda = 0}$. An analogous result holds true for the advanced M\o ller map $\mathscr{A}_{\lambda V}$, with the advanced and retarded Green operators interchanged. 
\end{proposition}

\begin{example}
For the conformally coupled Klein--Gordon field in Equation \eqref{Eq: KG equation} and for a quadratic interacting potential of the form 
\begin{equation*}
    \lambda V_h(\varphi) = \frac{\lambda}2 \int_{\mcM} d\mu_x \, h(x) \varphi^2(x), \qquad \varphi \in \mathcal{E}(\mcM), h \in \mathcal{D}(\mcM). 
\end{equation*}
Equation \eqref{Eq: expansion of the moller map} implies that the first-order classical correction to the linear field observable
\begin{equation*}
   \Phi_f \in \mathcal{F}_{\mathrm{loc}}(\mcM), \qquad \Phi_f (\varphi) = \int_\mcM d\mu_x \, f(x) \varphi(x), \; \; \varphi \in \mathcal{E}(\mcM), f \in \mathcal{D}(\mcM), 
\end{equation*}
reads 
\begin{equation}
    \label{Eq: first order classical correction to Phi}
    \mathscr{R}_{\lambda V_h} \Phi_f = \Phi_f - \lambda \Phi_{h \Delta_{\mathrm A}^{S_0}(f)} + \mathcal{O}(\lambda)^2. 
\end{equation}
\end{example}

The perturbative construction of the quantum interacting theory requires, in addition to the deformed product of the free theory, a \emph{time-ordered product}. Its extension to local observables entails a renormalization problem, whose admissible solutions are characterized axiomatically in \cite{HollandsWald2002}, with the residual freedom encoded by local and covariant counterterms. We shall not review this construction here and refer to \cite{Goi_2026} for further details. The singular structure of time ordering is governed by the \emph{Feynman propagator} $\Delta_F$ associated with the Hadamard two-point function $\Delta_+$ and by the corresponding Feynman parametrix $H_F$, defined by
\begin{equation*}
    \Delta_{\mathrm F} \doteq \Delta_+ + i\Delta_{\mathrm A},
    \qquad
    H_F \doteq H + i\Delta_{\mathrm A}.
\end{equation*}
Since $\Delta_{\mathrm F}-H_F \in C^\infty(\mcM \times \mcM)$, they share the same local singular structure, characterized by the Feynman wavefront set. We therefore fix a renormalized time-ordering prescription
\begin{equation*}
    T \doteq \{T_n\}_{n\geq 0}, \qquad
    T_n:
    \mathcal{A}_{\mathrm{loc}}(\mcM)^{\otimes n}
    \longrightarrow
    \mathcal{A}_{\mu c}(\mcM)[[\hbar]],
\end{equation*}
satisfying the standard axioms of perturbative quantum field theory. For regular polynomial functionals, the time-ordered product is explicitly given by
\begin{equation}
\label{Eq: time-ordered product}
    T^{\hbar H_F}(F_1\otimes\cdots\otimes F_n)
    \doteq
    F_1\cdot_T\ldots\cdot_T F_n
    =
    \mathrm M\circ
    e^{\sum_{\ell<j} D_{\hbar H_F}^{\ell j}}
    (F_1\otimes\cdots\otimes F_n),
\end{equation}
where
\begin{equation*}
    D_{\hbar H_F}^{\ell j}
    \doteq \left\langle \hbar H_F, \frac{\delta}{\delta\varphi_\ell} \otimes \frac{\delta}{\delta\varphi_j}\right\rangle ,
\end{equation*}
while $\mathrm M$ denotes the pullback along the diagonal of the configuration space. The operators $D_{\hbar H_F}^{\ell j}$ are symmetric under the exchange of $\ell$ and $j$. For local functionals, the products of Feynman parametrices arising in Equation \eqref{Eq: time-ordered product} are in general ill-defined on the thin diagonal in $\mcM^n$, although they have finite scaling degree. The renormalized maps $T_n$ are obtained by choosing a suitable extension of these distributions to the diagonal. In the following, we shall still denote this extension by $\cdot_T$, leaving the renormalization understood. 

The building blocks which allow us to encode the the information of the non-linear potential $\lambda V \in \mathcal{F}_{\loc}(\mcM)$ into the perturbative framework detailed above are then the S-matrix and the Bogoliubov map. We call {\bf S-matrix} the map
\begin{equation}
    \label{Eq: S-matrix}
    S: \lambda \mathcal{F}_{\loc }(\mcM) \to \mathcal{F}_{\mu c}(\mcM) [[\lambda]], \qquad S(\lambda V) = \sum_{n=0}^\infty \frac1{n!} \left(\frac{i \lambda}{\hbar} \right)^n \underbrace{V \cdot_T \ldots \cdot_T V}_{n \, \text{times}}. 
\end{equation}
The S-matrix admits a formal inverse with respect to the deformed product $\star_H$, which can be explicitly represented as
\begin{equation}
    \label{Eq: inverse S-matrix}
      S(\lambda V)^{\star - 1} = \sum_{n=0}^\infty \frac1{n!} \left(-\frac{i \lambda}{\hbar} \right)^n \underbrace{V \cdot_{\overline{T}} \ldots \cdot_{\overline{T}} V}_{n \, \text{times}}, 
\end{equation}
where $H_{AF} \doteq H - i \Delta_{\mathrm{R}}$ is the formal adjoint of the Feynman parametrix, called \emph{Dyson} (or \emph{anti-Feynman}) \emph{parametrix}, while $\cdot_{\overline{T}}$ denotes the anti-time ordered product deformed with respect to $H_{AF}$. With these ingredients, we can now introduce the {\bf Bogoliubov map} defined as 
\begin{equation}
    \label{Eq: Bogoliubov map}
    R_{\lambda V}: \mathcal{F}_{\loc}(\mcM) \to \mathcal{F}_{\mu c}(\mcM) [[\lambda]], \qquad R_{\lambda V} (F) \doteq S(\lambda V)^{\star - 1} \star_H [S(\lambda V) \cdot_T F], \; \; \forall F \in \mathcal{F}_\loc(\mcM). 
\end{equation}
The Bogoliubov map is the key ingredient in the construction of interacting quantum observables as formal power series in the coupling parameter of the theory. A more detailed discussion, including illustrative examples, can be found in \cite[Sec. 2.2.1]{Goi_2026}. We stress that $R_{\lambda V}$ is nothing but the quantisation of the retarded M\o ller map $\mathscr{R}_{\lambda V}$ in Definition \ref{Def: Moller maps on functionals}. It was shown in \cite[Prop. 4.3]{BDR24} and \cite[Prop. 6.1]{Rejzner_2016} that the formal expansion of $R_{\lambda V}(F)$, for any local functional $F$, contains no negative powers of $\hbar$. This entails that the classical limit $\hbar \to 0$ of the Bogoliubov map exists and, in particular, that
\begin{equation*}
    R_{\lambda V}(F) \vert_{\hbar = 0} = \mathscr{R}_{\lambda V}(F), \qquad \forall F \in \mathcal{F}_\loc(\mcM), 
\end{equation*}
see \cite[Prop. 6.2]{Rejzner_2016}. 

A particularly relevant property for the purposes of this work is the
\emph{Principle of Perturbative Agreement} (PPA), first introduced in
\cite{Hollands_2005} as an axiom constraining the renormalization freedom in the extension of the time-ordered product to local functionals, and later generalized in \cite{Drago_2016}. Heuristically, for interactions at most quadratic in the field, the PPA states that the perturbative construction based on the Bogoliubov map agrees, up to a canonical isomorphism, with the quantization of the theory in which the quadratic contribution is incorporated directly into the free action. Before stating the relevant definition, we introduce \emph{multilocal functionals} as finite linear combinations of pointwise products of local functionals, namely elements of
$$\mathcal{F}_{\mathrm{mloc}}(\mcM)\doteq\operatorname{span}_{\mathbb{C}}\left\{F_1 \cdots F_n : F_i \in \mathcal{F}_{\mathrm{loc}}(\mcM),\;
n \in \mathbb{N}_0 \right\},$$
where the empty product is identified with the constant functional $1$.

\begin{definition}
    \label{Def: PPA}
Let $g_1, g_2$ be two Lorentzian metrics on $\mcM$ such that $g_2 > g_1$, \textit{i.e.}, $J^\pm_1(p) \subset J^\pm_2(p)$ for any $p \in \mcM$. Consider the at most quadratic actions 
\begin{equation*}
    S_{i}[f](\varphi) \doteq -\frac12 \int_{\mcM} d\mu_{g_i} \, f(x) \left( g^{\mu \nu}_i(x) \nabla_\mu \varphi (x) \nabla_{\nu} \varphi(x) + m_i^2(x) \varphi^2(x) - 2j_i(x) \varphi(x)\right), \; \; i =1,2,
\end{equation*}
where $f \in \mathcal{D}(\mcM)$, $m^2_i\geq 0$, while $j_i\in C^\infty(\mcM)$, $i=1,2$. Set $Q(\varphi) \doteq S_{2}(\varphi) - S_{1}(\varphi)$ which, when expressed with
respect to the volume measure $\mu_{g_1}$, defines a local functional on
$(\mcM,g_1)$. Let $R_{1, Q}$ be the Bogoliubov map associated with the free action $S_1$ and interaction $Q$ and let $\mathscr{R}_{1, Q}$ denote the corresponding classical retarded M\o ller map. Furthermore, let $\star_i$ be the $\star$-products associated to $S_i$, which are related by $F \star_2 G = \mathscr{R}_{1, Q}^{-1} (\mathscr{R}_{1,Q}(F) \star_1 \mathscr{R}_{1,Q}(G))$ for any $F,G \in \mathcal{F}_{\mu c}(\mcM)_2$. We say that the time-ordered prescription satisfies the {\bf Principle of Perturbative Agreement (PPA)} if, writing
$T_i \equiv T_i(g_i,m_i^2,j_i,\star_i)$, $i=1,2$, the following relation
holds on the space of multilocal functionals $\mathcal{F}_{\mathrm{mloc}}(\mcM)$:
\begin{equation*}
    T_2 = \beta_{1,Q} \circ T_1, \qquad \beta_{1,Q} \doteq \mathscr{R}_{1,Q}^{-1} \circ R_{1, Q}.
\end{equation*}
\end{definition}

\noindent We shall exploit this principle in Section \ref{Sec: Projection of the massive scalar field}, where the mass term is treated as a quadratic perturbation of the massless theory, that is, we set $j_1=j_2=0$ while $m^2_1=0$ and $m^2_2>0$.

\subsection{The bulk--to--boundary correspondence} 
\label{Sec: The bulk to boundary correspondence}
The goal of this section is to establish a bulk-to-boundary correspondence between states and observables in the bulk of the physical spacetime and their counterparts on a partition of the conformal null boundary $\mathscr{I}^+$. To this avail, we shall first introduce the quantum algebra of observables at $\mathscr{I}^+$ of an asymptotically flat, physical spacetime $(\mcM, g)$ in the sense of Definition \ref{Def: asymptotically flat spacetimes}. We require also the Assumption \ref{Ass: V} holds true and we will follow slavishly the discussion in \cite{DMP_2017}. 

Adopting the notation of Section \ref{Sec: Asymptotically Flat Spacetimes}, let $(\tilde{\mcM}, \tilde{g})$ denote the unphysical spacetime and, with a slight abuse of notation, let us identify $\mcM$ with its image under the conformal embedding $\iota: \mcM \hookrightarrow \tilde{\mcM}$. We denote the conformal boundary of $\mcM$ by $\mathscr{C} \doteq \partial J^-(i^+; \tilde{\mcM})$, where $J^-(i^+; \tilde{\mcM})$ is the causal past of future time infinity $i^+$ with respect to the causal structure of $\tilde{\mcM}$. Note that it can be decomposed as $\mathscr{C} = \mathscr{I}^+ \cup \{i^+\}$.  Furthermore, it is convenient for the sake of the ensuing discussion to choose the conformal factor in such a way that the metric on $\mathscr{I}^+$ takes the Bondi form and, thus, the following identification holds $\mathscr{I}^+ \simeq \mathbb{R} \times \mathbb{S}^2$.

\paragraph{Bulk--to--boundary correspondence for states --} Let us introduce the real vector space
\begin{equation}
\label{Eq: space S}
    \mathscr{S}(\mathscr{I}^+) \doteq \{\psi \in C^\infty(\mathscr{I}^+) \, : \, \psi, \partial_u \psi \in L^2(\mathscr{I}^+)\}, 
\end{equation}
where $L^2(\mathscr{I}^+)$ refers to the measure $du d\mu_{\mathbb{S}^2}$. Then, the map $\sigma_\mathscr{I} : \mathscr{S}(\mathscr{I}^+) \times \mathscr{S}(\mathscr{I}^+) \to \mathbb{R}$ defined by
\begin{equation*}
   \sigma_\mathscr{I} (\psi, \psi') = \int_{\mathbb{R} \times \mathbb{S}^2} (\psi \partial_u \psi' - \psi' \partial_u \psi) \, du d\mu_{\mathbb{S}^2}, \qquad \forall \psi, \psi' \in \mathscr{S}(\mathscr{I}^+),
\end{equation*}
is a weakly non-degenerate symplectic form, see \cite[Prop. 3.2.1]{DMP_2017}. We can thus associate to the symplectic space $(\mathscr{S}(\mathscr{I}^+), \sigma_\mathscr{I})$, the complexified tensor algebra over $\mathscr{S}(\mathscr{I}^+)$,
\begin{equation*}
\mathcal{T}(\mathscr{I}^+) = \mathbb{C} \otimes_{\mathbb{R}} \bigoplus_{n=0}^\infty (\mathscr{S}(\mathscr{I}^+))^{\otimes n}, \qquad (\mathscr{S}(\mathscr{I}^+))^{\otimes 0} \equiv \mathbb{R}, 
\end{equation*}
equipped with an associative product, a unit element $\mathbb{I}$ and an anti-linear involutive $\ast$-operation. The {\bf boundary algebra on $\mathscr{I}^+$} is defined by
\begin{equation}
   \label{Eq: boundary algebra}
   \mathcal{A}(\mathscr{I}^+) \doteq \frac{\mathcal T (\mathscr{I}^+)}{\mathcal{I}(\mathscr{I}^+)}, 
\end{equation}
where $\mathcal{I}(\mathscr{I}^+)$ is the two-sided $\ast$-ideal of $ \mathcal{T}(\mathscr{I}^+)$ generated by $\psi \otimes \psi' - \psi' \otimes \psi - i \sigma_\mathscr{I}(\psi, \psi') \mathbb{I}$. The interplay between this algebra and the boundary symmetries encoded in the BMS group is explored in \cite[Prop. 3.2.2]{DMP_2017}. Observe that $\mathcal{A}(\mathscr{I}^+)$ carries no dynamical information on the underlying equation of motion. To encode such information into the boundary algebra, we can exploit \cite[Thm. 3.3.1]{DMP_2017} which entails the existence of a well-defined injective vector space homomorphism 
\begin{equation}
 \label{Eq: bulk to boundary map 1}
    \Gamma_\mathscr{I}: \mathcal{E}^{\mathrm{obs}}(\mcM) \to \mathscr{S}(\mathscr{I}^+), \qquad [f] \mapsto \Gamma_{\mathscr{I}}[f] \doteq \tilde{\Delta}^\mathscr{V}_{\mathrm S} (\Omega^{-3} f) \vert_{\mathscr{I}^+} \in C^\infty(\mathscr{I}^+), 
\end{equation}
where $\mathscr{S}(\mathscr{I}^+)$ is as per Equation \eqref{Eq: space S}, while $\tilde{\Delta}^\mathscr{V}_\mathrm{S}$ is the causal propagator on the submanifold $\mathscr{V}$ of the the unphysical spacetime, see Assumption \ref{Ass: V} and Remark \ref{Rem: conformal invariance of P0}.  In addition, $\Gamma_\mathscr{I}$ preserves the natural symplectic forms of $\mathcal{E}^{\mathrm{obs}}(\mcM)$ and $\mathscr{S}(\mathscr{I}^+)$, \textit{i.e.}, for any $\varphi, \varphi' \in \mathcal{E}^{\mathrm{obs}}(\mcM)$, it holds that 
\begin{equation*}
   \sigma_{\mcM}(\varphi, \varphi') = \sigma_{\mathscr{I}}(\Gamma_\mathscr{I}(\varphi), \Gamma_\mathscr{I}(\varphi')), 
\end{equation*}
where $\sigma_\mcM$ is a symplectic form on $\mathcal{E}^{\mathrm{obs}}(\mcM)$. Hence, $\Gamma_\mathscr{I}$ extends uniquely to an injective unital $\ast$-homomorphism
\begin{equation}
   \label{Eq: bulk-to-boundary algebra homomorphism}
   \iota_{\mathscr I}:
  \mathcal{A}^{\mathrm{obs}}(\mcM) \longrightarrow \mathcal{A}(\mathscr{I}^+), \qquad   \iota_{\mathscr I}\bigl(\Phi_f\bigr)
   = [\Gamma_\mathscr{I}[f]], \; \; f\in C^\infty_0(\mcM), 
\end{equation}
which is completely determined by its action on the generators, see Remark \ref{Rem: relation with Aobs}. Indeed, the symplectic property of $\Gamma_\mathscr{I}$ guarantees that the canonical commutation relations in the bulk are mapped into those of the boundary algebra. Equation \eqref{Eq: bulk-to-boundary algebra homomorphism} therefore provides the desired bulk-to-boundary correspondence for the massless, conformally coupled scalar field.

Following slavishly \cite[Sec. 4]{DMP_2017}, on $\mathcal{A}(\mathscr{I}^+)$ we can construct a quasi-free state $\omega_\mathscr{I}$, whose associated two-point function identifies an element in $\mathcal{D}'(\mathscr{I}^+ \times \mathscr{I}^+)$ and it reads at the level of integral kernels
\begin{equation}
\label{Eq: twopoint on I+}
    \omega_{2, \mathscr{I}}(x,x') = -\frac{1}{\pi} \lim_{\epsilon \to 0^+} \frac{\delta(s,s')}{(u-u'-i\epsilon)^2}, 
\end{equation}
where, given a Bondi reference frame, $x = (u,s)$, $x' = (u',s')$ with $u,u' \in \mathbb{R}$ and $s,s' \in \mathbb{S}^2$, while $\delta(s,s')$ is the Dirac delta distribution on the unit sphere $\mathbb{S}^2$ with respect to the standard rotational-invariant measure. We stress that the limit in Equation \eqref{Eq: twopoint on I+} has to be understood in a distributional sense. In its folium this quasi-free boundary state $\omega_\mathscr{I}$ is unique, this property being characterized by BMS invariance together with a positive-energy condition for Bondi-time translations, see \cite[Thm. 4.1.7]{DMP_2017}. Given the distinguished algebraic quasi-free state $\omega_{\mathscr{I}^+}$ on the boundary algebra $\mathcal{A}(\mathscr{I}^+)$, we can induce via pullback a counterpart on the algebra of observables $\mathcal{A}^{\mathrm{obs}}(\mcM)$, see Remark \ref{Rem: relation with Aobs},: 
\begin{equation*}
    \omega_\mcM : \mathcal{A}^{\mathrm{obs}}(\mcM) \to \mathbb{C}, \qquad \omega_\mcM \doteq \iota^*_{\mathscr{I}} \omega_\mathscr{I},
\end{equation*}
where $\iota_\mathscr{I}: \mathcal{A}^{\mathrm{obs}}(\mcM) \to \mathcal{A}(\mathscr{I}^+)$ is the injective unital $\ast$-algebra homomorphism in Equation \eqref{Eq: bulk-to-boundary algebra homomorphism}. The state $\omega_\mcM$ inherits the property of being quasi-free and the associated two-point function reads
\begin{equation*}
    \omega_{2, \mcM} (f,f') = \omega_\mathscr{I}([\Gamma_\mathscr{I}([f])] \otimes [\Gamma_\mathscr{I}([f'])]), \qquad f,f' \in C^\infty_0(\mcM), 
\end{equation*}
where $\Gamma_\mathscr{I}$ is as in Equation \eqref{Eq: bulk to boundary map 1}. Most notably, $\omega_\mcM$ is of Hadamard form as shown in \cite[Thm. 4.2.1]{DMP_2017} and it is invariant under the action of all isometries in the bulk. A complementary construction which focuses on pseudodifferential operators has been developed in \cite{Gerard:2014hla}. 

\begin{remark}
    Applying this construction to four-dimensional Minkowski spacetime $(\mathbb{M}^4, \eta)$, $\omega_\mcM$ coincides with the Gaussian Poincaré-invariant state. 
\end{remark}

\paragraph{Extension of the algebra on the boundary --}
\label{Sec: Extension of the algebra on the boundary} Adapting \cite[Def. 5.2.1]{DMP_2017} to the present setting, let us denote by $\mathcal{D}'(\mathscr{I}^+)$ the space of distributions on future null infinity and let us introduce the distributional spaces $\{\mathscr{F}^n\}_{n \in \mathbb{N}}$ defined as the set of distributions $F_n \in \mathcal{D}'((\mathscr{I}^+)^n)$ meeting the following requirements: 
\begin{itemize}
    \item[(i)] {\bf Support condition}. The closure of the distributional support $\operatorname{supp}(F_n)$ is compact in $\mathscr{C}^n$.
    \item[(ii)] {\bf Causal non-monotonic singular directions}. It holds that 
    \begin{equation*}
        \operatorname{WF}(F_n) \subset \{(x,\xi) \in \mathring{T}^* (\mathscr{I}^+)^n \, : \, (x, \xi) \notin \left( \overline{\mathcal V}_n^+ \cup \overline{\mathcal V}_n^-\right), (x,\xi) \notin S_n\}, 
    \end{equation*}
    where $\mathring{T}^* \mathscr{I}^+ \doteq T^* \mathscr{I}^+ \setminus \{0\}$ is the punctured cotangent space, while we adopt the short-hand notation $(x, \xi) = (x_1, \ldots, x_n; \xi_1, \ldots, \xi_n)$. Here we write $(x,\xi) \in \overline{\mathcal V}^+_n$ if, using Bondi coordinates on $\mathscr{I}^+$, either $\xi_i = 0$ or $(\xi_i)_u > 0$ for every $i = 1, \ldots, n$ and, analogously for $\overline{\mathcal V}^-_n$. In addition, $(x,\xi) \in S_n$ if there exists $i \in \{1, \ldots,n\}$ such that $\xi_i \ne 0$ and $(\xi_i)_u = 0$. 
    \item[(iii)] {\bf Smoothness condition.} There exists an open neighbourhood $\mathcal U \subset \widetilde{\mcM}$ of $i^+$ such that the following holds. For every $i=1,\ldots,n$, every permutation $P$ of $\{1,\ldots,n\}$, and every $\mathrm{u} \in C^\infty_0\bigl((\mathscr I^+)^{n-i}\bigr)$, let $\widetilde F_{n,P}(\mathrm{u})\in \mathcal D'\bigl((\mathscr I^+)^i\bigr)$ denote the partial smearing of $F_n$ in the variables $x_{P_{i+1}},\ldots,x_{P_n}$ against $\mathrm{u}$, as defined by the Schwartz kernel theorem. Then there exists $g_{P,\mathrm{u}}\in C^\infty(\mathcal U^i)$ such that
\begin{equation*}
    \widetilde F_{n,P}(\mathrm{u})
    \big|_{(\mathcal U\cap\mathscr I^+)^i}
    =
    g_{P,\mathrm{u}}
    \big|_{(\mathcal U\cap\mathscr I^+)^i}.
\end{equation*} 
\end{itemize}
A few remarks are in due order. The first condition is not restrictive as it encompasses the case of interest in which an observable supported in a compact set $\mathcal K \subset \mcM$ is mapped to a boundary counterpart whose support lies in $J^+(\mathcal K) \cap \mathscr{C}$ which is a closed subset of $J^+(\mathcal K) \cap J^-(i^+)$ and, hence, compactly supported in $\mathscr{C}$. Although the support may reach $i^+$, on account of item (iii), all singularities of the distributions $F_n$ are localized away from future timelike infinity $i^+$. This is particularly relevant for the construction of the bulk-to-boundary map below, as it implies that singular directions of a distribution supported in the bulk propagate along null curves, thereby reaching $\mathscr{I}^+$, rather than $i^+$. We emphasize that the above definition is independent from the choice of Bondi coordinates. Indeed, the support and smoothness conditions are formulated intrinsically in terms of the conformal completion, while the wavefront set restriction in item (ii) is invariant under change of Bondi coordinates. Indeed, under a BMS transformation, $u$ is mapped to $u^\prime=K_\Lambda(u+f(\zeta, \bar{\zeta}))$, where $K_\Lambda$ depends only on the angular coordinates on $\mathbb{S}^2$ through $(\zeta, \bar{\zeta})$, \textit{cf.}, Remark \ref{Rmk: action of BMS group}. Consequently, the component of any covector along $du$ transforms according to $(\xi')_{u'}=K_\Lambda^{-1}\xi_u$. Hence its sign and vanishing are preserved, implying that both $\overline{\mathcal V}_n^\pm$ and $S_n$ are invariant under the induced cotangent action of the BMS group.

We now possess all the ingredients to define the extended algebra of boundary observables on $\mathscr{I}^+$. 
\begin{definition}
\label{Def: algebra of boundary observable}
Denoting by $\mathscr{F}^n_s$ the subset of totally symmetric elements of $\mathscr{F}^n$, we call {\bf extended algebra of boundary observables} the unital $\ast$-algebra 
    \begin{equation*}
        \mathcal{A}_e(\mathscr{I}^+) = \bigoplus_{n = 0}^\infty \mathscr{F}^n_s, \qquad \mathscr{F}^0_s \equiv \mathbb{C}, 
    \end{equation*}
equipped with a $\ast$-involution which acts as follows
\begin{equation*}
    F_n^*(f_1 \otimes \ldots \otimes f_n) = \overline{F_n(f_n \otimes \ldots \otimes f_1)}, \qquad \forall f_i \in \mathcal{D}(\mathscr{I}^+), i=1, \ldots,n, \forall F \in \mathscr{F}^n_s.
\end{equation*}
For every element $F \in \mathcal{A}_e(\mathscr{I}^+)$, we can write
\begin{equation}
\label{Eq: decomposition of F psi}
    F(\psi) = \sum_{k=0}^{N_F} \frac{1}{k!} \langle F_k, \psi^{\otimes k}\rangle, \qquad F_k \in \mathscr{F}^k_s, \psi \in C^\infty(\mathscr{I}^+). 
\end{equation}
We can thus endow $\mathcal{A}_e(\mathscr{I}^+)$ with a $\star$-product $\star_\omega$ defined by
\begin{equation}
\label{Eq: star product at the boundary}
    (F \star_\omega G) (f) = \sum_{n = 0}^\infty \frac1{n!} \langle F^{(n)}(f), \omega^{\otimes n}_{2, \mathscr{I}} G^{(n)}(f)\rangle, 
\end{equation}
for all $F, G \in \mathcal{A}_e(\mathscr{I}^+)$ and $f \in \mathcal{D}(\mathscr{I}^+)$. Here the bidistribution $\omega_\mathscr{I} \in \mathcal{D}'(\mathscr{I}^+ \times \mathscr{I}^+)$ is the two-point function as per Equation \eqref{Eq: twopoint on I+}.
\end{definition}
\noindent Notice that, up to the standard intertwining isomorphism, the boundary algebra $\mathcal{A}(\mathscr{I}^+)$ is recovered as the subalgebra of $\mathcal{A}_e(\mathscr{I}^+)$ consisting of elements with smooth components, see Equation \eqref{Eq: decomposition of F psi} and \cite[Sec.~5.2, Eq.~(5.14)]{DMP_2017}.

\begin{remark}
The well-posedness of the $\star$-product in Equation \eqref{Eq: star product at the boundary} has been established in \cite[Prop. 3.5]{Dappiaggi_2011} and it is a byproduct of the microlocal restrictions defining $\mathscr{F}^n$, see also the discussion above, in particular, item (ii). Note, in addition, that if we were to consider the algebra of local functionals on $\mathscr{I}^+$, the wavefront set condition in item (ii) would not be satisfied. 
\end{remark}

We conclude this section by defining the bulk-to-boundary map which allows us to project observables supported in the bulk to future null infinity, see also \cite[Thm. 3.1]{Dappiaggi_2011}. 

\begin{definition}
\label{Def: bulk to boundary map}
    Let $(\mcM, g)$ be a four-dimensional globally hyperbolic spacetime which is asymptotically flat in the sense of Definition \ref{Def: asymptotically flat spacetimes} and denote by $\mathscr{C} = \mathscr{I}^+ \cup \{i^+\}$ the conformal boundary of $\mcM$ in the unphysical spacetime $(\tilde{\mcM}, \tilde{g})$. Furthermore, let $\mathcal{A}_H(\mcM)$ be the quantum microcausal algebra of polynomial functionals as per Equation \eqref{Eq: quantum microcausal algebra} with $\Delta_+$ replaced by $H$ and let $\mathcal{A}_e(\mathscr{I}^+)$ be the extended boundary algebra as per Definition \ref{Def: algebra of boundary observable}. Then, there exists a unit-preserving $\ast$-homomorphism 
    \begin{equation}
        \label{Eq: bulk-to-boundary map}
        \Gamma: \mathcal{A}_H(\mcM) \to \mathcal{A}_e(\mathscr{I}^+). 
    \end{equation}
    In addition, denoting by $\Delta_\mathrm{S}$ be the causal propagator for a conformally coupled Klein-Gordon field on $(\mcM, g)$ and by $\tilde{\Delta}^\mathscr{V}_\mathrm{S}$ the causal propagator on $(\mathscr{V}, \tilde{g}|_{\mathscr{V}})$ with $\mathscr{V}\subseteq\widetilde{\mcM}$, it holds that
    \begin{equation}
        \label{Eq: action of the bulk-to-boundary map}
        \Gamma(F) = \sum_{k=0}^N \frac1{k!} (\tilde{\Delta}^\mathscr{V}_\mathrm{S} \Omega^{-3})^{\otimes k} (F_k) \vert_{\mathscr{I}^+} \in \mathcal{A}_e(\mathscr{I}^+), 
    \end{equation}
where $\Omega \in \mathcal{E}(\tilde{\mcM})$ is the conformal factor. Here $F_k \in \mathcal{E}'(\mcM^k)$ are as in Remark \ref{Rem: functionals and integral kernels}.  
\end{definition}

\begin{remark}
Each term in Equation \eqref{Eq: action of the bulk-to-boundary map} has to be regarded as the distributional product 
\begin{equation}
    \label{Eq: dist product}
    K_k \doteq (\iota^{\times k})^*\left(\tilde{\Delta}^\mathscr{V}_\mathrm{S}\right)^{\otimes k} \cdot (1 \otimes (\Omega^{-3})^{\otimes k} F_k) \equiv (\iota^* \tilde{\Delta}^\mathscr{V}_\mathrm{S})^{\otimes k} \cdot (1 \otimes (\Omega^{-3})^{\otimes k} F_k), 
\end{equation}
where $\iota^*$ stands for the pullback with respect to the embedding $\iota: \mathscr{I}^+ \times \tilde{\mcM} \to \tilde{\mcM} \times \tilde{\mcM}$, while $1$ is the distribution generated by the constant function $1$ on $(\mathscr{I}^+)^k$. At the level of integral kernels we may write
\begin{equation*}
    \tilde{\Delta}^\mathscr{V}_\mathrm{S} (v_1, x_1) \Omega^{-3}(x_1) \ldots \tilde{\Delta}^\mathscr{V}_\mathrm{S} (v_k, x_k) \Omega^{-3}(x_k) F_k(x_1, \ldots, x_k), 
\end{equation*}
where $x_i \in \mcM$, $v_i \in \mathscr{I}^+$, for all $i=1, \ldots, k$. First and foremost, observe that the pullback $\iota^* \tilde{\Delta}^\mathscr{V}_\mathrm{S}$ is well-defined as a distribution on $\mathscr{I}^+ \times \tilde{\mcM}$ on account of the microlocal properties of the Pauli-Jordan propagator and of \cite[Thm. 8.2.4]{Hormander_1990}. Still by \cite[Thm. 8.2.4]{Hormander_1990}, $\operatorname{WF}((\iota^* \tilde{\Delta}^\mathscr{V}_\mathrm{S})^{\otimes k}) \subset \operatorname{WF}(\iota^* \tilde{\Delta}^\mathscr{V}_\mathrm{S})^{\times k}$ and, hence, any element in $\operatorname{WF}((\iota^* \tilde{\Delta}^\mathscr{V}_\mathrm{S})^{\otimes k})$ is of the form 
\begin{equation*}
    (u, \xi, x, \zeta) = (u_1, \ldots, u_k, \xi_1, \ldots, \xi_k, x_1, \ldots, x_k, \zeta_1, \ldots, \zeta_k), \qquad \text{with} \; (u_i, \xi_i) \sim (x_i, \zeta_i), \forall i=1, \ldots, n,
\end{equation*}
where the symbol $\sim$ entails that the two points $u$ and $x$ are connected by a null geodesic $\gamma$ with $\zeta$ tangent to $\gamma$ at $x$, while $\xi$ is the parallel transport of $\zeta$ along $\gamma$. Focusing on the second contribution to the above product, using that $F_k$ is compactly supported in $\mcM^k$, we can write 
$$(\Omega^{-3})^{\otimes k} F_k = [(\Omega^{-3})^{\otimes k} \eta] F_k, \qquad \eta \in \mathcal{D}(\mcM^k), \eta \vert_{\operatorname{supp}(F_k)} = 1.$$ 
Thus, since $\Omega$ is nowhere vanishing inside $\mcM$, the function $(\Omega^{-3})^{\otimes k} \eta$ is smooth, non-vanishing on the support of $F_k$ and compactly supported in $\mcM^k$. These observations combined with \cite[Thm. 8.2.9]{Hormander_1990} imply that 
$$\operatorname{WF}(1 \otimes (\Omega^{-3})^{\otimes k} F_k) \subset \{0\} \times \operatorname{WF}(F_k).$$ 
Every element in the wavefront set of $\operatorname{(1 \otimes (\Omega^{-3})^{\otimes k}} F_k)$ is of the form 
\begin{flalign*}
    (v,0, y, \beta) &= (v_1, \ldots, v_k, 0, \ldots, 0, y_1, \ldots, y_k, \beta_1, \ldots, \beta_k), \\ &(y_1, \ldots, y_k, \beta_1, \ldots, \beta_k) \in \operatorname{WF}(F_k), v_i \in \mathscr{I}^+, i =1, \ldots, k. 
\end{flalign*}
Consider the points $(u,\xi, x, \zeta) \in \operatorname{WF}((\iota^* \tilde{\Delta}^\mathscr{V}_\mathrm{S})^{\otimes k})$ and $(u,0, x, \beta) \in \operatorname{WF} (1 \otimes (\Omega^{-3})^{\otimes k} F_k)$, since the covectors $\xi_i$ are all non-vanishing, it holds that $(\xi, \zeta) + (0, \beta)$ can never sum to the zero section. This entails well-definiteness of the product in Equation \eqref{Eq: dist product}, see \cite[Thm. 8.2.10]{Hormander_1990}. Furthermore, denoting by $\pi_\mathscr{I}: (\mathscr{I}^+)^k \times \mcM^k \to (\mathscr{I}^+)^k$ the projection onto the first factor, the push-forward $(\pi_\mathscr{I})_\ast K_k$ is a well-defined element of $\mathcal{D}'(\mathscr{I}^k)$. More explicitly, given $\psi \in \mathcal{D}(\mcM^k)$ such that $\psi \vert_{\operatorname{supp}(F_k)} = 1$ and $h \in \mathcal{D}(\mathscr{I}^+)$, we set $\langle (\pi_\mathscr{I})_\ast K_k, h\rangle = \langle K_k, h \otimes \psi \rangle$, with the right-hand side independent of the choice of $\psi$ as $K_k$ is supported inside $\operatorname{supp}(F_k)$ in the bulk variables. 
\end{remark}

\begin{example}
    If we consider the linear functional $\Phi_f \in \mathcal{A}_{\mathrm{loc}}(\mcM)$, $f \in \mathcal{D}(\mcM)$, introduced in Remark \ref{Rem: relation with Aobs} and apply the bulk-to-boundary map, the projected functional reads $$\Gamma(\Phi_f) = \tilde{\Delta}^\mathscr{V}_\mathrm{S} (\Omega^{-3} f) \vert_{\mathscr{I}^+}.$$ 
\end{example}

\section{Projection of the massive scalar field}\label{Sec: Projection of the massive scalar field}
In this section, we consider a real, massive scalar field on a physical, asymptotically flat spacetime $(\mcM, g)$ as per Definition~\ref{Def: asymptotically flat spacetimes}, namely we work with $\varphi\in\mathcal E(\mcM)$, whose dynamics is governed by the action in Equation~\eqref{Eq: free action on M} and it satisfies the Cauchy problem in Equation~\eqref{Eq: KG equation}. We construct a projection of the associated observables on future null infinity $\mathscr I^+$. First of all, we establish a algorithmic procedure in the general setting, showing that observables can be realized on $\mathscr{I}^+$ as formal power series in the mass as expansion parameter. Subsequently we conisder four-dimensional Minkowski spacetime, proving herein convergence of the said series. 

Let $\Delta_{\mathrm R/\mathrm A}$ denote the retarded and advanced Green operators of $P_0$ on $(\mcM,g)$, and let $\widetilde\Delta^{\mathscr V}_{\mathrm R/\mathrm A}$ denote those of $\widetilde P_0$ on $(\mathscr{V},\widetilde g|_\mathscr{V})$ as per Assumption \ref{Ass: V}, see also Remark \ref{Rem: conformal invariance of P0}. We write $\Delta_{\mathrm S}\doteq\Delta_{\mathrm R}-\Delta_{\mathrm A}$ and $\widetilde\Delta^{\mathscr V}_{\mathrm S}\doteq\widetilde\Delta^{\mathscr V}_{\mathrm R}-\widetilde\Delta^{\mathscr V}_{\mathrm A}$ for the corresponding Pauli-Jordan (causal) propagators. In view of the principle of perturbative agreement \ref{Def: PPA}, we introduce the quadratic interaction on $\mcM$
\begin{equation}
    \label{Eq: massive potential}
    m^2 V_h(\varphi) \doteq - \frac{m^2}{2} \int_{\mcM} \varphi^2(x) h(x) \, d\mu_g(x), \qquad m^2 > 0,
\end{equation}
where $h \in C^\infty_0(\mcM)$ is a suitable cut-off needed to avoid infrared divergences. As discussed in Section~\ref{Sec: The perturbative algebraic framework}, the classical retarded M\o ller map acts on a linear observable $\Phi_f \in \mathcal{A}_{\mathrm{loc}}(\mcM)$, with $f\in C^\infty_0(\mcM)$, according to the formal expansion:
\begin{equation}
    \label{Eq: Neumann series}
    \mathscr{R}_{m^2 V}(\Phi_f) \doteq \Phi_f (r_{m^2V}(\varphi)) = \sum_{n \ge 0} (m^2)^n \Phi_{f_n}, \qquad f_n \doteq (h \circ \Delta_{\mathrm{A}})^n (f).
\end{equation}
Here $h$ acts by pointwise multiplication, and the expansion follows the sign conventions adopted for the retarded M\o ller map in Equation \eqref{Eq: retarded Moller map on the space of functional}. We emphasize that the expansion in Equation~\eqref{Eq: Neumann series} is understood formally. Its convergence has been established for compactly supported mass perturbations on Minkowski spacetime in \cite[Lem.~B.1]{Drago_2016}; see also Section~\ref{Sec: Four-dimensional Minkowski spacetime}.

\begin{remark}
    Since the field observable is linear in the field configuration, all its functional derivatives of order greater than one vanish. Consequently, the quantum corrections to the classical retarded M\o ller map vanish on linear observables, so that the corresponding Bogoliubov map agrees with its classical counterpart.
\end{remark}

In order to construct the projection to the future null infinity $\mathscr{I}^+$ of the massive field observable $\mathscr{R}_{m^2V}(\Phi_f)$, we reason as follows. We start by observing that each term $f_n \doteq (h \circ \Delta_\mathrm{A})^n f$ appearing on the right-hand side of Equation \eqref{Eq: Neumann series} is smooth and compactly supported. Hence, $\Phi_{f_n}$ lies in the local algebra $\mathcal{A}_{\mathrm{loc}}(\mcM)$ for any $n \ge 0$. Furthermore, it corresponds to a polynomial in the field of degree one. We can therefore apply the bulk-to-boundary map of Definition~\ref{Def: bulk to boundary map} and project the formal series separately for each coefficient, obtaining
\begin{equation}\label{Eq: projected massive field}
\Gamma(\mathscr{R}_{m^2V}(\Phi_f)) \doteq \sum_{n \ge 0} (m^2)^n \tilde{\Delta}^\mathscr{V}_\mathrm{S} (\Omega^{-3} (h \circ \Delta_{\mathrm{A}})^n (f)) \bigg\vert_{\mathscr{I}^+} \,, 
\end{equation}
where $\Omega \in \mathcal{E}(\tilde{\mcM})$ is the conformal factor corresponding to the embedding $\iota: \mcM \hookrightarrow \tilde{\mcM}$, while the conformally weighted sources $\psi_n \doteq \Omega^{-3} (h \circ \Delta_{\mathrm{A}})^n (f) \doteq \Omega^{-3} f_n$ are understood to be extended by zero to $\mathscr{V}$. Here, with a slight abuse of notation, we identify the observable $\Phi_{f_n}$ with its defining test function $f_n\in C^\infty_0(\mcM)$. In view of the support properties of the fundamental solutions, the image of the advanced propagator vanishes at future null infinity, \textit{i.e.}, $\tilde{\Delta}_\mathrm{A} \psi_n \vert_{\mathscr{I}^+} = 0$. This yields
\begin{equation}\label{Eq: projected massive field 1}
\Gamma(\mathscr{R}_{m^2V}(\Phi_f)) \doteq \sum_{n \ge 0} (m^2)^n \tilde{\Delta}^\mathscr{V}_\mathrm{R} (\Omega^{-3} (h \circ \Delta_{\mathrm{A}})^n (f)) \bigg\vert_{\mathscr{I}^+} \,.
\end{equation}
Each summand in the above formula is the restriction to $\mathscr{I}^+$ of the retarded solution $\psi_n$ of the following problem in $(\mathscr{V}, \tilde{g} \vert_\mathscr{V})$, 
\begin{equation}
\label{Eq: Goursat problem for psi}
    \begin{cases}
        \tilde{P}_0 (\tilde{\Delta}^\mathscr{V}_\mathrm{R} \psi_n) = \psi_n, \\
        \tilde{\Delta}^{\mathscr{V}}_\mathrm{R} \psi_n \vert_{\mathscr{I}^-} = 0,
    \end{cases}
\end{equation}
where initial data are assigned on a null hypersurface, rather than on a spacelike one. Although the formal expansion in Equation~\eqref{Eq: projected massive field 1} can be derived on a general asymptotically flat spacetime, establishing convergence of the boundary series requires additional estimates. We are not aware of a result holding true on any admissible background in our setting. Yet, in the next section, we establish them on four-dimensional Minkowski spacetime, which admits a conformal embedding into the Einstein static universe \textit{cf.}, Example~\ref{Ex: Minkowski and ESU}. 

\subsection{Projection of the massive scalar field on Minkowski spacetime}\label{Sec: Four-dimensional Minkowski spacetime}
In this section, we consider four-dimensional Minkowski spacetime $(\mathbb{M}^4, \eta)$, with signature $(-, +, +, +)$. With a mild abuse of notation, henceforth we shall simply write $\mathbb{M}^4$ in place of $(\mathbb{M}^4, \eta)$. On top of it, we consider a massless real scalar field $\varphi \in \mathcal{E}(\mathbb{M}^4)$, whose dynamics is governed by the action 
\begin{equation}
\label{Eq: free action}
    S_0 [f](\varphi) \doteq - \frac{1}{2} \int_{\mcM} \eta^{\mu \, \nu}(x) \nabla_\mu \varphi (x) \nabla_\nu \varphi (x) f(x) \, d^4 x, \qquad f \in C^\infty_0(\mathbb{M}^4). 
\end{equation}
The associated Cauchy problem takes the form 
\begin{equation}
\label{Eq: Cauchy problem on M}
    \begin{cases}
        P_0 \varphi \doteq \Box_\eta \varphi = 0, \\
        \varphi \vert_\Sigma = \varphi_0, \; \; n^\mu \nabla_\mu \varphi \vert_\Sigma = \varphi_1, 
    \end{cases}
\end{equation}
where $\varphi_0, \varphi_1 \in C^\infty_0(\Sigma)$, while $n$ is the outward-pointing unit normal vector field to the initial time Cauchy hypersurface $\Sigma$. Via the conformal embedding $\iota: \mathbb{M}^4 \hookrightarrow \tilde{\mathcal{M}}$ into the Einstein static universe $(\tilde{\mathcal{M}}, \tilde{g})$, we can equivalently formulate the Cauchy problem in Equation \eqref{Eq: Cauchy problem on M} on $(\tilde{\mcM}, \tilde{g})$ as follows: 
\begin{equation}
    \label{Eq: Cauchy problem on ESU}
    \begin{cases}
        \tilde{P}_0 \phi \doteq (\Box_{\tilde{g}} - 1) \phi = 0, \\
        \phi \vert_S = \phi_0, \; \; \tilde{n}^\mu \nabla_\mu \phi \vert_S = \phi_1, 
    \end{cases}
\end{equation}
where $\phi_0 \doteq \Omega^{-1}\varphi_0, \phi_1 \doteq \Omega^{-2}\varphi_1 \in C^\infty_0(S)$, while $\tilde{n} \doteq \Omega^{-1} n$ is the outward-pointing unit normal vector field to a constant Einstein-time Cauchy hypersurface $S$ in $\tilde{\mcM}$. We recall that $\Omega$ is the conformal factor individuated in Example \ref{Ex: Minkowski and ESU}, while the solutions of Equations \eqref{Eq: Cauchy problem on M} and \eqref{Eq: Cauchy problem on ESU} are related by the identity $\phi=\Omega^{-1}\varphi$ The initial value problem in Equation \eqref{Eq: Cauchy problem on ESU} is commonly referred to in the literature as the \emph{conformal wave equation}, owing to its conformal invariance. Since $(\widetilde{\mcM},\widetilde g)$ is globally hyperbolic in this setting, we may take $\mathscr V=\widetilde{\mcM}$ and henceforth omit the subscript $\mathscr V$ from the corresponding propagators. In this setting, convergence of the Neumann series on the right-hand side of Equation \eqref{Eq: Neumann series} has been proven in \cite[Lem. B1]{Drago_2016}. We recall here the statement of this result limited to those aspects which will be relevant for our analysis. 
\begin{lemma}
\label{Lem: Drago PPA}
Let $(\mathbb{M}^4, g)$ be four-dimensional Minkowski spacetime and consider a real Klein-Gordon field thereon, whose action is as per Equation \eqref{Eq: free action}. Furthermore, let $V \in \mathcal{F}_{\mathrm{loc}} (\mathbb{M}^4)$ be as in Equation \eqref{Eq: massive potential}. Then, the Neumann series in Equation \eqref{Eq: Neumann series} converges uniformly to $\mathscr{R}_{m^2 V}(\Phi_f)$ in the topology of $\mathcal{E}(\mathbb{M}^4)$. 
\end{lemma}

A separate problem that needs to be addressed concerns the convergence of the boundary series in Equation \eqref{Eq: projected massive field 1}. To this end, we shall apply suitable norm estimates obtained from the analysis of the Goursat problem in the Einstein static Universe. We shall outline succinctly the main definitions and results, referring to \cite{Baez_1990} for a complete overview of the topic. 

Let $S_\tau \doteq \{\tau\} \times \mathbb{S}^3 \subset \tilde{\mathcal{M}}$ be the constant-Einstein time spacelike Cauchy hypersurface in $(\tilde{\mcM}, \tilde{g})$, which is diffeomorphic to $\mathbb{S}^3$ for any $\tau \in \mathbb{R}$. The subscript $\tau$ will be omitted whenever its value is irrelevant for the ensuing discussion. For notational ease, we denote the Cauchy data by $\Phi(\tau) \doteq \phi(\tau)$, $\Phi'(\tau) \doteq \partial_\tau \phi(\tau)$ and by $\phi^S(\tau) \doteq \Phi(\tau) \oplus \Phi'(\tau)$ the full Cauchy datum. Instead of restricting the attention to smooth, compactly supported initial data on $S$, we may enlarge the class of admissible initial data for Equation \eqref{Eq: Cauchy problem on ESU} by considering those that give rise to finite-energy solutions. We can thus define the {\bf finite-energy space} by 
\begin{equation}
    \label{Eq: finite energy space}
    H(S) \doteq H^1(S) \oplus L^2(S). 
\end{equation}
This is a Hilbert space, equipped with the inner product 
\begin{equation}
    \label{Eq: inner product on finite energy space}
    \langle \Phi_1 \oplus \Phi_2, \Psi_1 \oplus \Psi_2\rangle_{H(S)} \doteq \langle \Phi_1, (-\Delta_{\mathbb{S}^3} + 1) \Psi_1 \rangle_{L^2(S)} + \langle \Phi_2, \Psi_2 \rangle_{L^2(S)},
\end{equation}
for all $\Phi_1, \Psi_1 \in H^1(S), \; \Phi_2, \Psi_2 \in L^2(S).$ The induced norm $\| \cdot \|_{H(S)}$ is called \emph{finite-energy norm}. In addition, let $H_\infty(S) \doteq C^\infty(S) \oplus C^\infty(S)$ be the space of smooth initial data on $S$. Note that the first order generator of the Cauchy evolution is the operator
\begin{equation}
\label{Eq: Cauchy evolution operator}
    A = \begin{pmatrix}
        0 & 1\\
        \Delta_{\mathbb{S}^3} - 1 & 0
    \end{pmatrix}. 
\end{equation}

\noindent We can now introduce the space of all finite-energy solution of Equation \eqref{Eq: Cauchy problem on ESU}. 

\begin{definition}
    \label{Def: finite energy solution}
Let $(\tilde{\mcM}, \tilde{g})$ be Einstein static universe with global coordinates $(\tau, \rho, \omega) \in \mathbb{R} \times [0, \pi] \times \mathbb{S}^2$ and let $H(S_\tau)$ be the finite-energy space in Equation \eqref{Eq: finite energy space}, the subscript $\tau$ denoting the choice of an arbitrary but fixed initial, Cauchy surface. We say that $\phi: \tilde{\mathcal{M}} \to \mathbb{R}$ is a {\bf finite-energy solution} of the conformal wave equation if it solves
\begin{equation}
\label{Eq: Goursat problem}
    \begin{cases}
        (\Box_{\tilde{g}} - 1) \phi = 0, \\
        \phi^S(\tau) \doteq \Phi(\tau) \oplus \Phi'(\tau) \in H(S_\tau), \quad \text{for all} \; \tau \in \mathbb{R}.
    \end{cases}
\end{equation}
We denote the space of finite-energy solutions by $H$ and we endow it with the inner product defined by
\begin{equation}
    \label{Eq: inner product on H}
    \langle \phi, \psi \rangle_E \doteq \langle \phi^S(\tau), \psi^S(\tau)\rangle_{H(S_\tau)}, \qquad \forall \phi, \psi \in H, 
\end{equation}
which in turn induces the \emph{energy norm}
\begin{equation}
\label{Eq: energy norm}
    \| \phi \|_E \doteq \| \phi^S(\tau) \|_{H(S_\tau)}, \qquad \forall \phi \in H, \tau \in \mathbb{R}. 
\end{equation}
Observe that, Equation \eqref{Eq: inner product on H} is independent from the chosen value of $\tau$. Furthermore, we set 
\begin{equation}
    \label{Eq: H infty}
    H_\infty \doteq \{ \phi \in H \, : \, \phi^S(\tau) \in H_\infty(S_\tau) \; \text{for some (equivalently, every)} \; \tau \in \mathbb{R} \}. 
\end{equation}
\end{definition}

Since our ultimate goal is to analyse the behaviour of the field at null infinity, it is more natural to reformulate the evolution problem in terms of data prescribed on a null hypersurface. We therefore switch from the standard Cauchy problem in Equation \eqref{Eq: Cauchy problem on ESU}, with initial data assigned on a constant-Einstein-time hypersurface $S$, to the corresponding Goursat problem with data prescribed on a lightcone: 
\begin{equation}
    \label{Eq: lightcone in ESU}
    C_t^\pm \doteq \{(\tau, \rho,\omega) \in \mathbb{R} \times [0, \pi] \times \mathbb{S}^2 \, : \, \tau \pm \rho = t\}, \qquad t \in \mathbb{R}. 
\end{equation}
As before, we will omit both the subscript $t$ and the superscript $\pm$ identifying the different null foliations, whenever no confusion can arise. We observe that every lightcone $C_t$ can be topologically identified with $\mathbb{S}^3$, although we stress that the embedding into $\tilde{\mcM}$ fails to be smooth at the endpoints $\rho = 0, \pi$. Henceforth, we shall also adopt the short-hand notation $\boldsymbol{\Phi}(t) \doteq \phi(t+\rho, \rho, \omega) \doteq \phi \vert_{C_t}$ to denote the restriction to the lightcone of a smooth finite-energy solution $\phi \in H_\infty$.

\begin{proposition}[see \protect{\cite[Prop. 4]{Baez_1990}}]
\label{Prop: 4 Baez}
    Let $(\tilde{\mcM}, \tilde{g})$ be the Einstein static Universe and let $H_\infty$ be the space of all smooth, finite-energy solutions as per Equation \eqref{Eq: H infty}. Furthermore, for some $t \in \mathbb{R}$, let $C_t$ be a lightcone in $\tilde{\mcM}$, see Equation \eqref{Eq: lightcone in ESU}. Then, for all $\phi, \psi \in H_\infty$ and for every $t \in \mathbb{R}$, it holds that 
    \begin{equation}
        \label{Eq: prop 4}
        \langle \phi, \psi \rangle_E = \langle\boldsymbol{\Phi}(t),\boldsymbol{\Psi}(t)\rangle_{H^1(C_t)}:=\int_{C_t} [\boldsymbol{\Phi}(t) (-\Delta_{\mathbb{S}^3} + 1) \boldsymbol{\Psi}(t)] \, d\mu_{C_t},
    \end{equation}
where $\langle \cdot, \cdot \rangle_E$ is the inner product on $H_\infty$ inherited from that on $H$ as per Equation \eqref{Eq: inner product on H}, while $d\mu_{C_t}$ denotes the standard volume measure on $C_t$ introduced via the identification $C_t \simeq \mathbb{S}^3$.  
\end{proposition}

A direct consequence of Proposition \ref{Prop: 4 Baez} is that the map $H_\infty \ni \phi \to \boldsymbol{\Phi} \in H^1(C_t)$ is one-to-one for any $t \in \mathbb{R}$. In addition, such map is isometric with respect to the inner product on $H$ as per Equation \eqref{Eq: inner product on H} and to the one on $H^1(C_t)$ defined by the right-hand side of Equation \eqref{Eq: prop 4}, \textit{cf.} \cite[Prop. 5]{Baez_1990}. Hence, let us denote by 
\begin{equation}
    \label{Eq: isometry}
    U_t: H \to H(C_t) \subseteq H^1(C_t), \qquad \text{for any} \; t \in \mathbb{R}
\end{equation}
the unique isometry induced by the map above, where $H(C_t)$ denotes the range of $U_t$. We equip $H(C_t)$ with the inner product on the right-hand side of Equation \eqref{Eq: prop 4} and we denote by $$H_\infty(C_t)\doteq U_t(H_\infty) \equiv H_\infty \vert_{C_t},$$ the space of Goursat data obtained by restricting smooth finite-energy solutions of the conformal wave equation to $C_t$. Notice that $H_\infty(C_t)\subset H^1(C_t)$. Since $H^1(C_t) = H(C_t)$ for every $t \in \mathbb{R}$, see \cite[Thm. 8]{Baez_1990}, the above argument actually entails that $U$ is a bijection onto $H^1(C_t)$ and, thus, that every function in $H^1(C_t)$ is the restriction to the lightcone of some finite-energy solution lying in $H$. Consequently, we can equivalently use the symbols $\langle \cdot, \cdot \rangle_E$ and $\| \cdot\|_E$ to denote, respectively, the inner product and norm on $H^1(C_t)$. On account of the above discussion, we can regard $H^1(C_t)$ as the natural functional space in which initial data for the Goursat problem in Equation \eqref{Eq: Goursat problem} lie. For this reason, in
the following we shall refer to it also as the \emph{space of Goursat data}. 

Let us now consider the subset 
\begin{equation}
    \label{Eq: D(t,t')}
    D(t; t') \doteq \{ (\tau, \rho, \omega) \in \mathbb{R} \times [0, \pi] \times \mathbb{S}^2 \, : \, \rho + t \le \tau \le \rho + t'\}, \qquad t, t' \in \mathbb{R},\; t < t', 
\end{equation}
whose boundaries correspond to the lightcones $C_t$ and $C_{t'}$. Moreover, let 
\begin{equation*}
    L_0: H_\infty(C) \to H_\infty(C), \qquad \boldsymbol{\Phi}(t) \mapsto L_0 \boldsymbol{\Phi}(t) = \frac{1}{2 \sin \rho} \int_0^\rho \sin \rho' (-\Delta_{\mathbb{S}^3} + 1) \boldsymbol{\Phi}(t) \, d\rho', 
\end{equation*}
where we have suppressed the subscript $t$ for notational brevity. Notice that $L_0$ is not defined at the tips of the lightcone $\rho = 0, \pi$. It is possible to show that $L_0 \boldsymbol{\Phi}(t) = \partial_t \boldsymbol{\Phi}(t) = U A \vert_{H_\infty} U^{-1}$, where $A$ is as per Equation \eqref{Eq: Cauchy evolution operator}. Furthermore $L_0$ is essentially skew-adjoint on $H(C)$, \textit{cf.}, \cite[Prop. 6 \& 7]{Baez_1990}. Therefore, we can denote by $L \doteq U A U^{-1}$ its closure.

The following lemma on the inhomogeneous wave equation hence gives a useful norm estimate to prove the convergence of the series in Equation \eqref{Eq: projected massive field 1}. 

\begin{lemma}[\textit{cf.}, \protect{\cite[Lem. 10]{Baez_1990}}]
\label{Lem: norm estimate 1}
Let $\tilde{f} \in C^\infty(\tilde{\mcM})$ be such that $\operatorname{supp} \tilde{f} \subset \operatorname{int} D(t;t')$ for some $t,t' \in \mathbb{R}$ with $t < t'$, see Equation \eqref{Eq: D(t,t')}. If $\boldsymbol{\Phi}_0 \in H_\infty(C_t)$, there exists a unique $\phi \in C^\infty(\tilde{\mcM}; \mathbb{R})$ such that 
\begin{equation*}
    \begin{cases}
        (\Box_{\tilde{g}} - 1)\phi = \tilde{f}, \\
        \boldsymbol{\Phi}(t) = \boldsymbol{\Phi}_0.
    \end{cases}
\end{equation*}
The Goursat evolution map 
\begin{equation}
    \label{Eq: Goursat evolution map}
    K_f(t; t'): H_\infty(C_t) \to H_\infty(C_{t'}), \qquad \boldsymbol{\Phi}(t) \mapsto K_f(t;t') \boldsymbol{\Phi}(t) \doteq \boldsymbol{\Phi}(t') 
\end{equation}
extends to an isometric affine map from $H(C_t)$ to itself and the assignment $f \mapsto  K_f(t; t')\boldsymbol{\Phi}(t)$ is continuous in the topology of $L^2(D(t;t'))$. In addition, there exists $c > 0$ such that
\begin{equation}
    \label{Eq: norm estimate}
    \| \boldsymbol{\Phi}(t') - e^{(t'-t) L} \boldsymbol{\Phi}(t)\|_E \le c \sqrt{(t'-t) + \pi} \|\tilde{f}\|_{L^2(\tilde{\mcM})}.
\end{equation}
\end{lemma}

\noindent We have all the ingredients to prove the first key result of this section.

\begin{theorem}
\label{Thm: convergence of the boundary series}
The series 
\begin{equation}
    \Gamma(\mathscr{R}_{m^2V}(\Phi_f)) \doteq \sum_{n \ge 0} (m^2)^n \tilde{\Delta}_\mathrm{R} (\Omega^{-3} f_n) \bigg\vert_{\mathscr{I}^+}, \qquad f_n \doteq (h \circ \Delta_{\mathrm{A}})^n (f), \; f \in C^\infty_0(\mathbb{M}^4) \,,
\end{equation}
see also Equation \eqref{Eq: projected massive field 1}, is convergent in the topology induced by the energy norm $\| \cdot \|_E$ on $H^1(C_\pi)$, see Equation \eqref{Eq: energy norm}.
\end{theorem}

\begin{proof}
We begin by observing that, under the conformal embedding $\iota: \mathbb{M}^4 \hookrightarrow \tilde{\mcM}$, the conformal boundary of $\iota(\mathbb{M}^4)$ consists of two subsets: $\partial \iota(\mathbb{M}^4) \doteq C_\pi^+ \cup C^+_{-\pi}$, where $C^+_{\pm \pi} \doteq \{\tau + \rho = \pm \pi\}$ and that $\mathscr{I}^+ \subset C^+_\pi$. Note that $C^+_{-\pi}$ is an auxiliary past cone, different from the past conformal boundary $C^-_{-\pi} = \mathscr{I}^- \cup \{i^-, i_0\}$. Furthermore, on $(\mathbb{M}^4, \eta)$, the retarded fundamental solution $\Delta_\mathrm{R}$ of the massless wave equation satisfies the sharp Huygens principle. Its distributional kernel is supported on pairs of points connected by future-directed null geodesics, rather than throughout the causal future. Under the conformal embedding $\iota$, these null geodesics terminate at future null infinity $\mathscr{I}^+$. Hence, for every $\chi \in C^\infty_0(\mathbb{M}^4)$, since $$\tilde{\Delta}_\mathrm{R} (\Omega^{-3} \chi) = \Omega^{-1} \Delta_\mathrm{R} \chi, \qquad \text{on} \; \; \iota(\mathbb{M}^4),$$ the support of $\tilde{\Delta}_\mathrm{R} (\Omega^{-3} \chi) \in \mathcal{E}(\tilde{\mcM})$ does not meet $i^+$, future timelike infinity. More precisely, its closure in the conformal completion intersects the future boundary only at $\mathscr{I}^+$. This entails that 
\begin{equation*}
   \operatorname{supp} \left(\tilde{\Delta}_\mathrm{R} (\Omega^{-3} \chi) \bigg\vert_{C^+_\pi}\right) \subset \mathscr{I}^+, 
\end{equation*}
namely, the trace on $C_\pi^+$ can be identified with the projected datum on $\mathscr{I}^+$ and, with a mild abuse of notation, we may write $\tilde{\Delta}_\mathrm{R} (\Omega^{-3} \chi) \vert_{C^+_\pi} = \tilde{\Delta}_\mathrm{R} (\Omega^{-3} \chi) \vert_{\mathscr{I}^+}$. Moreover, it holds that 
\begin{equation*}
    \tilde{\Delta}_\mathrm{R} (\Omega^{-3} \chi) \bigg\vert_{C^+_{-\pi}} = 0,
\end{equation*}
since the retarded Green operator propagates only in the causal future of $\operatorname{supp} \chi$. Thus, the $n$-th term in Equation \eqref{Eq: projected massive field 1} is nothing but the restriction to $\mathscr{I}^+$ of a smooth solution to the following Goursat problem: 
\begin{equation*}
    \begin{cases}
        (\Box_{\tilde{g}} - 1) \tilde{\Delta}_\mathrm{R} (\Omega^{-3} \chi_n) = \Omega^{-3} \chi_n, \\
        \tilde{\Delta}_\mathrm{R} (\Omega^{-3} \chi_n) \vert_{C_{-\pi}} = 0,
    \end{cases}
\end{equation*}
where we set $\chi_n \doteq (h \circ \Delta_{\mathrm{A}})^n f \in C^\infty_0(\mathbb{M}^4)$. Observe that the source term $\Omega^{-3} \chi_n \in C^\infty(\tilde{\mcM})$ and it is such that $\operatorname{supp} (\Omega^{-3} \chi_n) \subset \operatorname{int} D(-\pi; \pi)$. Therefore, on account of Lemma \ref{Lem: norm estimate 1}, we have that there exists $c > 0$ such that 
\begin{equation}
    \label{Eq: norm estimate 1}
    \| \tilde{\Delta}_\mathrm{R} (\Omega^{-3} \chi_n) \vert_{\mathscr{I}^+} \|_E \le c \sqrt{3\pi} \, \|\Omega^{-3} \chi_n\|_{L^2(\tilde{\mcM})}, \qquad \text{for all} \; n \ge 0. 
\end{equation}
We need to devise suitable upper bounds for the norm appearing on the right-hand side in the previous estimate. To this avail, fix a compact subset $K \subset \mathbb{M}^4$ such that $\operatorname{supp} f, \operatorname{supp} h \subset K$. We shall treat the cases $n=0$ and $n > 0$, separately. Setting $n = 0$, we obtain 
\begin{equation*}
    \|\Omega^{-3} f \|_{L^2(\tilde{\mcM})} \doteq \left(\int_{\tilde{\mcM}} |\Omega^{-3} f|^2 \, d\mu_{\tilde{g}} \right)^{\frac12} = \left(\int_{\mathbb{M}^4} \Omega^{-2} |f|^2 \, d^4 x \right)^{\frac12} \le \| f\|_{L^\infty (K)} \|\Omega^{-1}\|_{L^2(K; d^4x)}, 
\end{equation*}
where in the second equality we perform a change of coordinates from the globally-defined $(\tau, \rho, \omega)$ on the Einstein static universe to the standard Cartesian ones on $\mathbb{M}^4$. Note that the last term is finite since the conformal factor $\Omega$ is strictly positive on any compact set in $\mathbb{M}^4$, while $f$ is a bounded function, being smooth and compactly supported. 

Let us now turn to the case $n > 0$. As before, we can write
\begin{flalign}
\label{Eq: estimate n ge 0}
    \|\Omega^{-3} (h \circ \Delta_\mathrm{A})^n (f) \|_{L^2(\tilde{\mcM})} &= \left(\int_{\mathbb{M}^4} \Omega^{-2} |(h \circ \Delta_\mathrm{A})^n (f)|^2 \, d^4 x \right)^{\frac12} \notag \\&\le \| (\Delta_\mathrm{A} \circ h)^{n-1} \Delta_\mathrm{A} (f)\|_{L^\infty (K)} \|\Omega^{-1} h\|_{L^2(K; d^4x)}. 
\end{flalign}
Observe that both factors are finite and the particular factorization above, with $h$ appearing in the $L^2$-norm, is adopted solely for future convenience. To prove convergence of the boundary series, fix $N \ge 0$ and consider the partial sums
$$S_N \doteq \sum_{n=0}^N (m^2)^n \tilde{\Delta}_\mathrm{R}(\Omega^{-3} \chi_n) \bigg\vert_{C^+_\pi}.$$ Applying in turn the Goursat estimate in Equation \eqref{Eq: norm estimate 1} and the advanced analogue of the estimate derived in the proof of \cite[Lem. B1]{Drago_2016}, we obtain that, for all $N, M \ge 0$, with $N > M$, 
\begin{equation*}
    \|S_N - S_M\|_E \le c \sqrt{3\pi} \sum_{n=M+1}^N (m^2)^n \| \Omega^{-3} \chi_n\|_{L^2(\tilde{\mcM})} \le C \sum_{n=M+1}^N (m^2)^n \frac{A^{n-1}}{(n-1)!}, 
\end{equation*}
for some positive constants $C, A > 0$ independent of $n$. 
On account of this last estimate, we can conclude that $\{S_N\}_{N \ge 0}$ is a Cauchy sequence in $H^1(C^+_\pi)$ and, being $H^1(C^+_\pi)$ Hilbert, it converges therein to the right-hand side in Equation \eqref{Eq: projected massive field 1}. We can therefore establish the following estimate for the whole boundary series regarded now as an element of the boundary energy space $H^1(C^+_\pi)$: 
\begin{equation*}
    \| \Gamma(\mathscr{R}_{m^2V} \Phi_f)\|_E \le \tilde{c}_0 \|f\|_{L^\infty(K)} + \sum_{n \ge 1} \tilde{c}_n \|(\Delta_\mathrm{A} \circ h)^{n-1} \Delta_\mathrm{A}(f)\|_{L^\infty(K)}, 
\end{equation*}
where $$\tilde{c}_n \doteq (m^2)^n c \sqrt{3\pi} \max \{\| \Omega^{-1} h\|_{L^2(K, d^4x)}, \| \Omega^{-1}\|_{L^2(K, d^4x)}\}, \qquad \text{for any}\; n \ge 0.$$ We note that each term of the form $(\Delta_\mathrm{A} \circ h)^{n-1} \Delta_\mathrm{A}(f)$ coincides with the coefficient of order $n-1$ in the advanced M\o ller expansion of $a_{m^2V}(\Delta_{\mathrm A}f)$, see Definition \ref{Eq: classical retarded Moller map}. This observation provides an \emph{a posteriori} justification for placing the factor $h$ in the second term of the factorization in Equation \eqref{Eq: estimate n ge 0}. To conclude, we need to prove that we can further estimate the $L^\infty$-norm of this function on the compact set $K$. To this end, let us introduce a second compact set in $\mathbb{M}^4$ of the form $K' \doteq [t_0, t_1] \times N$, where $t_0, t_1 \in \mathbb{R}$ with $t_0 < t_1$ while $N \Subset \mathbb{R}^3$ is compact and chosen in such a way that $K \subset K'$. On $C^\infty_\mathrm{sc}(\mathbb{M}^4)$ we define the norm
\begin{equation}\label{Eq: norm 1 infty}
\|\widehat{\psi}\|_{1, \infty} \doteq \sup_{t \in [t_0, t_1]} \| \widehat{\psi}(t, \cdot)\|_{L^1(\mathbb{R}^3)}, \qquad \psi \in C^\infty_{\mathrm{sc}}(\mathbb{M}^4), 
\end{equation}
where $\widehat{(\cdot)}$ denotes the spatial Fourier transform. Note that, on account of Lemma \ref{Lem: Drago PPA}, the uniform norm of $\psi$ is bounded from above by the norm in Equation \eqref{Eq: norm 1 infty}, namely, $\|\psi\|_{L^\infty(K)} \le \|\widehat{\psi}\|_{1, \infty}$. More precisely, following slavishly the advanced analogue of the proof in \cite[App. B]{Drago_2016}, we have that 
\begin{flalign*}
    \| (\Delta_{\mathrm{A}} \circ h )^{n-1} \Delta_\mathrm{A} (f) \|_{L^\infty(K)} \le b_{n-1} \frac{(t_1 - t_0)^{2(n-1)} \|\hat{h}\|^{n-1}_{1, \infty}}{(n-1)!} \|\widehat{\Delta_\mathrm{A} (f)}\|_{1, \infty}, 
\end{flalign*}
where $b_{n-1} > 0$ are suitable powers of $2\pi$ arising from the spatial Fourier transform. Thus, we conclude that 
\begin{equation}
\label{Eq: norm estimate of the projected massive observable}
     \| \Gamma(\mathscr{R}_{m^2V} \Phi_f)\|_E \le \tilde{c}_0 \|f\|_{L^\infty(K)} + \sum_{n \ge 1} \tilde{c}_n b_{n-1} \frac{(t_1 - t_0)^{2(n-1)} \|\hat{h}\|^{n-1}_{1, \infty}}{(n-1)!} \|\widehat{\Delta_\mathrm{A} (f)}\|_{1, \infty}, 
\end{equation}
which is convergent since its tail is dominated by an exponential series. 
\end{proof}

Having proven convergence of the series defining the projected massive observable, we are left with one last problem to address. Observe that the result of Theorem \ref{Thm: convergence of the boundary series} does not by itself exclude the case in which the boundary series always converges to zero, regardless of $f \in C^\infty_0(\mathbb{M}^4)$. In the following we establish that this is exactly the case. This is tantamount to saying that, in the bulk, a field observable is trivially projected onto the future null infinity $\mathscr{I}^+$. At first glance this might look as a disaster, establishing the triviality of the boundary projection. Yet, we will show that this result is only telling us that taking a straightforward global adiabatic limit is not the correct procedure. On the contrary, if we add into the mix the time-slice axiom \ref{Prop: time slice}, we can bypass this hurdle projecting to $\mathscr{I}^+$ only the observables localized in a suitable neighbourhood of any but fixed Cauchy surface. Removing the cut-off in these subsets will turn out to be possible and to yield a non trivial projection.

\begin{theorem}
\label{Thm: convergence not to zero}
    Consider the projected massive field observable in Equation \eqref{Eq: projected massive field 1} and denote by $P_{m,h} \doteq P_0 + m^2 V^{(1)} = P_0 - m^2 h$, with $h \in C^\infty_0(\mathbb{M}^4)$, the Klein-Gordon operator on $(\mathbb{M}^4, \eta)$. Then, it holds that 
    \begin{equation}
        \Gamma(\mathscr{R}_{m^2 V} \Phi_f) = 0 \iff f = P_{m, h} \, q, \quad \text{for some} \; q \in C^\infty_0(\mathbb{M}^4), 
    \end{equation}
    or, equivalently, that the map 
    \begin{flalign}
    \label{Eq: quotient map}
    \notag
    \mathcal{T}_{m, h}:  \frac{C^\infty_0(\mathbb{M}^4)}{P_{m,h}[C^\infty_0(\mathbb{M}^4)]} &\to H_{\mathscr{I}^+} \\
   [f] &\mapsto  \Gamma(\mathscr{R}_{m^2 V} \Phi_f),
    \end{flalign}
    is well-defined and injective, where $H_{\mathscr{I}^+} \doteq \{\boldsymbol{\Psi} \in H^1(C^+_\pi) \, : \, \operatorname{supp}(\boldsymbol{\Psi}) \Subset \mathscr{I}^+\}$, \textit{cf.}, Theorem \ref{Thm: convergence of the boundary series}. 
\end{theorem}

\begin{proof}
Observe that the map $f \mapsto \Gamma(\mathscr{R}_{m^2 V} \Phi_f)$ descends uniquely to the quotient on account of its linearity and of the fact that its kernel is precisely $P_{m,h}[C^\infty_0(\mathbb{M}^4)]$. Hence, $\mathcal{T}_{m,h}$ in Equation \eqref{Eq: quotient map} is well-defined. 

To prove injectivity, let $K \Subset \mathbb{M}^4$ be a compact set such that $\operatorname{supp}(f), \operatorname{supp}(h) \subset K$. For any $N \in \mathbb{N}$, define
\begin{equation*}
  \mathsf g_{m^2}^N \doteq \sum_{n=0}^{N} (m^2)^n (h \circ \Delta_{\mathrm{A}})^n f \in C^\infty_K(\mathbb{M}^4) \subset C^\infty_0(\mathbb{M}^4),
\end{equation*}
where $C^\infty_K(\mathbb{M}^4) \doteq \{u \in C^\infty(\mathbb{M}^4) \, : \, \operatorname{supp}(u) \subset K\}$. By Theorem \ref{Thm: convergence of the boundary series}, we have that the sequence $\{\mathsf g_{m^2}^N\}_{N \ge 0}$ is convergent in the topology of $C^\infty_K(\mathbb{M}^4)$ and the limit is  
\begin{equation*}
    \mathsf g_{m^2} \doteq \sum_{n \ge 0} (m^2)^n (h \circ \Delta_{\mathrm{A}})^n f \in C^\infty_K(\mathbb{M}^4).
\end{equation*}
Let us now introduce the map 
\begin{equation}
    \label{Eq: massless radiation map}
    T_0: C^\infty_K(\mathbb{M}^4) \to H_\mathscr{I}^+, \qquad \mathsf g \mapsto T_0 \mathsf g \doteq \tilde{\Delta}_\mathrm{R}(\Omega^{-3} \mathsf g) \vert_{\mathscr{I}^+}. 
\end{equation}
Observe that, since the advanced contribution vanishes at future null infinity, the map $T_0$ coincides with the massless bulk-to-boundary map $\Gamma_{\mathscr I}$ introduced in Equation~\eqref{Eq: bulk to boundary map 1} and, hence, it preserves the corresponding symplectic forms. In addition, on account of Equation \eqref{Eq: norm estimate 1}, it is continuous in the relevant topologies. Applying it to the partial sums $\mathsf g_{m^2}^N$, we obtain 
\begin{equation*}
    T_0 \mathsf g_{m^2}^N \doteq \sum_{n=0}^N (m^2) \tilde{\Delta}_\mathrm{R}(\Omega^{-3} (h \circ \Delta_A)^n (f)) \vert_{\mathscr{I}^+} \doteq S_N, \qquad \forall N \ge 0
\end{equation*}
where $S_N$ are the partial sums defined in the proof of Theorem \ref{Thm: convergence of the boundary series}. Convergence of the sequence $\{S_N\}_{N \ge 0}$ in the energy norm established in the previous theorem yields 
\begin{equation*}
    S_N \doteq T_0 \mathsf g_{m^2}^N \to \Gamma(\mathscr{R}_{m^2V} \Phi_f) \qquad \text{in} \; H_{\mathscr{I}^+} \subset H^1(C_\pi^+), 
\end{equation*}
and, by continuity of $T_0$ and uniqueness of the limit, $T_0 \mathsf g_{m^2} = \Gamma(\mathscr{R}_{m^2V} \Phi_f)$. Hence, the problem reduces to characterizing the kernel of the map $T_0$. We show that $\operatorname{ker}(T_0) = P_0[C^\infty_0(\mathbb{M}^4)]$, proving each of the two inclusions separately. To this end, it is convenient to switch back to Equation \eqref{Eq: projected massive field} and replace $\tilde{\Delta}_\mathrm{R}$ by the causal propagator $\tilde{\Delta}^\mathscr{V}_\mathrm{S}$ in the definition of $T_0$. 

We start from the direct inclusion $\operatorname{ker}(T_0) \subset P_0 [C^\infty_0(\mathbb{M}^4)]$. Let $\mathsf g \in \operatorname{ker}(T_0) \subset C^\infty_0(\mathbb{M}^4)$, \textit{i.e.}, $T_0 \mathsf g = 0$. By the sharp Huygens principle then the corresponding Goursat datum $\tilde{\Delta}^\mathscr{V}_\mathrm{S} (\Omega^{-3} \mathsf g)$ must vanish on $C^+_\pi$. Since the Goursat evolution operator as per Equation \eqref{Eq: Goursat evolution map} is isometric, we have that $\tilde{\Delta}^\mathscr{V}_\mathrm{S} (\Omega^{-3} \mathsf g) = 0$ and by conformal covariance it also holds that $\Delta_\mathrm{S} \mathsf g = 0$. Using the standard exact sequence for normally hyperbolic operators, we can conclude that 
\begin{equation*}
    \operatorname{ker}(\Delta_{\mathrm{S}}: C^\infty_0(\mathbb{M}^4) \to C^\infty_{\mathrm{tc}}(\mathbb{M}^4)) = P_0 [C^\infty_0(\mathbb{M}^4)]. 
\end{equation*}

The reverse inclusion, namely, $P_0 [C^\infty_0(\mathbb{M}^4)] \subset \operatorname{ker}(T_0)$, follows straightforwardly. As a matter of fact, if $\mathsf g = P_0 q$ for some $q \in C^\infty_0(\mathbb{M}^4)$, then $\Delta_\mathrm{S} \mathsf g \vert_{\mathscr{I}^+} = 0$ implies that $T_0 \mathsf g = 0$ and, hence, $\mathsf g \in \operatorname{ker}(T_0)$. We can now transport this kernel condition to $f$ via the action of the M\o ller operator. More precisely, since the series defining $\mathsf g_{m^2}$ is convergent, we can write 
\begin{equation*}
    \mathsf g_{m^2} = (\operatorname{id} - m^2 (h \circ \Delta_\mathrm{A}))^{-1} f \Longrightarrow f = (\operatorname{id} - m^2 (h \circ \Delta_\mathrm{A})) \mathsf g_{m^2}. 
\end{equation*}
Assume that there exists $q \in C^\infty_0(\mathbb{M}^4)$ such that $\mathsf g_{m^2} = P_0 q$, then $f = (P_0 - m^2 h) q \doteq P_{m,h} q$, where we used that $\Delta_{\mathrm{A}} P_0 q = q$. Conversely, if $f = P_{m,h} q$ for $q \in C^\infty_0(\mathbb{M}^4)$, then a few algebraic manipulations of the above formula yield $\mathsf g_{m^2} = P_0 q$. Combining everything together, we have that 
\begin{equation*}
    \Gamma(\mathscr{R}_{m^2V} \Phi_f) = 0 \iff T_0 \mathsf g_{m^2} = 0 \iff \mathsf g_{m^2} \in P_0 C^{\infty}_0(\mathbb{M}^4) \iff f \in P_{m,h} [C^{\infty}_0(\mathbb{M}^4)], 
\end{equation*}
which concludes the proof.
\end{proof}

\begin{remark}
Notice that the result established in Theorem \ref{Thm: convergence not to zero} uses genuinely different techniques with respect to the ones adopted in \cite{Dappiaggi_2008}. Therein the massive data at null infinity are constructed directly using harmonic analysis on the mass hyperboloid $H_m$ rather than this perturbative M\o ller construction. For completeness, we mention that in the aforementioned reference, and particularly in Theorem 2.2, it is shown that every norm-finite massive Klein-Gordon solution on Minkowski spacetime yields a triplet $(\Psi, U', \tilde{T})$ on future null infinity, where $\Psi$ is the boundary datum of an auxiliary massless field, $U'$ is the quasi-regular Lorentz representation on $L^2(C)$ and $\tilde{T}: L^2(C) \to L^2(H_m) \oplus L^2(H_m)$ intertwines the light-cone representation with the massive hyperboloid one, thereby retaining the information needed to reconstruct the massive sector.
\end{remark}

Theorems \ref{Thm: convergence of the boundary series} and \ref{Thm: convergence not to zero} entail that the massive field observable $\mathscr{R}_{m^2 V} \Phi_f$ admits a well-defined projection to future null infinity and, therefore, that the projected massive observable identifies a boundary counterpart lying in $\mathscr{A}_e(\mathscr{I}^+)$. Observe that the estimate in Equation \eqref{Eq: norm estimate of the projected massive observable} fails if we replace the infrared cut-off $h$ by the constant function $1$ on $\mathbb{M}^4$. Indeed, a direct inspection tells us that the estimate depends critically on the cut-off through quantities such as $\| \Omega^{-1} h\|_{L^2}$, $(t_1 - t_2)$ and $\|\hat{h}\|_{1, \infty}$. Hence, the whole procedure outlined above is not stable under the adiabatic limit $h \to 1$ and prompts the natural question of whether such cut-off can be removed by weakening the type of convergence required and/or restricting the admissible class of cut-offs. The answer is affirmative and, for the sake of readability, we shall proceed one step at a time. 

\vskip .2cm

\textbf{Step 1: $\mathsf g_R \to \mathsf g_\infty$ in $C^\infty_{\mathrm{loc}}(\mathbb{M}^4)$ and the convergence is eventually stable on every compact set.} 
Let $K_R \Subset \mathbb{M}^4$ be an exhaustion of $\mathbb{M}^4$ by compact sets, namely, a family of compact subsets of $\mathbb{M}^4$ such that $K_R \subset \operatorname{int} K_{R'}$ for every $R' > R$ and $\bigcup_{R \in \mathbb{N}} K_R = \mathbb{M}^4$. Let $h_R \in C^\infty_0(\mathbb{M}^4)$ such that $0 \le h_R \le 1$ with $h_R \equiv 1$ on $K_R$ and $\operatorname{supp}(h_R) \subset K_{R+1}$. We stress that, although at this stage we are free to select an arbitrary exhaustion as above, to derive the sought-after estimates in Step 4 we will need to consider one which is specifically adapted to the causal structure of the underlying spacetime. Using the resolvent identity, define
\begin{equation*}
    \mathsf g_R \doteq f + m^2 (h_R \circ \Delta_{\mathrm A}^{m, h_R}) f, \qquad \mathsf g_\infty \doteq f + m^2 \Delta_{\mathrm A}^{m} f = P_0 \Delta_\mathrm{A}^m f, 
\end{equation*}
where $\Delta_{\mathrm{A}}^{m, h_R}$ is the advanced Green operator of $P_{m, h_R} \doteq P_0 - m^2 h_R$, while $\mathsf g_\infty \in C^\infty_{\mathrm{sc}}(\mathbb{M}^4)$ is the natural candidate bulk limit as $R \to \infty$. Fix an arbitrary compact set $L \Subset \mathbb{M}^4$ and consider the seminorms
\begin{equation*}
 p_{L,N} (\mathsf g_R - \mathsf g_\infty) = \max_{|\alpha| \le N} \sup_{x \in L} |\partial^{\alpha}(\mathsf g_R - \mathsf g_\infty)(x)|, \qquad \forall N \in \mathbb{N}. 
\end{equation*}
We want to prove that $p_{L, N}(\mathsf g_R - \mathsf g_\infty) \to 0$ as $R \to \infty$. To this end, denote by $u_R \doteq \Delta_\mathrm{A}^{m,h_R} f$ and by $u \doteq \Delta_\mathrm{A}^{m} f$. Consequently, we can rewrite $$\mathsf g_R = f + m^2 h_R u_R, \qquad \text{and} \qquad \mathsf g_\infty = f + m^2 u.$$ To control derivatives, let $L_1 \Subset \mathbb{M}^4$ be such that $L \subset \operatorname{int} L_1$. Since $f$ has compact support, choose $\Sigma_T \doteq \{t=T\}$ lying strictly in the future of both $L_1$ and $\operatorname{supp}(f)$ and consider the compact causal slab
\begin{equation*}
    D_L \doteq J^+(L_1) \cap J^-(\Sigma_T).
\end{equation*}
Observe that, since $\{K_R\}_{R \in \mathbb{N}}$ exhausts $\mathbb{M}^4$, there exists $R_L \in \mathbb{N}$ such that $h_R \vert_{D_L} \equiv 1$ for all $R \ge R_L$. Therefore, on $D_L$, $P_{m, h_R} = P_m$. Moreover, on account of the causal support properties of the advanced Green operators, both $u_R$ and $u$ vanish on $J^+(\operatorname{supp}(f))$. Hence, in particular, they have vanishing Cauchy data on $\Sigma_T$, \textit{i.e.}, $u_R \vert_{\Sigma_T} = u \vert_{\Sigma_T} = 0$ and $\partial_t u_R \vert_{\Sigma_T} = \partial_t u \vert_{\Sigma_T} = 0$. On the domain $D_L$ they therefore solve the same Cauchy problem with the same future Cauchy data. Uniqueness then gives $u_R = u$ on a neighbourhood of $L$ and, thus, $\mathsf g_R = \mathsf g_\infty$ therein. We can conclude that, for every $N \in \mathbb{N}$, it holds that $p_{L,N}(\mathsf g_R - \mathsf g_\infty) = 0$ for all sufficiently large $R$. The sought result descends from the arbitrariness of $L$. 

\vskip .2cm

\textbf{Step 2: An explicit formula for $T_0^* \chi$, $\chi \in C^\infty_0(\mathscr{I}^+)$.} Let us fix $R \in \mathbb{N}$ and consider the map $T_0$ as per Equation \eqref{Eq: massless radiation map} acting on $\mathsf g \equiv \mathsf g_R \in C^\infty_0(\mathbb{M}^4)$: $T_0 \mathsf g_R \doteq \tilde{\Delta}_\mathrm{R} (\Omega^{-3} \mathsf g_R) \vert_{\mathscr{I}^+} \in H^+_{\mathscr{I}^+}$. We start by observing that, inside $\iota(\mathbb{M}^4)$, it holds $\tilde{\Delta}_\mathrm{R} (\Omega^{-3} \mathsf g) = \Omega^{-1} \Delta_\mathrm{R} \mathsf g$, which allows us to compute the boundary value using the retarded propagator $\Delta_\mathrm{R}$ on Minkowski spacetime. Let $x = r \omega$ with $\omega \in \mathbb{S}^2$ and set $u = t-r$. Note that, in the limit $r \to \infty$, we are approaching future null infinity $\mathscr{I}^+$. From  
\begin{equation}
\label{Eq: retarded propagator in null coordinates}
    \Delta_{\mathrm{R}} \mathsf g (r+u, r\omega) = - \frac{1}{4\pi} \int_{\mathbb{R}^3} \frac{\mathsf g(r+ u - |r\omega - y|, y)}{|r\omega - y|} \, d^3 y, 
\end{equation}
since $\mathsf g$ is compactly supported, its spatial projection has support lying within a compact set $B \Subset \mathbb{R}^3$. Consequently, only the points $y \in B$ contribute to the integral on the right-hand side of Equation \eqref{Eq: retarded propagator in null coordinates}. Expanding $|r\omega - y| = r - \omega \cdot y + \mathcal{O}(r^{-1})$, while $r+u - |r\omega -y| = u + \omega \cdot y + \mathcal{O}(r^{-1})$, it holds $\frac{r}{|r\omega -y|} \xrightarrow[r \to \infty]{} 1$. Hence, multiplying by $r$ both sides of Equation \eqref{Eq: retarded propagator in null coordinates}, we have  
\begin{equation}
    \label{Eq: auxiliary 1 retarded}
    r  (\Delta_{\mathrm{R}} \mathsf g) (r+u, r\omega) = - \frac{1}{4\pi} \int_{B} \frac{r}{|r\omega - y|} \mathsf g(r+ u - |r\omega - y|, y) \, d^3 y. 
\end{equation}
Observe that the integrand converges pointwise to $\mathsf g(u+ \omega \cdot y, y)$. Since $B$ is a compact set while $\mathsf g$ is smooth and compactly supported on $\mathbb{M}^4$ with $\operatorname{supp} \mathsf g \subset B$, we can apply the dominated convergence theorem, thus obtaining 
\begin{equation}
    \label{Eq: ordinary radiation field}
    \lim_{r \to \infty}  r  \Delta_{\mathrm{R}} \mathsf g (r+u, r\omega) = - \frac{1}{4\pi} \int_{\mathbb{R}^3} \mathsf g(u + \omega \cdot y, y) \, d^3 y. 
\end{equation}
In order to recover the action of $T_0$, however, we need to include in the preceding equation the conformal factor $\Omega^{-1}$ and get rid of the multiplying factor $r$ in front of $\Delta_{\mathrm{R}} \mathsf g (r+u, r\omega)$. This can be done by observing that $$\lim_{r \to \infty} \frac{\Omega^{-1}}{r} = \lim_{r \to \infty} \sqrt{1 + u^2} \frac{\sqrt{1 + (u+2r)^2}}{2r} = \sqrt{1+u^2}.$$ Thus, 
\begin{equation}
    \label{Eq: explicit action of T0 1}
   T_0 \mathsf g (u, \omega) \doteq  \tilde{\Delta}_\mathrm{R} (\Omega^{-3} \mathsf g) \vert_{\mathscr{I}^+} = \lim_{r\to \infty} \frac{\Omega^{-1}}{r} r \Delta_{\mathrm{R}} \mathsf g (r+u, r\omega) = - \frac{\sqrt{1+u^2}}{4\pi} \int_{\mathbb{R}^3} \mathsf g(u + \omega \cdot y, y) d^3 y.
\end{equation}
To eliminate the prefactor $\sqrt{1+u^2}$, we perform a further conformal rescaling to a Bondi frame, in which $u$ is an affine parameter along the null generators of $\mathscr{I}^+$. Set $\mathfrak w \doteq \sec U$ and consider the conformal factor $\Omega_B = \mathfrak w \Omega = \frac{2}{\sqrt{1 + (u+2r)^2}}$. Note that, also in this case, $\lim_{r \to \infty} \frac{\Omega_B^{-1}}{r} = 1$ and, therefore, in the Bondi conormal frame
\begin{equation}
    \label{Eq: explicit action of T0}
   T_0 \mathsf g (u,\omega) = - \frac{1}{4\pi} \int_{\mathbb{R}^3} \mathsf g(u + \omega \cdot y, y) d^3 y.
\end{equation}
Despite the fact that $\sec U$ blows up as $U \to \pm \frac{\pi}2$, this does not constitute a problem as we work on the open manifold $\mathscr{I}^+ \simeq \mathbb{R}_u \times \mathbb{S}^2$ and we will smear $T_0 \mathsf g$ against $\chi \in C^\infty_0(\mathscr{I}^+) \equiv C^\infty_0(\mathbb{R} \times \mathbb{S}^2)$. We now possess all the ingredients to derive an explicit formula for the adjoint of $T_0$. Let $\chi \in C^\infty_0(\mathscr{I}^+)$ and use the Bondi measure $du \, d\omega$, where $\omega \in \mathbb{S}^2$. Then, Fubini theorem entails that 
\begin{flalign*}
   \langle T_0 \mathsf g, \chi \rangle & = - \frac{1}{4\pi} \int_\mathbb{R} du \int_{\mathbb{S}^2} d\omega \chi(u, \omega) \int_{\mathbb{R}^3} d^3y \, \mathsf g(u + \omega \cdot y, y) \\ &= - \frac{1}{4\pi}  \int_{\mathbb{R}^3} d^3y \int_\mathbb{R} dt \, \mathsf g(t, y) \int_{\mathbb{S}^2} d\omega \,  \chi (t-\omega \cdot y, \omega) = \langle \mathsf g, T_0^* \chi \rangle,
\end{flalign*}
where we adopt the change of variables $u \mapsto t = u + \omega \cdot y$, $du = dt$ for fixed $\omega \in \mathbb{S}^2\hookrightarrow\mathbb{R}^3$ and $y \in \mathbb{R}^3$. From the above formula we deduce that 
\begin{equation}
\label{Eq: expression for T0*}
    T_0^*: C^\infty_0(\mathscr{I}^+) \to C^\infty(\mathbb{M}^4), \qquad T_0^* \chi (t, y) \doteq -\frac{1}{4\pi} \int_{\mathbb{S}^2} \chi (t-\omega \cdot y, \omega) \, d\omega. 
\end{equation}
We list here some notable properties of $T_0^*\chi$ which will be of key relevance in establishing Step 3: 
\begin{itemize}
    \item[(i)] As a direct consequence of the above discussion, $T_0^* \chi$ is smooth on $\mathbb{M}^4$. However, it is not in general compactly supported.
    \item[(ii)] $T_0^*\chi$ solves the homogeneous massless wave equation. Indeed, set $s \doteq t-\omega \cdot y$ and observe that, for fixed $\omega \in \mathbb{S}^2$, $\partial_t^2 \chi (s, \omega) = \partial_s^2 \chi (s, \omega)$, while $\Delta_y \chi (s, \omega) = \sum_{j=1}^3 \omega_j^2 \partial_s^2 \chi(s, \omega) = \partial_s^2 \chi(s, \omega)$, since $|\omega| = 1$. Hence, $(-\partial_t^2 + \Delta_y) \chi (t- \omega \cdot y, \omega) = 0$ and, by integration, also $P_0 (T_0^* \chi) = 0$. 
    \item[(iii)] Assuming that $\operatorname{supp} \chi \subset [a,b] \times \mathbb{S}^2$, with $a < b$, then $T_0^*\chi (t, y) \ne 0$ when $t \in [a-|y|, b+|y|]$. For $y \ne 0$, let $y = r \theta$, where $|r| = y$ and $\theta \in \mathbb{S}^2$. Then for fixed $t, r, \theta$, on account of the previous consideration, the integrand in Equation \eqref{Eq: expression for T0*} is different from zero whenever $t-r \omega \cdot \theta \in [a, b]$ and, thus, $\omega \cdot \theta$ is restricted to an interval of length almost $(b-a) r^{-1}$. Since, choosing spherical coordinates on $\mathbb{S}^2$ with polar axis lying along $\theta$, locally $d\omega = d\mu \, d\varphi$, we have that $|T_0^*\chi (t,y)| \le C_\chi \langle y \rangle^{-1}$ for $C_\chi > 0$ depending on $\operatorname{supp} \chi$. The same argument applies after differentiation in both $t$ and $y$, hence, for every multi-index $\alpha \in \mathbb{N}^{\mathbb{N}_0}$, we have 
    \begin{equation}
    \label{Eq: decay of T0chi}
    |\partial^\alpha T_0^*\chi (t,y)| \le C_{\alpha, \chi} \langle y \rangle^{-1}.
    \end{equation}
\end{itemize}

\vskip .2cm

\textbf{Step 3: The pairing $\langle \mathsf g_\infty, T_0^* \chi \rangle$ is well-defined for any $\chi \in C^\infty_0(\mathscr{I}^+)$ --} In order for the pairing $\langle \mathsf g_\infty, T_0^*\chi\rangle$ to be well-defined, we need to  prove that $\mathsf g_\infty$ has suitable decay properties, compatible with the ones of $T_0^* \mathsf g_R$ in item (iii) above. We begin by observing that, on account of item (iii) above, the overlap $\operatorname{supp} \mathsf g_\infty \cap \operatorname{supp} (T_0^*\chi)$ is confined for large $r$ in a bounded interval of the advanced null coordinate, \textit{i.e.}, $v \in [a, C_f]$ for some $C_f > 0$ depending on $f \in C^\infty_0(\mathbb{M}^4)$. The goal is hence to establish in every bounded $v$-strip an estimate of the form 
\begin{equation*}
    |\partial^\alpha \Delta_\mathrm{A}^m f(r, r\theta) | \le C_{\alpha, N} \langle r \rangle^{-N}, \qquad v = t+r \in [a, C_f], 
\end{equation*}
where we set $y = r \theta$, for every $N \in \mathbb{N}$. More precisely, we prove the following lemma.

\begin{lemma}
\label{Lem: massive null decay lemma}
    Let $I \Subset \mathbb{R}$ be a compact interval. For every multi-index $\alpha$ and for every $N \in \mathbb{N}_0$, there exists $C_{\alpha, N, I} > 0$ such that 
    \begin{equation}
            \label{Eq: sought estimate for ginfty}
            \sup_{\theta \in \mathbb{S}^2} \sup_{v \in I} |\partial^\alpha \Delta_\mathrm{A}^m f (-r+v, r\theta)| \le C_{\alpha, N,I} \langle r \rangle^{-N}, \qquad r \ge 0.
    \end{equation}
In particular, $\Delta_\mathrm{A}^m f$ and all its derivatives decay faster than any polynomial along bounded strips of the advanced null coordinate $v = t+r$. 
\end{lemma}

\begin{proof}
    For notational ease, let us set $u \doteq \Delta_\mathrm{A}^m f$ and denote by $K \doteq \operatorname{supp} f$. Fix $t_0 \in \mathbb{R}$ such that $t_0 < \inf \{t \in \mathbb{R} \, : \, (t,x) \in K\}$. Since $f$ vanishes for $t < t_0$, it holds that $P_m u = 0$ for $t< t_0$ and, by the causal support property of $\Delta_\mathrm{A}^m$, $\operatorname{supp} u \subset J^-(K)$. Moreover, on the Cauchy hypersurface $\Sigma_{t_0}$, $u_0 \doteq u \vert_{\Sigma_{t_0}}$ and $u_1 \doteq \partial_t u \vert_{\Sigma_{t_0}}$ are compactly supported since, for $i=0,1$, $\operatorname{supp} u_i \subset \Sigma_{t_0} \cap J^{-}(K)$ which is compact being $K \Subset \mathbb{M}^4$. Since $u$ is smooth $u_0, u_1 \in C^\infty_0(\mathbb{R}^3)$ and, hence, taking the Fourier transform $\hat{u}_0 \hat{u}_1 \in \mathcal{S}(\mathbb{R}^3)$. 

    Let $\omega_m(\xi) \doteq \sqrt{|\xi|^2 + m^2}$. In the source-free region, namely, for $t< t_0$, the solution $u$ admits the representation in terms of an oscillatory/Fourier integral operator
    \begin{equation}
    \label{Eq: u as an oscillatory integral}
        u(t,x) = \sum_{\sigma = \pm 1} \int_{\mathbb{R}^3} e^{i x \cdot \xi + i \sigma(t-t_0) \omega_m (\xi)} a_\sigma (\xi) \, d\xi, \qquad a_\sigma \in \mathcal{S}(\mathbb{R}^3). 
    \end{equation}
    Set $x = r \theta$, $t=-r +v$ with $\theta \in \mathbb{S}^2$, $v \in I$, and replacing it into Equation \eqref{Eq: u as an oscillatory integral}, each term becomes $$\int_{\mathbb{R}^3} e^{ir \phi_{\sigma, \theta}(\xi)} A_{\sigma, v}(\xi) \, d\xi,$$ where $\phi_{\sigma, \theta}(\xi) = \theta \cdot \xi - \sigma \omega_m (\xi)$ and $A_{\sigma, v} (\xi) = e^{i \sigma (v-t_0)\omega_m(\xi)} a_\sigma(\xi)$. For $v \in I$, the family $\{A_{\sigma, v} \, : \, v \in I\}$ is bounded in $\mathcal{S}(\mathbb{R}^3)$. The same conclusion remains true after applying the spacetime derivative $\partial^\alpha$ to $u$. We observe that 
    \begin{equation}
    \label{Eq: estimate for the gradient}
        |\nabla_\xi \phi_{\sigma, \theta} (\xi) | = \left|\theta - \sigma \frac{\xi}{\omega_m(\xi)} \right| \ge 1 - \frac{|\xi|}{\omega_m(\xi)} = \frac{m^2}{\omega_m(\xi) (\omega_m(\xi) + |\xi|)}. 
    \end{equation}
   Hence, there exists $c_m > 0$ such that $$|\nabla_\xi \phi_{\sigma, \theta} (\xi) | \ge c_m \langle \xi \rangle^{-2},$$ and the estimate is uniform in $\theta \in \mathbb{S}^2$ as well as in $\sigma = \pm$. This entails, in particular, that $\nabla_\xi \phi_{\sigma, \theta}(\xi)$ cannot vanish for any finite $\xi$. Define 
   \begin{equation*}
       B_{\sigma, \theta}(\xi) \doteq \frac{\nabla_\xi \phi_{\sigma, \theta}(\xi)}{|\nabla_\xi \phi_{\sigma, \theta}(\xi)|^2}, \qquad B_{\sigma, \theta}(\xi) \cdot \nabla_\xi \phi_{\sigma, \theta}(\xi) = 1.
   \end{equation*}
Equation \eqref{Eq: estimate for the gradient} implies that its growth at infinity is controlled by
$|B_{\sigma, \theta}(\xi)| \le C_m \langle \xi \rangle^2,$ for $C_m > 0$. Furthermore, by direct inspection, every derivative of $B_{\sigma, \theta}$ has at most polynomial growth in $\xi$, namely, there exists $C_{\gamma, m}, M_\gamma > 0$ such that the estimate
   \begin{equation}
       \label{Eq: polynomial growth of derivatives of B}
       |\partial^\gamma_\xi B_{\sigma, \theta} (\xi) | \le C_{\gamma, m} \langle \xi \rangle^{M_\gamma}, 
   \end{equation}
  holds true uniformly in $\theta$. Consequently, the operator
   \begin{equation*}
       \mathcal{L}_{\sigma, \theta} \doteq \frac{1}{ir} B_{\sigma, \theta}(\xi) \cdot \nabla_\xi,  
   \end{equation*}
   satisfies 
   \begin{equation}
       \label{Eq: equation for L sigma theta}
        \mathcal{L}_{\sigma, \theta} e^{ir \phi_{\sigma, \theta}} = e^{ir \phi_{\sigma, \theta}}.
   \end{equation}
  Therefore, using Equation \eqref{Eq: equation for L sigma theta} and integrating by parts $N$ times, we obtain
   \begin{flalign*}
       \int_{\mathbb{R}^3} e^{ir\phi_{\sigma, \theta}(\xi)} A_{\sigma, v}(\xi) \, d\xi &=  \int_{\mathbb{R}^3} (\mathcal{L}_{\sigma, \theta} e^{ir\phi_{\sigma, \theta}(\xi)}) A_{\sigma, v}(\xi) \, d\xi \\ &=  \int_{\mathbb{R}^3} e^{ir\phi_{\sigma, \theta}(\xi)} \mathcal{L}_{\sigma, \theta}^* A_{\sigma, v}(\xi) \, d\xi \\&= r^{-N}  \int_{\mathbb{R}^3} e^{ir\phi_{\sigma, \theta}(\xi)} Q_{N, \sigma, \theta} A_{\sigma, v}(\xi) \, d\xi,
   \end{flalign*}
   where $Q_{N, \sigma, \theta}$ is a differential operator whose coefficients and all their derivatives have polynomial growth. Moreover, since $A_{\sigma, v}$ ranges over a bounded subset of the Schwartz space, $Q_{N, \sigma, \theta} A_{\sigma, v} \in L^1(\mathbb{R}^3)$, uniformly for $v \in I$ and $\theta \in \mathbb{S}^2$. Thus, 
   \begin{equation*}
       \sup_{\theta\in\mathbb{S}^2, v\in I} \bigg\vert \int_{\mathbb{R}^3} e^{ir \phi_{\sigma, \theta}(\xi)} A_{\sigma, v}(\xi) \, d\xi \bigg\vert \le C_{N,I} \, r^{-N}, \qquad r \ge 1,
   \end{equation*}
   and, for every multi-index $\alpha$, it holds 
   \begin{equation*}
    |\partial^{\alpha} u(-r+v, r\theta)|\le C_{\alpha, N, I} \, r^{-N}, \qquad r \ge 1, \, v \in I, 
   \end{equation*}
   for $C_{\alpha, N, I} > 0$. For $r \ge 1$, $r^{-N} \le 2^{\frac{N}2} \langle r \rangle^{-N}$ and, hence, in this region the desired estimate descends. For $0 \le r \le 1$, the set 
   $$\{(-r+v, r\theta) \, : \, 0 \le r \le 1, v \in I, \theta \in \mathbb{S}^2\},$$
is compact. Hence smoothness of $u$ entails that $\sup |\partial^\alpha u| < \infty$ thereon. Since $\langle r \rangle^{-N} \ge 2^{-\frac{N}2}$ on that region, the constant can be increased so to encompass also the case $0 \le r \le 1$, thereby obtaining Equation \eqref{Eq: sought estimate for ginfty}.  
\end{proof}
Lemma \ref{Lem: massive null decay lemma} entails that the pairing with $\mathsf g_\infty$ is absolutely convergent. Indeed, away from the compact support of $f$, $\mathsf g_\infty = m^2 \Delta_\mathrm{A}^m f$ and the spacetime measure in $(v,r,\theta)$ coordinates is $dx^4 = dv \, r^2 \, dr \, d\theta$. Therefore, combining Equations \eqref{Eq: decay of T0chi} and \eqref{Eq: sought estimate for ginfty}, we have that 
\begin{equation*}
    |\mathsf g_\infty T_0^*\chi| \, d^4 x \lesssim r^{1-N} dv \, dr \, d\theta, \qquad \forall N \in \mathbb{N}, 
\end{equation*}
thus $N > 2$ already suffices for integrability. Consequently, the expression
\begin{equation}
    \langle \mathsf g_\infty, T_0^* \chi \rangle \doteq \int_{\mathbb{M}^4} \mathsf g_\infty(x) T_0^*\chi(x) \, d^4x,
\end{equation}
is genuinely well-defined. 

\vskip .2cm

\textbf{Step 4: $\langle \mathsf g_R, T_0^* \chi\rangle \to \langle \mathsf g_\infty, T_0^*\chi\rangle$ for all $\chi \in C^\infty_0(\mathscr{I}^+)$.} To establish the desired convergence result, instead of working with an arbitrary exhaustion of $\mathbb{M}^4$ as in Step 1, we select one which is adapted to the causal structure of Minkowski spacetime. More precisely, let $\rho(x) \doteq \sqrt{1+|x|^2}$ and define 
\begin{equation}
    \label{Eq: sigma pm}
    \sigma_\pm(t, x) \doteq t \pm \rho(x) \in C^\infty(\mathbb{M}^4).
\end{equation}
Note that $\sigma_\pm$ are smooth and non-decreasing along any future-directed causal curve. Choose $\eta \in C^\infty(\mathbb{R}, [0,1])$ such that $\eta(s) = 1$ if $s \le 0$ and $\eta(s) = 0$ if $s \ge 1$ and, for fixed $R \in \mathbb{N}$, set 
\begin{equation*}
    \eta_R(s) \doteq \eta(s-R) \eta(-s-R) = \begin{cases}
        1 \qquad \text{if} \; |s| \le R \\
        0 \qquad \text{if} \; |s| \ge R+1
    \end{cases}. 
\end{equation*}
Therefore, we can define the relevant cut-off by
\begin{equation}
    \label{Eq: hR causally adapted}
    h_R(t,x) \doteq \eta_R(\sigma_-(t,x)) \eta_R(\sigma_+(t,x)). 
\end{equation}
Indeed, $0 \le h_R \le 1$ and $h_R \equiv 1$ on $K_R \doteq \{(t,x) \, : \, |\sigma_-| \le R, |\sigma_+| \le R\}$. Moreover, it holds that $\operatorname{supp}(h_R) \subset K_{R+1}$ and, on account of the fact that $t = \frac{\sigma_+ + \sigma_-}{2}$ and $\rho = \frac{\sigma_+- \sigma_-}{2}$, each $K_R$ is compact and the family $\{K_R\}_{R \in \mathbb{N}}$ is an exhaustion of $\mathbb{M}^4$. In this setting, it is thus convenient to consider the following splitting
\begin{equation}
    \label{Eq: splitting gR - ginfty}
    \langle \mathsf g_R - \mathsf g_\infty, T_0^*\chi \rangle = \underbrace{m^2 \langle (h_R -1) u, T_0^*\chi \rangle}_{(1)} + \underbrace{m^2 \langle h_R (u_R - u), T_0^* \chi \rangle}_{(2)}, 
\end{equation}
where we set $u \doteq \Delta_\mathrm{A}^m f$ and $u_R \doteq \Delta_\mathrm{A}^{m, R} f$. For notational ease, we shall henceforth denote by $\psi \doteq T_0^*\chi$, $P_m \doteq P_0 - m^2$, $P_{m,R} \doteq P_0-m^2 h_R$ and by $w_R \doteq u_R - u$. Notice that $w_R$ solves the following equation 
\begin{equation}
    \label{Eq: wR}
    P_{m,R} w_R = - m^2(1-h_R)u \doteq F_R,
\end{equation}
and that $P_0 w_R = \mathsf g_R - \mathsf g_\infty$. 

On account of the properties of $\psi$ listed in items (i)--(iii) above, we have that $$|\partial^\alpha \psi (t,x) | \le C_{\alpha, \chi} \langle r \rangle^{-1},$$ for some positive constant $C_{\alpha, \chi} > 0$ depending on $\chi \in C^\infty_0 (\mathscr{I}^+)$ with $\operatorname{supp}(\chi) \subset [a,b] \times \mathbb{S}^2$, $a, b \in \mathbb{R}, a< b$. Recall also the support properties of the advanced solution, \textit{i.e.}, $\operatorname{supp}(u) \subset J^-(\operatorname{supp} f)$. Since $\sigma_+$ is non-decreasing along future null causal curves, there exists 
\begin{equation*}
    B_f \doteq \max_{q \in \operatorname{supp} f} \sigma_+(q),
\end{equation*}
such that $\sigma_+(p) \le B_f$ for any $p  \in \operatorname{supp} u$. Furthermore, if $\psi(p) \doteq \psi(t,x) \ne 0$, there exists $\omega \in \mathbb{S}^2$ such that $t- \omega \cdot x \in [a,b]$ which implies $v \doteq t+r \ge a$. Since $\rho \ge r$ and, hence, $\sigma_+ = t + \rho \ge t + r \ge a$, it holds that 
\begin{equation*}
    \operatorname{supp} u \, \cap \, \operatorname{supp} \psi \subset \{(t,x) \, : \, a \le \sigma_+(t,x) \le B_f\}. 
\end{equation*}
So, the relevant region for the pairing is $\{\sigma_+ \in J\}$, where $J \doteq [a, B_f]$ is the fixed compact interval above. Notice that on $\{\sigma_+ \in J\}$, the null coordinate $v \doteq t + r = \sigma_+ - (\rho - r) = \sigma_+ -\frac{1}{\rho+r} \le \sigma_+$ also ranges in a fixed bounded interval. In other words, a bounded $v$-strip is always contained in a, possibly slightly enlarged, $\sigma_+$-strip and also the converse is true. This observation guarantees that Lemma \ref{Lem: massive null decay lemma} applies in such region. Hence, let us consider on 
\begin{equation*}
    \mathcal{N}_J \doteq \{(t,x) \, : \, \sigma_+(t,x) \in J\}, 
\end{equation*}
the weighted $L^2$-space 
\begin{equation}
    \label{Eq: weighted L2}
    L^2_\delta (\mathcal{N}_J) \doteq \{v \, : \, \langle r \rangle^{\delta} v \in L^2(\mathcal{N}_J; d^4 x) \}, \qquad \delta \in \mathbb{R}, 
\end{equation}
equipped with the norm $\| v\|_{L^2_\delta} \doteq \| \langle r \rangle^{\delta} v \|_{L^2}$ under the usual $L^2-$ pairing $(L^2_\delta)' = L^2_{-\delta}$. Notice that $\psi \in L^2_{-\delta}(\mathcal{N}_J)$ for any $\delta > \frac12$. Indeed, on $\{\sigma_+ \in J\}$, the four-volume behaves at large $r$, \textit{e.g.}, for all $r > r_0 >> 0$, like $d^4 x \asymp r^2 dr \, d\sigma_+ d \omega$, while $|\psi| \lesssim r^{-1}$. Thus, 
\begin{equation*}
    \Vert \psi \Vert^2_{L^2_{-\delta}} \simeq \int_{\mathbb{S}^2} d\omega \int_{J} d \sigma_+ \int^{\infty}_{r_0} r^2 r^{-2\delta} r^{-2} \, dr \simeq \int^{\infty}_{r_0} r^{-2\delta} \, dr, 
\end{equation*}
which is convergent provided that $\delta > \frac12$.

We begin by examining the first contribution $(1)$ in Equation \eqref{Eq: splitting gR - ginfty} on the region relevant for the pairing, namely, $\{\sigma_+ \in J\}$. Since $J$ is a compact interval, for sufficiently large $R > 0$, $|\sigma_+| < R$ and, hence, $\eta_R(\sigma_+) \equiv 1$ thereof. Thus, $h_R \ne 1$ only if $\sigma_- < -R$, which implies 
\begin{equation*}
    \operatorname{supp}((1-h_R) u) \cap \operatorname{supp} \psi \subset \{(t,x) \, : \, \sigma_+(t,x) \in J, \; \text{and} \; \sigma_-(t,x) \le -R\}. 
\end{equation*}
Furthermore, in this region it holds that $\rho \gtrsim  R$ and $r \gtrsim R$. Lemma \ref{Lem: massive null decay lemma} entails that, for every $M > 0$, there exists $C_M > 0$ such that $|u| \le C_M \langle r \rangle^{-M}$ uniformly on $\{\sigma_+ \in J\}$. This in turn implies that $u \in L^2_\delta (\mathcal{N}_J)$ for all $\delta \in \mathbb{R}$. Moreover, 
\begin{equation*}
    \Vert \boldsymbol{1}_{\{r \gtrsim R\}} u \Vert_{L^2_\delta}^2 \lesssim \int_R^\infty r^{2\delta} r^{-2M} r^2 \, dr = \int_R^\infty r^{2\delta - 2M + 2} \, dr,
\end{equation*}
which converges provided that $M > \delta + \frac32 > 0$. Choosing such $M$, by the weighted Cauchy-Schwartz inequality, we obtain
\begin{equation}
\label{Eq: estimate term 1}
  |(1)| \doteq |m^2\langle (h_R-1) u, \psi \rangle | \le \Vert (h_R -1) u \Vert_{L^2_\delta} \Vert \psi \Vert_{L^2_{-\delta}} \longrightarrow 0, \qquad \text{as} \; R \to \infty. 
\end{equation}

We now turn our attention to term $(2)$ in Equation \eqref{Eq: splitting gR - ginfty}. Recall that $w_R \doteq u_R - u$ is the advanced solution of Equation \eqref{Eq: wR} with source term $F_R \doteq - m^2(1-h_R)u$. Hence, $w_R =\Delta_\mathrm{A}^{m,R} F_R$. We want to show that $w_R$ is identically zero outside a suitable shell of fixed width in $\sigma_-$, relevant for the distributional pairing. 

For $p = (t,x) \in \operatorname{supp} u \cap \operatorname{supp} \psi$ and $\sigma_-(p) \ge - R$, let $q = (t',y) \in J^+(p) \cap \operatorname{supp} u$. Because $\sigma_\pm$ are non-decreasing along future directed causal curves, we have that $\sigma_-(q) \ge \sigma_-(p) \ge - R$. In addition, $q \in \operatorname{supp} u \subset J^-(\operatorname{supp} f)$, since $u$ is an advanced solution of $P_m u = f$. This implies that $\sigma_\pm (q) \le \max_{\operatorname{supp} f} \sigma_\pm < R$ for large enough $R > 0$ and, also, $|\sigma_\pm (q)| \le R$. Hence, for such $R$, $h_R (q) = 1$ and $F_R \doteq - m^2(1-h_R) u = 0$ throughout the part of the causal future of $p$ which contributes to the advanced solution. By uniqueness, $w_R(p) \doteq u_R(p) - u(p) = 0$ for any $p \in \operatorname{supp} u \cap \operatorname{supp} \psi$ such that $\sigma_-(p) \ge - R$. Observe that $u_R$ and $u$ satisfy the same Cauchy problem, with the same initial conditions and advanced support. Therefore,
\begin{equation}
    \operatorname{supp} (h_R w_R) \cap \operatorname{supp} \psi \subset \Sigma_R \doteq \{(t,x) \; : \; -R-1 < \sigma_-(t,x) < -R, \; \text{and} \; \sigma_+ \in J\}.
\end{equation}
Since $\rho = \frac{\sigma_+ - \sigma_-}{2}$, on $\Sigma_R$ it holds that $\rho = \frac{R}2 + \mathcal{O}(1)$ and also $r \asymp R$. In addition, since $t = \frac{\sigma_+ + \sigma_-}{2}$, we have that for points in $\Sigma_R$, $t$ ranges over an interval of length $\frac{B_f - a -1}{2}$, independent of $R$. In other words, the $t$-extent of $\Sigma_R$ is bounded independently of $R$.

As next step, we shall derive suitable estimates on the source $F_R$ in the equation for $w_R$. For it to be non-zero in the relevant causal future region, we must have $-R-1 < \sigma_-(q) < -R$ for any $q \in J^+(p)$ and $\sigma_+(q) \in J_\ast$, where $J_\ast$ is a fixed compact depending on $J$ and $\operatorname{supp} f$, but not on $R$. Hence, all source points $q$ which can influence $\Sigma_R$ lie in the set
\begin{equation}
    \label{Eq: Omega R}
    \Omega_R = \{(t',y) \, : \, -R-1 < \sigma_-(t',y) < -R, \; \text{and} \; \sigma_+(t',y) \in J_\ast\}, 
\end{equation}
and $r \asymp R$ throughout the shell $\Omega_R$ and its thickness is bounded uniformly in $R$, the estimates being analogous to those for $\Sigma_R$. By Lemma \ref{Lem: massive null decay lemma}, we know that, for all $M > 0$, $|F_R (t,x)| \le C_M R^{-M}$ on $\Omega_R$. On any fixed $t$-slice, the spatial part of $\Omega_R$, namely, $\Omega_R(t) \doteq \{x \; : \; (t,x) \in \Omega_R\}$, is a spherical shell of thickness $\mathcal{O}(1)$ and radius $\mathcal{O}(R)$. Hence, its spatial volume is of order $\mathcal{O}(R^2)$. This together with the fact that the time interval has $R$-independent length entails that 
\begin{equation}
\label{Eq: estimate FR}
    \Vert F_R \Vert_{L^1_t L^2_x (\Omega_R)} \le C_M R^{1-M}, \qquad \text{for every} \; M > 0.
\end{equation}

To estimate the $L^2$-norm of $w_R$, we shall resort to a uniform finite-time energy estimate. By finite propagation, to determine $w_R$ on $\Sigma_R$, we may replace the source $F_R$ by a smooth truncation supported in an enlargement of $\Omega_R$. In this way, the ensuing problem is defined on a time interval $I_R$ such that $|I_R| \le T_0$ for some fixed $T_0$ which does not depend on $R$. Since we are dealing with the advanced solution, we can apply the ordinary energy estimates backwards in time. Indeed, let us consider the energy functional associated to the operator $-P_{m,R}$,
\begin{equation}
    \label{Eq: energy functional}
    E_R(t) \doteq \frac12 \left( \Vert \partial_t w_R(t) \Vert^2_{L^2} + \Vert \nabla w_R (t) \Vert^2_{L^2} + \Vert w_R(t) \Vert^2_{L^2} \right). 
\end{equation}
A direct computation, using Equation \eqref{Eq: wR}, entails that 
\begin{equation*}
    \frac{d}{dt} E_R(t) = \langle F_R, \partial_t w_R\rangle - \langle (1-m^2h_R) w_R, \partial_t w_R\rangle.
\end{equation*}
Since $|h_R| < 1$, we can estimate
\begin{equation*}
    \bigg\vert \frac{d}{dt} E_R(t) \bigg\vert \le 2(1+m^2) E_R(t) + \sqrt{2} \Vert F_R(t) \Vert_{L^2} E_R(t)^{\frac12}, 
\end{equation*}
or, equivalently, 
\begin{equation*}
       \frac{d}{dt} E_R^{\frac12}(t) \le (1+m^2) E_R^\frac12(t) + \frac{1}{\sqrt{2}} \Vert F_R(t) \Vert_{L^2}.
\end{equation*}
Applying Gr\"onwall inequality yields 
\begin{equation*}
    \sup_{t \in I_R} E_R(t)^\frac12 \le C_{m, T_0} \Vert F_R\Vert_{L^1_t L^2_x (\Omega_R)}, 
\end{equation*}
and, using Equation \eqref{Eq: estimate FR} together with $|I_R| \le T_0$, we obtain 
\begin{equation}
    \label{Eq: estimate wR on IR}
   \Vert w_R \Vert_{L^2 (\Sigma_R)} \le C'_M R^{1-M}, \qquad \forall M > 0, 
\end{equation}
and for some positive constant $C'_M > 0$ depending on $m, T_0$, but not on $R$. Hence, on $\Sigma_R$ where $r \asymp R$ we have the following estimate
\begin{equation*}
    \Vert h_R w_R \Vert_{L^2_\delta (\Sigma_R)} \le C R^\delta  \Vert w_R \Vert_{L^2(\Sigma_R)} \le \tilde{C}_M R^{\delta + 1 -M},
\end{equation*}
for all $M > 0$. From $\psi = \mathcal{O}(R^{-1})$ and $\operatorname{Vol}(\Sigma_R) = \mathcal{O}(R^2)$, we have that $\Vert \psi \Vert_{L^2(\Sigma_R)} = \mathcal{O}(1)$ and, hence, $\Vert \psi \Vert_{L^2_{-\delta}(\Sigma_R)} \le C R^{-\delta}$ for  $\delta > \frac12$. The weighted Cauchy-Schwartz inequality thus gives 
\begin{equation}
\label{Eq: weighted CS}
   |(2)| \doteq |\langle h_R w_R, \psi \rangle | \le \Vert h_R w_R \Vert_{L^2_\delta (\Sigma_R)} \Vert \psi \Vert_{L^2_{-\delta}(\Sigma_R)} \le C_M R^{1-M}, \qquad \forall M > 0. 
\end{equation}
Choosing $M > 1$, we have that it converges to zero as $R \to \infty$. Since the $\delta-$weight cancels in the above estimate, we can consistently choose any $\delta > \frac12$. Therefore, Equations \eqref{Eq: estimate term 1} and \eqref{Eq: weighted CS} yield 
\begin{equation*}
    \langle \mathsf g_R - \mathsf g_\infty, T_0^*\chi \rangle \rightarrow 0, \qquad \text{as} \; R \to \infty, \forall \chi \in C^\infty_0(\mathscr{I}^+),
\end{equation*}
which in turn implies $T_0 \mathsf g_R \to T_0^{\mathrm{ext}} \mathsf g_\infty$ in $\mathcal{D}'(\mathscr{I}^+)$, where $T_0^{\mathrm{ext}}$ is an extension of $T_0$ by duality, \textit{i.e.}, $\langle T_0^{\mathrm{ext}} \mathsf g_\infty, \chi \rangle \doteq \langle \mathsf g_\infty, T_0^*\chi\rangle$. Finally, on account of item (ii) above, as well as of Equation \eqref{Eq: sought estimate for ginfty}, we can conclude that
\begin{equation*}
    \langle \mathsf g_\infty, T_0^*\chi \rangle = \langle P_0 \Delta_\mathrm{A}^m f, T_0^* \chi \rangle  = \langle \Delta_\mathrm{A}^m f, P_0 (T_0^*\chi) \rangle = 0,
\end{equation*} 
where the boundary terms vanish due to the rapid decay of $u$ and the bounds on $T_0^*\chi$. The weak adiabatic limit is thus zero. Since the explicit form of $h_R$ in Equation \eqref{Eq: hR causally adapted} has never actually been used in the preceding argument, the above reasoning adapts slavishly to any cut-off lying within a suitable controlled class, as clarified by the following theorem. 

\begin{theorem}\label{Thm: convergence to zero in D'}
Let $\sigma_\pm(t,x) \doteq t \pm \sqrt{1+|x|^2}$ and let $\{h_R\}_{R \ge R_0 > 0} \subset C^\infty_0(\mathbb{M}^4)$ satisfy $0 \le h_R \le 1$. Assume in addition that there exists $C>0$, independent of $R$, such that 
\begin{equation}
\label{Eq: hR within a controlled class}
    h_R \equiv 1 \qquad \text{on} \; K_R \doteq \{ |\sigma_-| \le R, |\sigma_+| \le R\}, \quad \operatorname{supp}(h_R) \subset K_{R+C}, \; \forall R \ge R_0 > 0.
\end{equation}
For every $f \in C^\infty_0(\mathbb{M}^4)$, denoting by $u_R \doteq \Delta_\mathrm{A}^{m, h_R} f$ and by $\mathsf g_R \doteq f + m^2h_Ru_R$, it holds that 
\begin{equation}
   \Gamma(\mathscr{R}_{m^2 V_R} \Phi_f) \equiv T_0 \mathsf g_R \longrightarrow 0 \qquad \text{in} \; \mathcal{D}'(\mathscr{I}^+), 
\end{equation}
independently of the chosen cut-off, where $T_0 \mathsf g_R \doteq \tilde{\Delta}_\mathrm{R} (\Omega^{-3} \mathsf g_R) \vert_{\mathscr{I}^+}$ is defined in the proof of Theorem \ref{Thm: convergence not to zero}.
\end{theorem}

\begin{remark}
    The only place in which the constant $C$ in Equation \eqref{Eq: hR within a controlled class} enters the proof detailed above is when estimating $\operatorname{supp} (h_R w_R) \cap \operatorname{supp}\psi$ in Step 4. Nevertheless, the independence of the shell $\Sigma_R$ radial thickness on $R$ and on the individual cut-off $h_R$ is preserved. Consequently, also all estimates remain valid.
\end{remark}

\begin{remark}
The fact that the weak adiabatic limit is zero should not come as a surprise. Indeed, the bulk limit $\mathsf g_\infty$ is a $P_0$-exact source built from a genuinely massive solution $u \doteq \Delta_\mathrm{A}^m f$. Consequently, ordinary massless radiation data disappear from future null infinity, while the massive information migrates towards $i^+$. We can conclude that the non-zero $\mathscr{I}^+$-data in the presence of a cut-off $h_R$ are a cut-off-induced massless radiation effect. This unveils an intrinsic obstruction in encoding the full massive theory at future null infinity. 
\end{remark}

\paragraph{Removing adiabatically the cut-off --}\label{Sec: Local algebraic limit} Theorem~\ref{Thm: convergence to zero in D'} shows that removing the infrared cut-off through the weak adiabatic limit leads to a trivial projection of the massive field observables onto future null infinity. To circumvent this obstruction, we can exploit a key structural property of the theory, namely the time-slice axiom, see Proposition~\ref{Prop: time slice}. This allows us to implement a local adiabatic limit in a suitable neighbourhood of a Cauchy hypersurface while retaining the information encoded in the bulk observables. We recall that, according to the time-slice axiom, if $\mathcal{N}\subset \mcM$ is a causally convex open subset of a globally hyperbolic spacetime $(\mcM, g)$, which is in turn globally hyperbolic if endowed with the metric $g \vert_\mathcal{N}$ and which contains a Cauchy surface of $\mcM$, then the inclusion $\mathcal{N}\hookrightarrow\mcM$ induces an algebra isomorphism $\mathscr{A}^{\mathrm{obs}}(\mathcal{N}; \mcM)\simeq \mathscr{A}^\mathrm{obs}(\mcM).$ Thus, it is sufficient to construct the theory in an arbitrarily small neighbourhood of a Cauchy surface in order to recover the full algebra of observables on $\mcM$. Confining our attention to the four-dimensional Minkowski spacetime $(\mathbb{M}^4,\eta)$, the above reasoning entails that there exists an isomorphism between the massive bulk algebras on $\mathbb{M}^4$ and on any globally hyperbolic, causally convex Cauchy neighbourhood. This suggests removing the cut-off only in such a neighbourhood, rather than taking a global weak adiabatic limit.

\begin{theorem}\label{Thm: convergence not to zero on a Cauchy neighbourhood}
    Let $\mathcal{N} \subset \mathbb{M}^4$ be an open, causally convex, globally hyperbolic neighbourhood of a Cauchy surface $\Sigma \subset \mathbb{M}^4$ and let $\mathsf{g}\in C^\infty_\mathrm{pc}(\mathbb{M}^4)$ be such that $\mathsf{g}=1$ on $\mathcal{N}$. For $f \in C^\infty_0(\mathcal{N})$ define $G_{\mathsf{g}} f\doteq f + m^2 \mathsf{g}\Delta_\mathrm{A}^{m,\mathsf{g}} f$. Then, $G_\mathsf{g}f \in C^\infty_0(\mathbb{M}^4)$ and 
    \begin{flalign}
        \label{Eq: Gamma N}
        \notag \Gamma_{\mathsf{g}, \mathcal N}: \frac{C^\infty_0(\mathcal{N})}{P_m[C^\infty_0(\mathcal{N})]} &\hookrightarrow H_{\mathscr{I}^+} \\
        [f] &\mapsto \Gamma_{\mathsf{g},\mathcal N}[f] \doteq T_0 G_{\mathsf{g}} f, 
    \end{flalign}
    where $P_m \doteq P_0 - m^2$, is an injective symplectic map. Furthermore, for every compact approximation $h_R \in C^\infty_0(\mathbb{M}^4)$ of $\mathsf{g}$ which eventually agrees with $\mathsf{g}$ on $J^-(\operatorname{supp}f) \cap \operatorname{supp}\mathsf{g}$, it holds that $$\Gamma_{h_R} [f] = \Gamma_{\mathsf{g}, \mathcal N}[f],$$ for sufficiently large $R$. 
\end{theorem}

\begin{proof}
Let $\mathsf{g}\in C^\infty_{\mathrm{pc}}(\mathbb M^4)$ satisfy $\mathsf{g}|_{\mathcal N}=1$, and fix $f\in C^\infty_0(\mathcal N)$. Setting $K \doteq \operatorname{supp}f$ and $u_{\mathsf{g}} \doteq \Delta_{\mathrm A}^{m,\mathsf{g}}f$, the causal support property of the advanced Green operator gives $\operatorname{supp}u_{\mathsf{g}}\subseteq J^-(K)$. Since $\operatorname{supp}{\mathsf{g}}$ is past compact, the intersection $\operatorname{supp}\mathsf{g}\cap J^-(K)$ is compact. Hence, 
$$G_{\mathsf{g}} f \doteq f + m^2 \mathsf{g} u_{\mathsf{g}} \in C^\infty_0(\mathbb{M}^4).$$ 
Consider the map in Equation \eqref{Eq: Gamma N} and observe that its action on $[f]$ is well-defined on account of the preceding discussion. Take any $h \in C^\infty_0(\mathbb{M}^4)$ such that $h=\mathsf{g}$ on $J^{-}(K)$. By uniqueness, denoting by $u_h \doteq \Delta_\mathrm{A}^{m,h} f$, it holds that $u_{\mathsf{g}}= u_h$, which in turn implies that $G_h f = G_{\mathsf{g}} f$ and, hence, $\Gamma_{h}[f] = \Gamma_{g, \mathcal{N}}[f]$. If $\{h_R\}_{R > 0} \subset C^\infty_0(\mathbb{M}^4)$ is any sequence such that $h_R = \mathsf{g}$ on $J^-(K)$ for all sufficiently large $R$, then eventually $T_0 G_{h_R} f = \Gamma_{\mathsf{g}, \mathcal N}[f]$ for any $f \in C^\infty_0(\mathcal{N})$. Thus, the local adiabatic limit 
$$\lim_{R \to \infty} T_0 G_{h_R} f = \Gamma_{g, \mathcal N}[f],$$ 
since the sequence is eventually constant for each local observable. We now prove that the induced local boundary projection is injective. Suppose that there exists $f \in C^\infty_0(\mathcal{N})$ such that $\Gamma_{\mathsf{g}, \mathcal{N}}[f] = 0$. By definition, this entails that $T_0 G_\mathsf{g} f = 0$. On account of Theorem \ref{Thm: convergence not to zero}, this is possible if and only if $G_\mathsf{g} f = P_0 q$ for $q \in C^\infty_0(\mathbb{M}^4)$. A direct computation entails that $P_0 u_\mathsf{g} = G_\mathsf{g} f= P_0 q$ which, combined with the future compactness of the support of $u_\mathsf{g}$, implies $u_\mathsf{g} = q$ and, thus, $f = P_{m,\mathsf{g}} q$. Since the operator $P_{m,\mathsf{g}}$ is normally hyperbolic and since $\mathcal N$ contains a Cauchy surface, the time-slice isomorphism applies to $P_{m,\mathsf{g}}$. More precisely, denoting such isomorphism by $\tilde{\iota}$, the fact that $[f] = 0$ in $\frac{C^\infty_0(\mathbb{M}^4)}{P_{m,\mathsf{g}}[C^\infty_0(\mathbb{M}^4)]}$ together with the injectivity of $\tilde{\iota}$ implies $[f] = 0$ in $\frac{C^\infty_0(\mathcal N)}{P_{m,\mathsf{g}}[C^\infty_0(\mathcal N)]}$. By definition, there exists $q_{\mathcal{N}} \in C^\infty_0(\mathcal N)$ such that $f = P_{m,\mathsf{g}} q_\mathcal{N}$. Moreover, since $\mathsf{g}=1$ on $\mathcal N$ and $\operatorname{supp} q_\mathcal{N} \subset \mathcal N$, it holds that $P_m \equiv P_{m,\mathsf{g}}$ on $\mathcal N$. Therefore, $f = P_m q_\mathcal{N} \in P_m[C^\infty_0(\mathcal{N})]$ and $\Gamma_{\mathsf{g}, \mathcal{N}}$ is injective. 

It remains to be shown that $\Gamma_{\mathsf{g},\mathcal N}$ preserves the symplectic forms. To this avail, let $f \in C^\infty_0(\mathcal N)$ and recall that $G_\mathsf{g}f =P_0\Delta_{\mathrm A}^{m,\mathsf{g}}f.$ Since $G_\mathsf{g} f \in C^\infty_0(\mathbb{M}^4)$ and $\Delta_\mathrm{A}^{m,\mathsf{g}} f$ has advanced support, uniqueness of advanced solutions yields $\Delta_{\mathrm A}G_\mathsf{g} f =\Delta_{\mathrm A}^{m,g}f.$ Choose a constant-time Cauchy hypersurface $\Sigma_{-}$ lying strictly in the past of the supports of $f$ and $G_\mathsf{g} f$. Since the corresponding retarded solutions vanish in a neighbourhood of $\Sigma_-$, we obtain 
$$\Delta_{\mathrm S}G_\mathsf{g}f = -\Delta_{\mathrm{A}} G_\mathsf{g} f = - \Delta_\mathrm{A}^{m,\mathsf{g}} f = \Delta_\mathrm{S}^{m,\mathsf{g}} f.$$ Hence, the massless solution $v_f \doteq \Delta_{\mathrm S}G_\mathsf{g}f$ and the corresponding solution $u_f \doteq \Delta_\mathrm{S}^{m,\mathsf{g}} f$ of $P_{m,\mathsf{g}}$ have identical Cauchy data on $\Sigma_-$. The same argument applies \emph{verbatim} to any other test function $f' \in C^\infty_0(\mathcal{N})$, provided $\Sigma_-$ is selected in such a way that it lies strictly in the past of $f'$ and of $G_{\mathsf{g}} f'$ as well. Since the corresponding symplectic currents agree and are conserved on solutions of their respective equations, we obtain
\begin{flalign*}
    \sigma_{0,\mathbb M^4}([G_{\mathsf{g}}f],[G_{\mathsf{g}}f']) &= \int_{\Sigma_-} (v_f \nabla_n v_{f'} - v_{f'} \nabla_n v_{f}) \, d\mu_{\Sigma} = \int_{\Sigma_-} (u_f \nabla_n u_{f'} - u_{f'} \nabla_n u_{f}) \, d\mu_{\Sigma} \\&= \sigma_{m,\mathsf{g},\mathbb M^4}([f],[f']) = \sigma_{m,\mathcal N}([f],[f']),
\end{flalign*}
where the last equality follows from the time-slice identification and from the restriction $P_{m,\mathsf{g}}|_{\mathcal N}=P_m|_{\mathcal N}$. To conclude, using that the massless bulk-to-boundary map $T_0$ is a symplectomorphism, see Equation \eqref{Eq: massless radiation map}, we find 
\begin{equation}
    \sigma_{\mathscr I}\bigl(\Gamma_{\mathsf{g},\mathcal N}[f],\Gamma_{\mathsf{g},\mathcal N}[f']\bigr) = \sigma_{0,\mathbb M^4}([G_\mathsf{g}f],[G_\mathsf{g}f']) = \sigma_{m,\mathcal N}([f],[f']),
\end{equation}
which proves the claim. 
\end{proof}

A direct consequence of the preceding theorem is that the vanishing of the boundary projection of a non-trivial massive local observable is independent of the choice of admissible cut-off. Let  
\begin{equation}
    \label{Eq: GN}
        \mathscr{G}_\mathcal{N} \doteq \{\mathsf{g} \in C^\infty_{\mathrm{pc}}(\mathbb{M}^4) \; : \; \mathsf{g}\vert_{\mathcal{N}} = 1, \; 0 \le \mathsf{g} \le 1\},
    \end{equation} 
denote the class of admissible cut-off functions introduced in Theorem \ref{Thm: convergence not to zero on a Cauchy neighbourhood}. Observe that the class $\mathscr{G}_\mathcal{N}$ may be empty for certain choices of $\mathcal{N}$, \textit{e.g.}, for $\mathcal N = \mathbb{M}^4$. Henceforth, we shall restrict the attention to those Cauchy neighbourhoods admitting at least one such switching functions.

\begin{remark}
\label{Rem: GN and Cauchy neighbourhoods}
For later convenience, we observe the following. Consider $\mathcal N, \tilde{\mathcal{N}} \subset \mathbb{M}^4$ to be two open, causally convex neighbourhoods containing the same Cauchy hypersurface $\Sigma$ of $\mathbb{M}^4$ and assume that $\mathscr{G}_\mathcal{N} \ne \emptyset$, $\mathscr{G}_{\tilde{\mathcal{N}}} \ne \emptyset$. Choose $\mathsf{g}_1 \in \mathscr{G}_\mathcal{N}$, $\mathsf{g}_2 \in \mathscr{G}_{\tilde{\mathcal{N}}}$ and set 
$$\mathsf{g}\doteq 1 - (1-\mathsf{g}_1) (1-\mathsf{g}_2) = \mathsf{g}_1+ \mathsf{g}_2 - \mathsf{g}_1\mathsf{g}_2.$$ Then, $\mathsf{g}$ is smooth, $0 \le\mathsf{g}\le 1$, and $\mathsf{g} \vert_{\mathcal{N}} = \mathsf{g}\vert_{\tilde{\mathcal{N}}} = 1$. Furthermore, it holds $\operatorname{supp} \mathsf{g}\subseteq \operatorname{supp} \mathsf{g}_1 \cup \operatorname{supp}\mathsf{g}_2$. Since a finite union of past-compact closed sets is past compact, $\mathsf{g}$ is an admissible cut-off for both neighbourhoods, namely, $\mathsf{g}\in \mathscr{G}_{{\mathcal{N}}} \cap \mathscr{G}_{\tilde{\mathcal{N}}}$ and, thus, $\mathscr{G}_{{\mathcal{N}}} \cap \mathscr{G}_{\tilde{\mathcal{N}}} \ne \emptyset$. This entails that, given two Cauchy neighbourhoods as in Theorem \ref{Thm: convergence not to zero on a Cauchy neighbourhood}, we can always select a common switching function, whenever both $\mathscr{G}_{\mathcal{N}}$ and $\mathscr{G}_{\tilde{\mathcal{N}}}$ are non-empty.
\end{remark}

\noindent Therefore, assuming $\mathscr{G}_\mathcal{N} \ne \emptyset$, for every $f\in C^\infty_0(\mathcal N)$ and for any arbitrary admissible cut-off $\mathsf{g}_1, \mathsf{g}_2 \in \mathscr{G}_\mathcal{N}$, it holds 
\begin{equation*}
   \Gamma_{\mathsf{g}_1,\mathcal N}[f]=0 \quad\Longleftrightarrow\quad \Gamma_{\mathsf{g}_2,\mathcal N}[f]=0 \quad\Longleftrightarrow\quad [f]=0 \quad \text{in} \quad \frac{C^\infty_0(\mathcal N)}{P_m[C^\infty_0(\mathcal N)]}.
\end{equation*}
Most notably, the boundary projections associated with different admissible cut-off provide symplectically equivalent realizations of the same collection of observables.

\begin{corollary}[Equivalence of local boundary projections]
\label{Cor: cut-off independence}
Let $\mathcal N\subset\mathbb M^4$ be as in Theorem \ref{Thm: convergence not to zero on a Cauchy neighbourhood}, and let $\mathsf{g}_1,\mathsf{g}_2\in \mathscr{G}_\mathcal{N}$
be two admissible cut-off functions, where $\mathscr{G}_\mathcal{N}$ is as per Equation \eqref{Eq: GN}. Then there exists a unique symplectic isomorphism
\begin{equation}
\label{Eq: cut-off intertwining}
    U_{\mathsf{g}_2\mathsf{g}_1}:
\operatorname{Im}\Gamma_{\mathsf{g}_1,\mathcal N}
\longrightarrow
\operatorname{Im}\Gamma_{\mathsf{g}_2,\mathcal N} \qquad \text{such that} \qquad
U_{\mathsf{g}_2\mathsf{g}_1}\circ\Gamma_{\mathsf{g}_1,\mathcal N}
= \Gamma_{\mathsf{g}_2,\mathcal N}.
\end{equation}
\end{corollary}

\begin{proof}
By Theorem \ref{Thm: convergence not to zero on a Cauchy neighbourhood}, both maps $\Gamma_{\mathsf{g}_i, \mathcal{N}}$ are injective and symplectic. Consequently, their restrictions to their respective images are bijective. We may therefore define
\begin{equation*}
    U_{\mathsf{g}_2\mathsf{g}_1} \doteq \Gamma_{\mathsf{g}_2,\mathcal N}\circ \Gamma_{\mathsf{g}_1,\mathcal N}^{-1},
\end{equation*}
where the inverse has $\operatorname{Im}\Gamma_{\mathsf{g}_1,\mathcal N}$ as domain. By construction, $U_{\mathsf{g}_2\mathsf{g}_1}$ is a linear isomorphism satisfying Equation~\eqref{Eq: cut-off intertwining}. Moreover, for any $[f],[f']$ corresponding to observables for the massive theory, being the boundary projections symplectomorphisms, the following chain of equalities holds true:
\begin{align*}
&\sigma_{\mathscr I}
\bigl(U_{\mathsf{g}_2\mathsf{g}_1}\Gamma_{\mathsf{g}_1,\mathcal N}[f],
U_{\mathsf{g}_2\mathsf{g}_1}\Gamma_{\mathsf{g}_1,\mathcal N}[f']
\bigr) =
\sigma_{\mathscr{I}}
\bigl(
\Gamma_{\mathsf{g}_2,\mathcal N}[f],
\Gamma_{\mathsf{g}_2,\mathcal N}[f']
\bigr) =
\sigma_{m,\mathcal N}([f],[f'])\\
&\qquad =
\sigma_{\mathscr I}
\bigl(
\Gamma_{\mathsf{g}_1,\mathcal N}[f],
\Gamma_{\mathsf{g}_1,\mathcal N}[f']
\bigr),
\end{align*}
where $\sigma_{m, \mathcal{N}}$ is the symplectic form associated with the massive Klein--Gordon field on $\mathcal{N}$. Thus, $U_{\mathsf{g}_2\mathsf{g}_1}$ preserves the symplectic form and it is therefore a symplectic isomorphism. At last, uniqueness follows from surjectivity of $\Gamma_{\mathsf{g}_1,\mathcal N}$ onto its image.
\end{proof}

The local adiabatic construction detailed above admits a further notable property. Although the identification of individual massive observables with their boundary representatives depends on the cut-off, the resulting boundary space is independent of this choice. More precisely, the image of the massive local bulk-to-boundary map coincides with that of the massless theory, as established by the following corollary.

\begin{corollary}[cut-off independence of the boundary image]
\label{Cor: independence of boundary image}
In the same setting of Theorem~\ref{Thm: convergence not to zero on a Cauchy neighbourhood}, let $g \in \mathscr{G}_N$ be any admissible cut-off function, where the set $\mathscr{G}_\mathcal{N}$ is defined in Equation \eqref{Eq: GN}. Then,
\begin{equation}
    \operatorname{Im}\Gamma_{g,\mathcal N} = \operatorname{Im} T_0, 
\end{equation}
where, with a mild abuse of notation, we denote again by $T_0$ the map induced by Equation \eqref{Eq: massless radiation map} on the respective quotient spaces, \textit{i.e.}, 
\begin{equation*}
    T_0: \frac{C^\infty_0(\mathbb{M}^4)}{P_0[C^\infty_0(\mathbb{M}^4)]} \longrightarrow \mathscr{S}(\mathscr{I}^+), \qquad [f] \mapsto T_0 [f] \doteq T_0 f. 
\end{equation*}
\end{corollary}

\begin{proof}
Arguing as in the proof of Theorem \ref{Thm: convergence not to zero on a Cauchy neighbourhood}, one can show that the map 
$$G_\mathsf{g}: C^\infty_0(\mathbb{M}^4) \to C^\infty_0(\mathbb{M}^4), \qquad G_\mathsf{g} f \doteq f + m^2 \mathsf{g} \Delta^{m,\mathsf{g}}_\mathrm{A} f = P_0 \Delta_\mathrm{A}^{m,\mathsf{g}} f,$$ 
is well-defined. We claim that the map $$H_{\mathsf g}: C^\infty_0(\mathbb{M}^4) \to C^\infty_0(\mathbb{M}^4), \qquad H_\mathsf{g} f \doteq f - m^2\mathsf{g} \Delta_\mathrm{A} f,$$ is an inverse of $G_\mathsf{g}$. Indeed, fix $f \in C^\infty_0(\mathbb{M}^4)$ and note that $H_\mathsf{g} f = P_{m,\mathsf{g}} \Delta_\mathrm{A} f$. Denoting by $u \doteq \Delta_\mathrm{A} f$, we have that $P_0 u = f$, which implies in turn $P_{m,\mathsf{g}} u = f - m^2\mathsf{g} u \doteq H_\mathsf{g} f$. Moreover, $u$ has advanced support. By uniqueness of the advanced solution, it holds $u = \Delta_{\mathrm A}^{m,\mathsf{g}} H_\mathsf{g} f$. Applying now $G_\mathsf{g}$ on the right, we obtain 
$$G_\mathsf{g} (H_\mathsf{g} f) = H_\mathsf{g} f + m^2\mathsf{g} \Delta_\mathrm{A}^{m,\mathsf{g}} H_\mathsf{g} f = P_{m,\mathsf{g}} u + m^2\mathsf{g} u = P_0 u = f.$$
This proves that $G_\mathsf{g} H_\mathsf{g} = \operatorname{id} \vert_{C^\infty_0(\mathbb{M}^4)}$. Similarly, we show that $H_\mathsf{g} G_\mathsf{g} = \operatorname{id} \vert_{C^\infty_0(\mathbb{M}^4)}$. Thus, $G_\mathsf{g}$ is a linear bijection of $C^\infty_0(\mathbb{M}^4)$ with inverse $G_{\mathsf{g}}^{-1} \doteq H_\mathsf{g} = \operatorname{id} \vert_{C^\infty_0(\mathbb{M}^4)} - m^2\mathsf{g} \Delta_\mathrm{A}$. Using the intertwining identities on $C^\infty_0(\mathbb{M}^4)$
\begin{equation*}
    G_\mathsf{g} P_{m,\mathsf{g}} = P_0, \qquad H_\mathsf{g} P_0 = P_{m,\mathsf{g}}, 
\end{equation*}
it follows that $G_\mathsf{g}$ induces a vector space isomorphism 
\begin{equation*}
    \overline{G_\mathsf{g}}: \frac{C^\infty_0(\mathbb{M}^4)}{P_{m,\mathsf{g}}[C^\infty_0(\mathbb{M}^4)]} \xrightarrow[]{\simeq} \frac{C^\infty_0(\mathbb{M}^4)}{P_0[C^\infty_0(\mathbb{M}^4)]}, \qquad [f] \mapsto \overline{G_\mathsf{g}}[f] \doteq [G_\mathsf{g} f], 
\end{equation*}
with inverse induced by $H_\mathsf{g}$. 

Fix $\mathcal{N}$ as in the statement of Theorem \ref{Thm: convergence not to zero on a Cauchy neighbourhood} and let $\mathsf{g}\in \mathscr{G}_\mathcal{N}$ be an admissible cut-off as per Equation \eqref{Eq: GN}. Since $g\vert_{\mathcal N} \equiv 1$, it holds $P_{m,\mathsf{g}} \vert_{\mathcal{N}} = P_m \vert_{\mathcal N}$, where $P_m \doteq P_0 - m^2$. Because $\mathcal N$ is causally convex and it contains a Cauchy hypersurface of $\mathbb{M}^4$, the time-slice axiom as per Proposition \ref{Prop: time slice} entails that there exists an isomorphism 
\begin{equation*}
  j^\mathsf{g}_\mathcal{N}: \frac{C^\infty_0(\mathcal N)}{P_{m,\mathsf{g}}[C^\infty_0(\mathcal N)]}  \xrightarrow[]{\simeq} \frac{C^\infty_0(\mathbb{M}^4)}{P_{m,\mathsf{g}}[C^\infty_0(\mathbb{M}^4)]}. 
\end{equation*}
Composing $\overline{G_\mathsf{g}}$ with $j^\mathsf{g}_\mathcal{N}$ yields an isomorphism 
\begin{equation*}
    \overline{G_\mathsf{g}} \circ j^\mathsf{g}_\mathcal{N}: \frac{C^\infty_0(\mathcal N)}{P_{m,\mathsf{g}}[C^\infty_0(\mathcal N)]}  \xrightarrow[]{\simeq} \frac{C^\infty_0(\mathbb{M}^4)}{P_{0}[C^\infty_0(\mathbb{M}^4)]}. 
\end{equation*}
In particular, this map is surjective. By definition, $\Gamma_{\mathsf{g}, \mathcal{N}} = T_0 \circ \overline{G_\mathsf{g}} \circ j^\mathsf{g}_\mathcal{N}$, from which it descends $\operatorname{Im} \Gamma_{\mathsf{g}, \mathcal N} = \operatorname{Im} T_0$, independently of the admissible cut-off $g$.
\end{proof}

The preceding results establish an independence from the choice of cut-off, namely, for any fixed Cauchy neighbourhood $\mathcal N$, every admissible cut-off $g$ gives rise to an injective symplectic map whose image coincides with $\operatorname{Im}T_0$. Thus, the boundary space obtained from the massive local theory is not merely symplectically isomorphic to the massless boundary image, but it coincides with it as a subspace of $\mathcal S(\mathscr I^+)$. This does not, however, imply that the individual embeddings $\Gamma_{\mathsf{g},\mathcal N}$ are cut-off-independent. Indeed, for any two admissible cut-offs $\mathsf{g}_1,\mathsf{g}_2$, the map $U_{\mathsf{g}_2\mathsf{g}_1}$ defines a symplectic automorphism of $\operatorname{Im}T_0$, satisfying $\Gamma_{\mathsf{g}_2,\mathcal N} = U_{\mathsf{g}_2\mathsf{g}_1}\circ\Gamma_{\mathsf{g}_1,\mathcal N}.$ In particular, while the space of boundary observables is independent from the cut-off, their identification with individual massive bulk observables may depend on this choice. This will be further investigated at the end of the section, when we discuss the construction of a boundary-induced algebraic state for the algebra of observables of a massive Klein--Gordon field.

\vskip .2cm 

The cut-off independence of the boundary image established above suggests a further question concerning the compatibility of the construction with the bulk time-slice axiom. More precisely, given two causally convex neighbourhoods of the same Cauchy hypersurface, one may ask whether their corresponding observables for the massive theory admit compatible realizations at future null infinity. To answer this question, we need a preliminary result, which yields a useful intertwining property for the maps $\Gamma_{\mathsf{g}, \mathcal N}$. 

\begin{proposition}\label{Prop: boundary time-slice}
    Let $\mathcal N, \tilde{\mathcal{N}} \subset \mathbb{M}^4$ be two open, causally convex neighbourhoods containing the same Cauchy hypersurface $\Sigma$ of $\mathbb{M}^4$. Let $P_m \doteq P_0 - m^2$ be the massive Klein--Gordon operator on $\mathbb{M}^4$ and denote by $$\mathcal{E}_{m} \doteq \frac{C^\infty_0(\mathbb{M}^4)}{P_m[C^\infty_0(\mathbb{M}^4)]}, \qquad \mathcal{E}_m(\mathcal N) \doteq \frac{C^\infty_0(\mathcal N)}{P_m[C^\infty_0(\mathcal N)]},$$ and similarly for $\mathcal{E}_m(\tilde{\mathcal N})$, where $\mathcal{E}_m(\mathcal N)$ and $\mathcal{E}_m(\tilde{\mathcal N})$ are the local observable spaces associated with $P_m \vert_\mathcal{N}$ and $P_m \vert_{\tilde{\mathcal{N}}}$, respectively. Consider the canonical symplectic isomorphism 
    \begin{equation}\label{Eq: canonical symp isomorphism}
        \tau_{\tilde{\mathcal N}, \mathcal N} \doteq \iota^{-1}_{\tilde{\mathcal N}} \circ \iota_\mathcal{N}: \mathcal{E}_m(\mathcal N) \to \mathcal{E}_m(\tilde{\mathcal N}), 
    \end{equation}
    where $\iota_{\mathcal N}: \mathcal{E}_m(\mathcal N) \xrightarrow[]{\simeq} \mathcal{E}_m$ and $\iota_{\tilde{\mathcal N}}: \mathcal{E}_m(\tilde{\mathcal N}) \xrightarrow[]{\simeq} \mathcal{E}_m$ are the symplectic isomorphism induced by the time-slice property for the massive Klein--Gordon operator, see Proposition \ref{Prop: time slice} for the algebraic counterpart. Then, it holds
    \begin{equation}
        \label{Eq: intertwining property}
        \Gamma_{\mathsf{g}, \tilde{\mathcal N}} \circ \tau_{\tilde{\mathcal N}, \mathcal N} = \Gamma_{\mathsf{g}, \mathcal N},
    \end{equation}
    for any arbitrary admissible cut-off $\mathsf{g} \in \mathscr{G}_\mathcal{N} \cap \mathscr{G}_{\tilde{\mathcal{N}}}$. 
    
\end{proposition}

\begin{proof}
    On account of Remark \ref{Rem: GN and Cauchy neighbourhoods}, we may fix a common cut-off $\mathsf{g}$ such that $\mathsf{g}\vert_\mathcal{N} = \mathsf{g} \vert_{\tilde{\mathcal{N}}} = 1$. Hence, we have $P_{m,\mathsf{g}} \vert_\mathcal{N} = P_m \vert_{\mathcal{N}}$ and $P_{m,\mathsf{g}} \vert_{\tilde{\mathcal{N}}} = P_m \vert_{\tilde{{\mathcal{N}}}}$. Let 
$$j^\mathsf{g}_\mathcal{N}: \mathcal{E}_m(\mathcal{N}) \xrightarrow[]{\simeq} \mathcal{E}_{m,\mathsf{g}} (\mathbb{M}^4), \qquad j^\mathsf{g}_{\tilde{\mathcal{N}}}: \mathcal{E}_m({\tilde{\mathcal{N}}}) \xrightarrow[]{\simeq} \mathcal{E}_{m,\mathsf{g}} (\mathbb{M}^4), $$ 
be the inclusion-induced time-slice isomorphisms for $P_{m,\mathsf{g}}$ as in the proof of Corollary \ref{Cor: independence of boundary image}, where 
$$\mathcal{E}_{m,\mathsf{g}} \doteq \frac{C^\infty_0(\mathbb{M}^4)}{P_{m,\mathsf{g}}[C^\infty_0(\mathbb{M}^4)]}.$$ 
Identifying the on-shell observable spaces $\mathcal{E}_{m,\mathsf{g}}(\mathbb{M}^4)$, $\mathcal{E}_m(\mathcal N)$ and $\mathcal{E}_m(\tilde{\mathcal N})$ with compactly supported Cauchy data on the common Cauchy hypersurface $\Sigma$, we obtain the relation: 
\begin{equation}\label{Eq: intermediate intertwining identity}
j^\mathsf{g}_{\tilde{\mathcal N}}\circ\tau_{\tilde{\mathcal N},\mathcal N}=j^\mathsf{g}_{\mathcal N}.
\end{equation}
Indeed, since $P_{m,\mathsf{g}}$ coincides with $P_m$ on both neighbourhoods, their respective Cauchy-data identifications agree. Moreover, causal convexity and uniqueness of Green operators ensures that, for any $f \in C^\infty_0(\mathcal N)$, $\Delta_\mathrm{S}^{m,\mathsf{g}} f \vert_\mathcal{N} = \Delta_\mathrm{S}^{m,\mathcal N} f = \Delta_\mathrm{S}^{m} f \vert_\mathcal{N}$ and similarly for $\tilde{\mathcal N}$. In particular, the corresponding solutions have identical Cauchy data on $\Sigma$ and, since these uniquely determine the global solutions of both operators, the two inclusion-induced identifications are compatible with the map $\tau_{\tilde{\mathcal N},\mathcal N}$. Recalling that $\Gamma_{\mathsf{g}, \mathcal{N}} = T_0 \circ \overline{G_\mathsf{g}} \circ j^\mathsf{g}_\mathcal{N}$ and similarly for $\tilde{\mathcal N}$, we have 
$$\Gamma_{\mathsf{g}, \tilde{\mathcal{N}}} \circ \tau_{\tilde{\mathcal N}, \mathcal N} = T_0 \circ \overline{G}_\mathsf{g} \circ j^\mathsf{g}_{\tilde{\mathcal N}} \circ \tau_{\tilde{\mathcal N}, \mathcal N} = T_0 \circ \overline{G}_\mathsf{g} \circ j^\mathsf{g}_\mathcal{N} = \Gamma_{\mathsf{g},\mathcal N}.$$ 
This concludes the proof.
\end{proof}

\noindent The counterpart of Proposition \ref{Prop: boundary time-slice} at the level of the local observable algebras is as follows. 

\begin{corollary}[Compatibility with the time-slice axiom at the CCR level]
\label{Cor: boundary CCR time slice}
Let $\mathcal N,\tilde{\mathcal N}\subset\mathbb M^4$ satisfy the hypotheses of Proposition~\ref{Prop: boundary time-slice}, and let $\mathsf{g}\in\mathscr G_{\mathcal N}\cap \mathscr G_{\tilde{\mathcal N}}$. Denote by $\mathfrak A_m^{\mathrm{obs}}(\mathcal N)$
and $\mathfrak A_m^{\mathrm{obs}}(\tilde{\mathcal N})$ the CCR algebras associated with the symplectic spaces $\mathcal E_m(\mathcal N)$ and $\mathcal E_m(\tilde{\mathcal N})$, respectively. Furthermore, let
$$\mathfrak B (\mathscr{I}^+) \doteq \mathrm{CCR}(\operatorname{Im} T_0) \subseteq \mathcal{A} (\mathscr{I}^+),$$ be the unital subalgebra generated by $\operatorname{Im}T_0$. Then, the symplectic isomorphisms $\Gamma_{\mathsf{g},\mathcal N}$ and $\Gamma_{\mathsf{g},\tilde{\mathcal N}}$ induce unital $\ast$-isomorphisms
\begin{equation*}
\widehat\Gamma_{\mathsf{g},\mathcal N}:\mathfrak A_m^{\mathrm{obs}}(\mathcal N)\xrightarrow{\simeq}\mathfrak B(\mathscr I^+),
\qquad
\widehat\Gamma_{\mathsf{g},\tilde{\mathcal N}}:\mathfrak A_m^{\mathrm{obs}}(\tilde{\mathcal N})\xrightarrow{\simeq}\mathfrak B(\mathscr I^+), 
\end{equation*}
such that the following diagram commutes:
\begin{equation*}
\begin{tikzcd}[column sep=large,row sep=large]
\mathfrak A_m^{\mathrm{obs}}(\mathcal N)
\arrow[r,"\widehat\tau_{\tilde{\mathcal N},\mathcal N}"]
\arrow[d,"\widehat\Gamma_{\mathsf{g},\mathcal N}"']
&
\mathfrak A_m^{\mathrm{obs}}(\tilde{\mathcal N})
\arrow[d,"\widehat\Gamma_{\mathsf{g},\tilde{\mathcal N}}"]
\\
\mathfrak B(\mathscr I^+)
\arrow[r,"\mathrm{id}"']
&
\mathfrak B(\mathscr I^+).
\end{tikzcd}
\end{equation*}
Here $\widehat\tau_{\tilde{\mathcal N},\mathcal N}$ is the $\ast$-isomorphism induced by the canonical symplectic time-slice identification $\tau_{\tilde{\mathcal N},\mathcal N}$.
\end{corollary}

\begin{proof}
By Equation \eqref{Eq: intertwining property}, $\Gamma_{\mathsf{g},\tilde{\mathcal N}}\circ \tau_{\tilde{\mathcal N},\mathcal N} = \Gamma_{\mathsf{g},\mathcal N}.$ Moreover, by Corollary~\ref{Cor: independence of boundary image}, both boundary projections are symplectic isomorphisms onto $\operatorname{Im}T_0$. The universal property of the CCR algebra therefore yields the corresponding unital
$\ast$-isomorphisms, together with the desired intertwining identity
\begin{equation}\label{Eq: boundary intertwining of Gamma}
    \widehat{\Gamma}_{\mathsf{g}, \tilde{\mathcal N}} \circ \widehat{\tau}_{\tilde{\mathcal N}, \mathcal N} = \widehat{\Gamma}_{\mathsf{g}, \mathcal N}. 
\end{equation}
which proves the claim.
\end{proof}

Hence, for a common admissible switching function, the time-slice identification between the two bulk theories is intertwined by their respective boundary projections. Notice that this statement concerns the CCR algebras of linear observables. Extending it to the full on-shell microcausal algebra or to Wick polynomials requires an additional construction compatible with the relevant products.

\begin{remark}
One might wonder whether it is still possible to establish an analogue of Equation \eqref{Eq: boundary intertwining of Gamma} if, instead of a common cut-off $g$, we choose two different ones $\mathsf{g}_1 \in \mathscr{G}_\mathcal{N}$ and $\mathsf{g}_2 \in \mathscr{G}_{\tilde{N}}$. The answer is positive, the only difference being that, in place of $\operatorname{id} \vert_{\operatorname{Im} T_0}$, one has to consider the symplectic automorphism 
\begin{equation*}
U_{\mathsf{g}_2,\mathsf{g}_1}^{\widetilde{\mathcal N},\mathcal N}
= \Gamma_{\mathsf{g}_2,\widetilde{\mathcal N}} \circ\tau_{\widetilde{\mathcal N}\mathcal N} \circ\Gamma_{\mathsf{g}_1,\mathcal N}^{-1}.
\end{equation*}
At the algebraic level, the symplectic automorphism induces a $\ast$-automorphism of $\mathfrak B(\mathscr I^+)$, which replaces the identity in the bottom arrow of the commutative diagram. Hence, the correct intertwining relation becomes
\begin{equation*}
   \widehat U_{\mathsf{g}_2,\mathsf{g}_1}^{\tilde{\mathcal N},\mathcal N}
\circ\widehat\Gamma_{\mathsf{g}_1,\mathcal N} =
\widehat\Gamma_{\mathsf{g}_2,\tilde{\mathcal N}}
\circ\widehat\tau_{\tilde{\mathcal N},\mathcal N}. 
\end{equation*}
\end{remark}

\paragraph{Hadamard states for the massive theory --}
The preceding construction allows us to induce a state for the massive Klein--Gordon field starting from the distinguished BMS-invariant state $\omega_{\mathscr I}$ on future null infinity, whose two-point function is given in Equation \eqref{Eq: twopoint on I+}. To this end, let $\mathcal N \subset \mathbb M^4$ be a globally hyperbolic, causally convex neighbourhood of a Cauchy hypersurface and let $\mathsf{g} \in \mathcal G_{\mathcal N}$ be an admissible cut-off as per Equation \eqref{Eq: GN}. By Theorem \ref{Thm: convergence not to zero on a Cauchy neighbourhood} and Corollary \ref{Cor: boundary CCR time slice}, the local bulk-to-boundary map $\Gamma_{\mathsf g,\mathcal N}$ induces an injective unital $*$-homomorphism $\widehat{\Gamma}_{\mathsf{g},\mathcal N}$ of the corresponding CCR algebras. Hence, the pullback
$$\omega_{m,\mathcal N}^{\mathsf{g}}\doteq\omega_{\mathscr{I}}\circ \widehat{\Gamma}_{\mathsf{g},\mathcal{N}}$$
defines a quasifree state for the massive Klein--Gordon field on $\mathcal N$, whose two-point function reads
$$\omega_{2,m,\mathcal{N}}^{\mathsf{g}}(f,h)=\omega_{2,\mathscr{I}}\bigl(\Gamma_{\mathsf{g},\mathcal{N}}[f],\Gamma_{\mathsf{g},\mathcal{N}}[h]\bigr)=\omega_{2,0}(G_\mathsf{g} f,G_\mathsf{g} h),$$
for all $f,h\in C_0^\infty(\mathcal{N})$, where $\omega_{2,0}$ denotes the two-point function of the massless Minkowski vacuum.

The state $\omega_{m,\mathcal{N}}^{\mathsf{g}}$ is of Hadamard form. Indeed, the map $G_\mathsf{g}$ as in Theorem \ref{Thm: convergence not to zero on a Cauchy neighbourhood} induces a quasifree state for the auxiliary Klein--Gordon operator $P_{m,\mathsf{g}}=P_0-m^2\mathsf{g}$ on $\mathbb{M}^4$. Since $\mathsf{g}$ is smooth and past compact, the induced state coincides with the massless vacuum in a suitable Cauchy neighbourhood lying to the past of $\operatorname{supp}\mathsf{g}$. Its Hadamard property therefore follows from the propagation of singularities, see \cite{Hormander_1990} but also \cite{Radzikowski_1996}, in agreement with the construction outlined in \cite{Dappiaggi:2015xba}. Since $P_{m,\mathsf{g}}|_{\mathcal N}=P_m|_{\mathcal N}$, the restriction to $\mathcal N$ is a Hadamard state for the massive theory. In addition, the time-slice axiom yields an isomorphism between the CCR algebra of the massive field on $\mathcal N$ and that on $\mathbb M^4$. Consequently, $\omega_{m,\mathcal N}^{\mathsf{g}}$ induces a unique global quasifree state $\omega_m^\mathsf{g}$ on the massive field algebra. Its Hadamard property on the whole spacetime follows again from the propagation of singularities.

Notice that the resulting state generally depends on the choice of the admissible cut-off $\mathsf g$, although different choices yield Hadamard two-point functions whose difference is hence smooth. In particular, the construction does not, in general, select the Poincar\'e-invariant massive ground state. The latter can instead be recovered through a suitable adiabatic limit of the induced states, under the hypotheses discussed in \cite{Dappiaggi:2015xba}. This limit is conceptually distinct from the global adiabatic limit of the projected observables investigated in Theorem \ref{Thm: convergence to zero in D'}.

\section{Projection of Interacting Observables}
\label{Sec: Projection of Interacting Observables}

The aim of this section is to investigate how to establish a bulk-to-boundary correspondence to self-interacting scalar field theories on asymptotically flat spacetimes. The construction developed in the preceding section for the massive scalar field indeed suggests a natural approach to deal with interacting observables, inspired by perturbative algebraic quantum field theory, see Section \ref{Sec: The perturbative algebraic framework}. As a matter of fact, within the perturbative framework, any interacting observable can be expressed as formal power series in the coupling parameter by means of the Bogoliubov map, see Equation \eqref{Eq: Bogoliubov map}. Hence, the individual expansion coefficients are elements of the algebra of observables of the free field theory and they can be projected to future null infinity via the bulk-to-boundary map in Equation \eqref{Eq: bulk-to-boundary map}. Unlike the case of a quadratic interaction on Minkowski spacetime detailed in Section \ref{Sec: Four-dimensional Minkowski spacetime}, where convergence of the projected series has been established under suitable assumptions, the construction considered here is understood at the level of formal power series. 

Let $(\mcM,g)$ be a four-dimensional globally hyperbolic spacetime, asymptotically flat in the sense of Definition~\ref{Def: asymptotically flat spacetimes}. We consider a massless, conformally coupled scalar field $\varphi\in\mathcal E(\mcM)$, whose dynamics is ruled by the free action
\begin{equation}
\label{Eq: KG action SET}
S_0[f](\varphi)\doteq -\frac12\int_{\mcM} \left[g^{\mu\nu}(x)\nabla_\mu\varphi(x)\nabla_\nu\varphi(x)
+\frac{R(x)}6\varphi^2(x) \right]f(x)\,d\mu_x,\qquad f\in\mathcal D(\mcM),
\end{equation}
where $d\mu_g(x)$ is the metric-induced volume measure, while $R$ denotes the scalar curvature of $(\mcM,g)$. Let $V\in\mathcal P_{\mathrm{loc}}(\mcM)$ be a polynomial functional describing a compactly supported interaction of the form
\begin{equation*}
    \lambda V_h(\varphi) = - \lambda \int_{\mcM} h(x) \varphi^n(x) \, d\mu_g(x), \qquad h \in \mathcal{D}(\mcM), \; n \ge 1,
\end{equation*}
where $\lambda\in\mathbb R$ is the coupling parameter. Henceforth, we suppress the dependence on $h$ for notational simplicity, while emphasizing that the cut-off remains fixed and no adiabatic limit is taken. Once a renormalized time-ordering prescription has been fixed, the Bogoliubov map associates with a local, polynomial observable $F \in \mathcal{P}_\mathrm{loc}(\mcM)$ the formal expansion: 
\begin{equation}
\label{Eq: interacting observable boundary expansion}
R_{\lambda V}(F) = \sum_{n = 0}^\infty \lambda^n F^{(n)}, \qquad F^{(n)} \in \mathcal{A}_e(\mcM), 
\end{equation}
where $F^{(n)}$ denotes the coefficient of order $n$ obtained by expanding the quantum Bogoliubov map in Equation \eqref{Eq: Bogoliubov map}.
For those coefficients belonging to the domain of the bulk-to-boundary map $\Gamma$, we may define their boundary projection order by order:
\begin{equation}
\label{Eq: interacting boundary projection}
\Gamma\bigl(R_{\lambda V}(F)\bigr)
\doteq
\sum_{n=0}^{\infty}\lambda^n 
\Gamma\bigl(F^{(n)}\bigr).
\end{equation}
This expression is to be interpreted as an element of the corresponding space of formal power series with coefficients in the extended boundary algebra $\mathcal A_e(\mathscr I^+)$. Its well-definiteness requires the individual coefficients to satisfy the support and microlocal conditions entering the definition of $\Gamma$, see Section \ref{Sec: Extension of the algebra on the boundary} and, in particular, Equation \eqref{Eq: action of the bulk-to-boundary map}. We stress that no convergence of the series in $\lambda$ is assumed. 

\begin{remark}
\label{Rem: two couplings}
The preceding construction can be extended formally to interactions depending on more than one coupling parameter. In particular, if the interaction consists of a quadratic mass term and a non-linear, polynomial self-interaction, the corresponding observables admit a joint formal expansion in $m^2$ and $\lambda$. Each coefficient may then be projected to future null infinity, provided it belongs to the domain of the bulk-to-boundary map. The generalized Principle of Perturbative Agreement, \textit{cf.}, Definition \ref{Def: PPA}, relates this description to the perturbative construction based on the massive free theory with an interacting potential, subject to a compatible choice of renormalized time-ordered products. We stress, however, that this equivalence does not by itself establish convergence of the resulting boundary expansion or the existence of its adiabatic limit. 
\end{remark}

\vskip .2cm 

The construction raises two related questions. First, we need to understand how the bulk-to-boundary map acts on Wick polynomials and whether their finite renormalization freedoms are left untouched under projection to future null infinity. Second, given a local observable $F$, we wish to investigate the boundary projection of its interacting counterpart, constructed through the Bogoliubov map as a formal power series in the coupling parameter. This requires projecting the individual coefficients of this expansion, which are determined by the chosen renormalized time-ordered products. In particular, the renormalization freedom inherent in the definition of Wick polynomials raises the question of how different admissible prescriptions are represented at future null infinity.

\paragraph{The interplay between the bulk-to-boundary map and finite renormalization freedoms --} We first consider the free theory. As established by Hollands and Wald~\cite{Hollands_2001}, in four spacetime dimensions two admissible prescriptions for the $n$-th Wick power differ by local and covariant lower-order terms. More precisely, after a suitable redefinition of the coefficients, for any $n \ge 2$
\begin{equation}
\label{Eq: finite Wick renormalization}
\widetilde{\Phi}^n_f-\Phi^n_f = \sum_{k=0}^{n-2} \binom{n}{k}\Phi^k_{C_{n-k}f}, \qquad \Phi^0_f\doteq
\left(\int_{\mcM}f(x)\,d\mu_g(x)\right)\boldsymbol 1.
\end{equation}
where each $C_j$ is a locally and covariant scalar field of the appropriate engineering dimension. Assuming that the Wick polynomials appearing in Equation~\eqref{Eq: finite Wick renormalization} belong to the domain of $\Gamma$, its linearity and unit-preserving property yield
\begin{equation}
\label{Eq: projected finite Wick renormalization}
\Gamma\bigl(\widetilde{\Phi}^n_f\bigr) -\Gamma\bigl(\Phi^n_f\bigr) = \sum_{k=0}^{n-2} \binom{n}{k} \Gamma\bigl(\Phi^k_{C_{n-k}f}\bigr).
\end{equation}
Thus, the finite renormalization relations between bulk Wick polynomials are preserved under projection. When $\Gamma$ is injective on the relevant algebra, distinct bulk observables remain distinct in its boundary image. The Wick square provides the simplest example of this fact. In four spacetime dimensions, for a scalar field of mass $m$ and curvature coupling fixed as above, two admissible prescriptions differ by
\begin{equation}
\label{Eq: Wick square renormalization SET}
\widetilde{\Phi^2}_f-\Phi^2_f
=
\left(
\int_{\mcM}
f(x)\bigl(c_1m^2+c_2R(x)\bigr)\,d\mu_g(x)
\right)\boldsymbol 1,
\qquad c_1,c_2\in\mathbb R.
\end{equation}
In any unit-preserving boundary realization of the corresponding massive observable algebra, this difference is therefore mapped to
\[
\left(
\int_{\mcM}
f(x)\bigl(c_1m^2+c_2R(x)\bigr)\,d\mu_g(x)
\right)\boldsymbol 1_{\mathscr I^+}.
\]
In particular, the ambiguity proportional to $m^2$ survives as a multiple of the boundary identity. For interacting observables, a further renormalization freedom arises from the extension of time-ordered products to coinciding points. A change of admissible renormalization prescription modifies the coefficients entering Equation~\eqref{Eq: interacting observable boundary expansion}, and hence their boundary projections. When the corresponding finite renormalization transformation preserves the domain of $\Gamma$, it induces a transformation on the image algebra by conjugation with the bulk-to-boundary embedding. This describes the renormalization freedom inherited from the bulk, without introducing an independent renormalization prescription at future null infinity.

\begin{remark}
It is important to stress that the bulk-to-boundary map does not, in general, preserve localization, as discussed in~\cite{DMP_2017}. As a consequence, the images of local bulk counterterms need not define local counterterms intrinsic to $\mathscr I^+$. Preserving all finite renormalization relations should therefore be understood within the image of the bulk observable algebra, rather than as a statement of intrinsic locality and covariance on the boundary.
\end{remark}

\vskip.2 cm

The preceding discussion provides a framework for investigating interacting local observables at future null infinity. In the following subsection, we focus on a notable example of local interacting observable: the stress-energy tensor associated with a quartic self-interaction, see \cite{Goi_2026}. After constructing its perturbative expansion through the Bogoliubov map, we examine the corresponding boundary projection order by order, retaining the distinction between the renormalization ambiguities already present in the free theory and those arising from the interacting perturbative expansion.

\subsection{The Example of the Stress-Energy Tensor}
\label{Sec: Projection of the Stress-Energy Tensor}
In this section, we apply the methodology developed above to determine the projection of the interacting stress-energy tensor to future null infinity $\mathscr{I}^+$, retaining contributions up to first order in the coupling parameter. The starting point is the classical improved stress-energy tensor of a free scalar theory on $(\mcM, g)$ which reads \cite{Moretti_2003}
\begin{flalign}
    \label{Eq: free SET}
    \notag
    T^{(\eta)}_{\mu \nu, f}(\varphi) = \int_\mcM &\left[\nabla_\mu \varphi(x) \nabla_\nu \varphi(x) - \frac12 g_{\mu \nu}(x) \nabla^\sigma \varphi(x) \nabla_\sigma \varphi(x) + \eta g_{\mu \nu}(x) \varphi(x) P_0 \varphi(x) +\right.\\ & \left.+ \frac16 (G_{\mu \nu}(x) + g_{\mu \nu}(x) \Box_g - \nabla_\mu \nabla_\nu) \varphi^2(x) \right] f(x) \, d\mu_x, \quad \eta \in \mathbb{R}, 
\end{flalign}
for every $\varphi \in \mathcal{E}(\mcM)$, where $G_{\mu \nu} \doteq R_{\mu \nu} - \frac12 g_{\mu \nu} R$ is the Einstein tensor, while $\nabla_\mu$ denotes the covariant derivative associated with the Levi-Civita connection of $g$. For any $\eta\in\mathbb{R}$, the classical improved stress-energy tensor is both divergence-free and traceless on shell. Requiring its quantum counterpart to remain divergence-free, however, fixes the improvement parameter uniquely to $\eta = \frac13$ \cite[Thm. 3.2]{Moretti_2003}. The price to pay is that its trace acquires an anomalous contribution, which is purely geometric and, in particular, proportional to the coinciding point limit of the first Hadamard coefficient $v_1$. Standard computations entail that Equation \eqref{Eq: free SET} can be rewritten as
\begin{flalign} 
\notag T^{(\frac13)}_{\mu \nu, f} = &(\nabla_\mu \Phi\nabla_\nu \Phi)_f - \frac{1}{2} (g_{\mu \nu} \nabla^\sigma \Phi \nabla_\sigma \Phi)_f + \frac13 (g_{\mu \nu} \Phi P_0 \Phi)_f + \\ &+ \frac16 \left[ (G_{\mu \nu} \Phi^2)_f + 2(g_{\mu \nu} \nabla_\sigma \Phi \nabla^\sigma \Phi + g_{\mu \nu} \Phi \Box_g \Phi)_f- 2(\nabla_\nu \Phi \nabla_\mu \Phi + \Phi \nabla_\mu \nabla_\nu \Phi)_f \right], \label{Eq: SET as a local functional}
\end{flalign}
where the field configuration has been omitted for brevity and $\eta$ is chosen equal to $\frac13$ on account of the previous discussion. Here we adopt the notation 
\begin{equation*}
    (\nabla_\mu \Phi\nabla_\nu \Phi)_f [\varphi] \doteq \int_\mcM \nabla_\mu \varphi(x) \nabla_\nu \varphi(x) \, f(x) \, d\mu_g(x), \qquad f \in \mathcal{D}(\mcM), \; \varphi \in \mathcal{E}(\mcM),
\end{equation*}
and, similarly, for the other terms appearing in the previous formula. By direct inspection, we can conclude that $T_{\mu \nu, f} \in \mathcal{F}_\mathrm{loc}(\mcM)$.  

In order to compute the interacting corrections to the stress-energy tensor, we introduce a quartic interaction potential $\lambda V_h \in \mathcal{F}_{\mathrm{loc}}(\mcM)$ of the form 
\begin{equation}\label{Eq: potential SET}
\lambda V_h(\varphi) = \frac{\lambda}{4!} \int_\mcM \varphi^4(x) h(x) \, d\mu_x, \qquad h \in \mathcal{D}(\mcM),\; \lambda \in \mathbb{R} \setminus \{0\}. 
\end{equation}
In turn, we shall exploit once again the action of the Bogoliubov map on the free observable in Equation \eqref{Eq: free SET}, truncating the expansion to order $\mathcal{O}(\lambda^2)$. We stress that a similar computation has been performed in \cite{Goi_2026} up to $\mathcal{O}(\lambda^3)$, albeit starting from a massive scalar theory. First and foremost, Equation \eqref{Eq: Bogoliubov map} involves time-ordered products of local functionals, which are ill-defined on coinciding configurations and therefore require renormalization. In what follows, we assume that a renormalization prescription has been fixed once and for all by choosing suitable extensions to the thin diagonal of $\mcM \times \mcM$ of the singular distributions built from the Feynman parametrix, in particular its powers. Such extensions exist because the corresponding scaling degrees, and hence the degrees of divergence, are finite. The expansion of the S-matrix and its inverse then yields, at second order in perturbation theory, 
\begin{flalign}
    \label{Eq: interacting SET at second order}
    \notag 
    R_{\lambda V_h} (T_{\mu \nu, f}[\varphi]) &= T_{\mu \nu, f}[\varphi] + \frac{i \lambda}{ \hbar} (V_h \cdot_T T_{\mu \nu, f} - V_h \star_H T_{\mu \nu, f}) [\varphi] + \frac{\lambda^2}{2\hbar^2} \left[2 V_h \star (V_h \cdot_T T_{\mu \nu, f}) \right. \\ &\left.- (V_h \cdot_T V_h) \cdot_T T_{\mu \nu, f} - (V_h \cdot_{AT} V_h) \star T_{\mu \nu, f}\right][\varphi] + \mathcal{O}(\lambda^3). 
\end{flalign}
To compute explicitly each term appearing in Equation \eqref{Eq: interacting SET at second order}, we shall exploit the following result, see \cite[Lem. 4.2]{Goi_2026}, which we report here for the reader's convenience:
\begin{lemma}
    Let $(\mcM, g)$ be a four-dimensional globally hyperbolic spacetime and let 
    \begin{equation*}
        \begin{cases}
            \mathrm D \doteq a(x) \nabla_{\mu_1} \ldots \nabla_{\mu_i}, \qquad a \in C^\infty(\mcM), \\
            \mathrm D' \doteq b(x) \nabla_{\nu_1} \ldots \nabla_{\nu_j}, \qquad b \in C^\infty(\mcM),
        \end{cases}
    \end{equation*}
    be finite-order differential operators built out of the covariant derivative of $g$. Consider the functional $F \in \mathcal{P}_\mathrm{loc}(\mcM)$, namely, for any $\varphi \in \mathcal{E}(\mcM)$
    \begin{equation*}
        F_f[\varphi] = \int_\mcM \mathrm D \varphi(x) \mathrm D' \varphi(x) \, f(x) \, d\mu_x, \qquad f \in \Gamma_0(T\mcM^{\otimes(i+j)}).
    \end{equation*}
    Then, for any $K \in \mathcal{D}'(\mcM \times \mcM)$, $f, h \in \mathcal{D}(\mcM)$, it holds that, at the level of integral kernels, 
    \begin{flalign*}
        (V_h \star_K F_f)[\varphi] &= V_h[\varphi] F_f[\varphi] + \frac{\hbar^2}{2} \int_{\mcM^2} \mathrm D K(x,y) \mathrm D' K(x,y) \varphi^{(n-2)}(x) \, h(x) f(y) \, d\mu_x \, d\mu_y \\ &+\frac{\hbar}{6} \int_{\mcM^2} [\mathrm D K(x,y) \mathrm D' \varphi(y) + \mathrm D \varphi(y) \mathrm{D}' K(x,y)]\varphi^{(n-1)}(x) \, h(x) f(y) \, d\mu_x \, d\mu_y,
    \end{flalign*}
    where $V_h$ is the interacting potential as per Equation \eqref{Eq: potential SET}.
\end{lemma}

\paragraph{Corrections at first perturbative order --} Let us begin by examining the first order contributions to Equation \eqref{Eq: interacting SET at second order}, thereby truncating the perturbative expansion at order $\mathcal{O}(\lambda^2)$. A direct computation hence entails that 
\begin{equation}
\label{Eq: V T tensor}
    (V_h \cdot_T T_{\mu \nu, f}) [\varphi] = V_h[\varphi] \, T_{\mu \nu, f}[\varphi] + A_{\mu \nu, f \otimes h}^{1,F} [\varphi] + A_{\mu \nu, f \otimes h}^{2,F} [\varphi], 
\end{equation}
where 
\begin{flalign}
\label{Eq: Amunu1}
\notag
    A_{\mu \nu, f \otimes h}^{1,F} & [\varphi] = \frac{\hbar}{6} \int_{\mcM^2} d\mu_x \, d\mu_y \, h(x) f(y) \varphi^3(x) \left[ \frac23 (\nabla_\mu^{(y)} H_F(x,y) \nabla_{\nu}^{(y)} \varphi(y) + \nabla_\nu^{(y)} H_F(x,y) \nabla_\mu^{(y)} \varphi(y) \right. \\ &\notag \left. -\frac13 g_{\mu \nu}(y) \nabla^{\sigma, (y)} H_F(x,y) \nabla_\sigma^{(y)} \varphi(y) +\frac13 \left(G_{\mu \nu} (y) - g_{\mu \nu}(y) \frac{R(y)}{3}\right) H_F(x,y) \varphi(y) \right. \\& \notag \left. + \frac23 g_{\mu \nu}(y) \left( \Box_g^{(y)} H_F(x,y) \varphi(y) + H_F(x,y) \Box_g^{(y)} \varphi(y)\right) \right.\\& \left. -\frac13 \left(H_F(x,y) \nabla_\mu^{(y)} \nabla_\nu^{(y)} \varphi (y) + \varphi(y) \nabla_{\mu}^{(y)} \nabla_\mu^{(y)} H_F(x,y) \right) \right]
\end{flalign}
while 
\begin{flalign}
    \label{Eq: Amunu2}
    \notag 
   A_{\mu \nu, f \otimes h}^{2,F} [\varphi] &= \frac{\hbar^2}{2}\int_{\mcM^2} d\mu_x d\mu_y \, h(x) f(y) \varphi^2(x) \left[\frac23 \nabla_\mu^{(y)} H_F(x,y) \nabla_\nu^{(y)} H_F(x,y) \right.\\& \notag \left.- \frac16 g_{\mu \nu}(y) \nabla^{\sigma, (y)} H_F(x,y) \nabla_\sigma^{(y)} H_F(x,y) + \frac16 \left(G_{\mu \nu}(y) - \frac13 g_{\mu \nu}(y) R(y) \right) H^2_F(x,y) \right. \\& \left.+ \frac23 g_{\mu \nu}(y) H_F(x,y) \Box_g^{(y)} H_F(x,y) - \frac13 H_F(x,y) \nabla_\mu^{(y)} \nabla_\nu^{(y)} H_F(x,y) \right].
\end{flalign}
An analogous expansion, albeit replacing everywhere $H_F$ with the Hadamard parametrix $H$, holds for the term $(V_h \star_H T_{\mu \nu, f})[\varphi]$, \textit{i.e.},
\begin{equation}
    \label{Eq: V AT tensor}
    (V_h \star_H T_{\mu \nu, f})[\varphi] = V_h[\varphi] T_{\mu \nu, f}[\varphi] +  A_{\mu \nu, f \otimes h}^{1} [\varphi] +  A_{\mu \nu, f \otimes h}^{2} [\varphi]. 
\end{equation}
To obtain the first-order correction to the stress-energy tensor, we need to compute the difference $V_h \cdot_T T_{\mu \nu, f} - V_h \star_H T_{\mu \nu, f}$. To this avail, we observe the following: 
\begin{itemize}
    \item[\ding{104}] all terms of order $\hbar^0$ cancel out; 
    \item[\ding{104}] the integrand in the terms of order $\hbar$ in Equations \eqref{Eq: V T tensor} and \eqref{Eq: V AT tensor} is linear in $H_F$ and $H$, respectively. By definition of Feynman parametrix, $H_F(x,y) - H(x,y) = i \Delta_{\mathrm{A}}(x,y)$, where $\Delta_\mathrm{A}(x,y)$ is the integral kernel of the advanced propagator of $P_0$. 
\end{itemize}
Hence, 
\begin{flalign*}
    \frac{i \lambda}{\hbar} &(V_h \cdot_T T_{\mu \nu, f} - V_h \star_H T_{\mu \nu, f})[\varphi] = -\frac{\lambda}{6} \int_{\mcM^2} d\mu_x d\mu_y \, h(x) f(y) \varphi^3(x) \left[ \frac23 \nabla_\mu^{(y)} \Delta_\mathrm{A}(x,y) \nabla_\nu^{(y)} \varphi(y) \right.\\ &\left. + \frac23 \nabla_\nu^{(y)} \Delta_\mathrm{A} (x,y) \nabla_\mu^{(y)} \varphi(y) - \frac13 g_{\mu \nu}(y) \nabla^{\sigma, (y)} \Delta_\mathrm{A} \nabla_\sigma^{(y)} \varphi(y) \right.\\ & \left.+ \frac13 \left(G_{\mu \nu}(y) - g_{\mu \nu}(y) \frac{R(y)}{3} \right) \Delta_\mathrm{A}(x,y) \varphi(y) + \frac23 g_{\mu \nu}(y) \varphi (y) \Box_g^{(y)} \Delta_\mathrm{A}(x,y) \right. \\ & \left.+ \frac23  g_{\mu \nu}(y) \Delta_\mathrm{A}(x,y) \Box_g^{(y)} \varphi(y) -\frac13 \varphi(y) \nabla_\nu^{(y)} \nabla_\mu^{(y)} \Delta_\mathrm{A} (x,y) -\frac13 \Delta_\mathrm{A}(x,y) \nabla_\nu^{(y)} \nabla_\mu^{(y)} \varphi(y) \right] \\ &+ \frac{i \hbar \lambda}{2} \int_{\mcM^2} d\mu_x d\mu_y \, h(x) f(y) \varphi^2(x) \left[ \frac23 \nabla_\mu^{(y)} H_F(x,y) \nabla_\nu^{(y)} H_F(x,y) \right.  \\ & \left. - \frac23 \nabla_\mu^{(y)} H(x,y) \nabla_\nu^{(y)} H(x,y) -\frac16 g_{\mu \nu}(y) \nabla^{\sigma, (y)} H_F(x,y) \nabla_\sigma^{(y)} H_F(x,y) \right. \\&\left. +\frac16 g_{\mu \nu}(y) \nabla^{\sigma, (y)} H(x,y) \nabla_\sigma^{(y)} H(x,y) + \frac16 \left(G_{\mu \nu}(y) - g_{\mu \nu}(y) \frac{R(y)}{3} \right) (H^2_F(x,y) - H^2(x,y)) \right. \\&\left. +\frac23 g_{\mu \nu}(y) H_F(x,y) \Box_g^{(y)} H_F(x,y) - \frac23 g_{\mu \nu}(y) H(x,y) \Box_g^{(y)} H(x,y) \right. \\&\left. -\frac13 H_F(x,y) \nabla_\mu^{(y)} \nabla_\nu^{(y)} H_F(x,y) + \frac13 H(x,y) \nabla_\mu^{(y)} \nabla_\nu^{(y)} H(x,y)\right].
\end{flalign*}

\begin{remark}
For later convenience let us define field-independent linear operators with distributional coefficients of the form 
\begin{equation*}
    D_{\mu \nu, k} (x,y) (\varphi) \doteq \langle D_{\mu \nu, k} (x,y; \cdot), \varphi(\cdot) \rangle, \qquad \text{for any} \; \varphi \in \mathcal{E}(\mcM), \quad k=2,4,
\end{equation*}
which correspond to the integral kernels in the above equation depending on $2$ and $4$ field configurations, respectively. Most notably, they act as differential operators only in the $y$-variable and, in addition, $D_{\mu\nu, 4}(x,y)$ is linear in the field configuration. We stress once again that, whenever required, all distributional products and extensions to the relevant diagonals are understood in the sense of the renormalization prescription fixed above, so that the expressions and operations appearing below are well-defined. 
\end{remark}

In order to compute explicitly the action of the bulk-to-boundary map on each of the contributions above, we need to express these functionals in terms on their integral kernels in the vein of Remark \ref{Rem: functionals and integral kernels}. Indeed, what really matters for the sake of our computations, is the number of field configurations appearing in the functionals. For instance, we could schematically write 
\begin{equation}
    \label{Eq: first order correction}
    \frac{i \lambda}{\hbar} (V_h \cdot_T T_{\mu \nu, f} - V_h \star_H T_{\mu \nu, f}) [\varphi] = -\frac{\lambda}{6} F_4[\varphi] + \frac{i \lambda \hbar}{2} F_2[\varphi], 
\end{equation}
where $F_2, F_4$ gather all the above contributions depending on $2$ or $4$ fields, respectively. Observe that, upon choosing a renormalization prescription for the singular products entering $F_4$, which preserves the microlocal spectrum condition, $F_2, F_4 \in \mathcal{F}_{\mu c}(\mcM)$, as the resulting compactly supported kernels satisfy the relevant wavefront set condition, see Equation \eqref{Eq: Microcausal Functionals}. Starting from $T^{(\frac13)}_{\mu \nu, f}$ as per Equation \eqref{Eq: SET as a local functional}, we can write
\begin{equation}
    \label{Eq: T mu nu intergal kernel}
    T^{(\frac13)}_{\mu \nu, f} [\varphi] = \int_{\mcM^2} d\mu_x d\mu_y \, f(x) \delta(x,y) D_{\mu \nu} \varphi(x) \varphi(y), 
\end{equation}
where $\delta(x,y)$ is the integral kernel of the Dirac delta distribution supported on the diagonal of $\mcM \times \mcM$, while $D_{\mu \nu}$ is the bi-differential operator given by
\begin{flalign}
    \label{Eq: bidifferential operator}
    \notag
    D_{\mu \nu} &\doteq \nabla_\mu^{(x)} \nabla_\nu^{(y)} - \frac12 g_{\mu \nu} \nabla_{\sigma}^{(x)} \nabla^{\sigma, (y)} + \frac16 (G_{\mu \nu} + 2 g_{\mu \nu}(\Box_g^{(y)} + \nabla_{\sigma}^{(x)} \nabla^{\sigma, (y)}) \\& \notag -2(\nabla_\mu^{(x)} \nabla_\nu^{(y)} + \nabla_\mu^{(y)} \nabla_\nu^{(y)}) + \frac13 g_{\mu \nu} P_0^{(y)} \\ & = \frac23  \nabla_\mu^{(x)} \nabla_\nu^{(y)} -\frac16 g_{\mu \nu} \nabla_{\sigma}^{(x)} \nabla^{\sigma, (y)} + \frac16 \left(G_{\mu \nu} - g_{\mu \nu} \frac{R}{3}\right) + \frac23 g_{\mu \nu} \Box_g^{(y)} - \frac13 \nabla_\mu^{(y)} \nabla_\nu^{(y)}.
\end{flalign}
Observe that, with a slight abuse of notation, we employ the symbol $D_{\mu\nu}$ although derivatives might act either on the first or on the second entry, denoted respectively by $x$ and $y$. On account of the above discussion, we can express the free stress-energy tensor as a compactly supported distribution whose integral kernel reads 
\begin{equation*}
    T_{\mu \nu, f}(x,y) = D^*_{\mu \nu} f(x) \delta(x,y), 
\end{equation*}
where $D_{\mu\nu}^*$ denotes the formal adjoint of $D_{\mu \nu}$ with respect to the dual pairing. Similarly, the integral kernels of $F_4 \in \mathcal{E}'(\mcM^4)$ and $F_2 \in \mathcal{E}'(\mcM^2)$ read, up to all possible permutations of $x_1, x_2, x_3, x_4$,
\begin{flalign}
    \label{Eq: F4}
    \notag
   F_4(x_1, x_2, x_3, x_4)& = \int_{\mcM^2} d\mu_x d\mu_y \, h(x) f(y) \delta(x,x_1) \, \delta(x,x_2) \, \delta(x,x_3) \notag \\ &\notag \left[-\frac23 \nabla_{\mu}^{(y)} \Delta_\mathrm{A}(x,y) \nabla_\nu^{(y)} \delta (y, x_4) - \frac23 \nabla_\nu^{(y)} \Delta_\mathrm{A} (x,y) \nabla_{\mu}^{(y)} \delta(y,x_4) \right.\\ &\notag \left. +\frac13 g_{\mu \nu}(y) \nabla{\sigma, (y)} \Delta_\mathrm{A}(x,y) \nabla_{\sigma}^{(y)} \delta(y,x_4) \right. \\ \notag & \left.+\frac13 \left(G_{\mu \nu}(y) - g_{\mu \nu}(y) \frac{R(y)}{3} \right) \Delta_\mathrm{A}(x,y) \delta(y,x_4) \right. \\ \notag & \left. +\frac23 g_{\mu \nu}(y) \left(\delta(y,x_4) \Box_g^{(y)} \Delta_{\mathrm A}(x,y) + \Delta_\mathrm{A} (x,y) \Box_g^{(y)} \delta(y, x_4) \right) \right. \\ \notag &\left. -\frac13 \left(\delta(y,x_4) \nabla_\nu^{(y)} \nabla_\mu^{(y)} \Delta_\mathrm{A}(x,y) + \Delta_\mathrm{A}(x,y) \nabla_\nu^{(y)} \nabla_\mu^{(y)} \delta (y,x_4) \right) \right] \\ &\doteq \int_{\mcM^2} d\mu_x d\mu_y \, h(x) f(y) \delta(x,x_1) \delta(x, x_2) \delta (x,x_3) D_{\mu \nu, 4} \delta(y,x_4)
\end{flalign}
and, similarly up to permutations of $x_1, x_2$,
\begin{flalign}
    \label{Eq: F2}
    \notag
    F_2(x_1, x_2) &= \int_{\mcM^2} d\mu_x d\mu_y \, \delta(x,x_1) \delta(x,x_2) h(x) f(y) \\ \notag &\left[\frac23 \left(\nabla_\mu^{(y)} H_F(x,y) \nabla_\nu^{(y)}H_F(x,y) - \nabla_\nu^{(y)} H (x,y) \nabla_\mu^{(y)} H(x,y)\right) \right. \\ \notag & \left. - \frac13 g_{\mu \nu}(y) \left(\nabla^{\sigma, (y)} H_F(x,y) \nabla_\sigma^{(y)} H_F(x,y) - \nabla^{\sigma, (y)} H(x,y) \nabla_\sigma^{(y)} H(x,y)\right) \right. \\& \notag \left. +\frac16 \left(G_{\mu \nu}(y) - g_{\mu \nu}(y) \frac{R(y)}{3} \right) (H_F^2(x,y) - H^2(x,y)) \right. \\&\notag \left. +\frac23 g_{\mu \nu}(y) \left(H_F(x,y) \Box_g^{(y)} H_F(x,y) - H(x,y) \Box_g^{(y)} H(x,y) \right) \right. \\&\notag \left. -\frac13 \left( H_F(x,y) \nabla_\mu^{(y)} \nabla_\nu^{(y)} H_F(x,y) - H(x,y) \nabla_\mu^{(y)} \nabla_\nu^{(y)} H(x,y) \right) \right] \\ &\doteq \int_{\mcM^2} d\mu_x d\mu_y \, \delta(x,x_1) \delta(x,x_2) h(x) f(y) \delta(x,y) D_{\mu \nu, 2}(x,y).
\end{flalign}
While $D_{\mu \nu, 4}$ is a genuine differential operator with distributional coefficients, $D_{\mu \nu, 2}$ is a zeroth-order operator acting by multiplication on the tensor product of field configurations. On account of the linearity of the bulk-to-boundary map as per Definition \ref{Def: bulk to boundary map}, we can project separately each of the above functionals onto $\mathscr{I}^+$. Let us denote by $(\tilde{\mcM}, \tilde{g})$ the unphysical spacetime as per Definition \ref{Def: asymptotically flat spacetimes}, by $\tilde{P}_0$ the conformally coupled Klein-Gordon operator thereon and by $\Omega \in C^\infty(\tilde{\mcM})$ the associated conformal factor. Furthermore, let $\iota^* \tilde{\Delta}^\mathscr{V}_\mathrm{S}$ be the pullback of the Pauli-Jordan propagator $\tilde{\Delta}_{\mathrm S}$ under the map $\iota: \mathscr{I}^+ \times \tilde{\mcM} \to \tilde{\mcM} \times \tilde{\mcM}$. The projection of the kernel $F_k\in\mathcal E'(\mcM^k)$ to $(\mathscr I^+)^k$ is the distribution $\Gamma[F_k]\in\mathcal D'((\mathscr I^+)^k)$ defined by
\begin{equation}
\label{Eq: Gamma k-th order}
    \left\langle \Gamma[F_k],\psi\right\rangle_k
    \doteq \frac{1}{k!} \left\langle (\iota^*\tilde{\Delta}_{\mathrm S})^{\otimes k} \cdot \left(1\otimes(\Omega^{-3})^{\otimes k}F_k\right), \psi\otimes\eta \right\rangle_k,
\end{equation}
for every $\psi\in\mathcal{D}((\mathscr I^+)^k)$ and $\eta\in\mathcal D(\mcM^k)$ such that $\eta \vert_{\operatorname{supp}(F_k)}\equiv1$. Here $1\in\mathcal{D}'((\mathscr{I}^+)^k)$ denotes the distribution generated by the constant function $1$. Let us work at the level of integral kernels and set $u_1, \ldots, u_4 \in \mathscr{I}^+$. Using Equation \eqref{Eq: Gamma k-th order}, the projected free stress-energy tensor thus reads 
\begin{flalign*}
    \Gamma[T_{\mu \nu, f}] (u_1, u_2) &= 
    \int_{\mcM^2} d\mu_x d\mu_y \, \tilde{\Delta}^\mathscr{V}_\mathrm{S}(u_1, x) \Omega^{-3}(x) \tilde{\Delta}^\mathscr{V}_\mathrm{S}(u_2, y) \Omega^{-3}(y) [D^*_{\mu \nu} f(x) \delta(x,y)]^\mathrm{sym} \\ &= \int_{\mcM^2} d\mu_x d\mu_y \, f(x) \delta(x,y) [D_{\mu \nu} (\tilde{\Delta}^\mathscr{V}_\mathrm{S}(u_1, x) \Omega^{-3}(x) \tilde{\Delta}^\mathscr{V}_\mathrm{S}(u_2, y) \Omega^{-3}(y))]^{\mathrm{sym}} \\ &= \int_\mcM d\mu_x \, f(x) \left[\frac13 \nabla_\mu^{(x)} (\tilde{\Delta}^\mathscr{V}_\mathrm{S}(u_1, x) \Omega^{-3}(x)) \nabla_\nu^{(x)} (\tilde{\Delta}^\mathscr{V}_\mathrm{S}(u_2, x) \Omega^{-3}(x)) \right. \\ &\left. + \frac13 \nabla_\nu^{(x)} (\tilde{\Delta}^\mathscr{V}_\mathrm{S}(u_1, x) \Omega^{-3}(x)) \nabla_\mu^{(x)} (\tilde{\Delta}^\mathscr{V}_\mathrm{S}(u_2, x) \Omega^{-3}(x))  \right. \\ & \left. - \frac16 g_{\mu \nu}(x) \nabla_{\sigma}^{(x)} (\tilde{\Delta}^\mathscr{V}_\mathrm{S}(u_1, x) \Omega^{-3}(x)) \nabla^{\sigma, (x)} (\tilde{\Delta}^\mathscr{V}_\mathrm{S}(u_2, x) \Omega^{-3}(x)) \right. \\ & \left. +\frac16 \left(G_{\mu \nu}(x) - \frac{R(x)}{3} g_{\mu \nu}(x) \right) \tilde{\Delta}^\mathscr{V}_\mathrm{S} (u_1, x) \tilde{\Delta}^\mathscr{V}_\mathrm{S} (u_2,x) \Omega^{-6}(x) \right. \\ &\left. +\frac13 g_{\mu \nu}(x) \tilde{\Delta}^\mathscr{V}_\mathrm{S}(u_1, x) \Omega^{-3}(x) \Box_g^{(x)} \tilde{\Delta}^\mathscr{V}_\mathrm{S}(u_2, x) \Omega^{-3}(x) \right.  \\ &\left. +\frac13 g_{\mu \nu}(x) \tilde{\Delta}^\mathscr{V}_\mathrm{S}(u_2, x) \Omega^{-3}(x) \Box_g^{(x)} \tilde{\Delta}^\mathscr{V}_\mathrm{S}(u_1, x) \Omega^{-3}(x) \right. \\ &\left. -\frac16 (\tilde{\Delta}^\mathscr{V}_\mathrm{S}(u_1, x) \Omega^{-3}(x)) \nabla_\mu^{(x)} \nabla_\nu^{(x)} (\tilde{\Delta}^\mathscr{V}_\mathrm{S}(u_2, x) \Omega^{-3}(x)) \right. \\ &\left. -\frac16 (\tilde{\Delta}^\mathscr{V}_\mathrm{S}(u_2, x) \Omega^{-3}(x)) \nabla_\mu^{(x)} \nabla_\nu^{(x)} (\tilde{\Delta}^\mathscr{V}_\mathrm{S}(u_1, x) \Omega^{-3}(x))\right]. 
\end{flalign*}
We observe that, although the bulk stress-energy tensor is a local functional, its projection to $\mathscr{I}^+$ is supported also outside of the thin diagonal of $\mathscr{I}^+ \times \mathscr{I}^+$. This supports the statement that the extended boundary algebra $\mathscr{A}_e(\mathscr{I}^+)$ cannot be identified with that of local functionals on $\mathscr{I}^+$. 

The boundary projections of the first-order correction terms $F_2$ and $F_4$, up to all possible permutations stemming from the computation of Gateaux derivatives, read 
\begin{flalign}
\label{Eq: Gamma F4}
\notag
    \Gamma[F_4](u_1, u_2, u_3, u_4) &= \int_{\mcM^2} d\mu_x d\mu_y \, h(x) f(y) \mathrm E (u_1, x) \mathrm E (u_2, x) \mathrm E (u_3, x) D_{\mu \nu, 4} \mathrm{E} (u_4, y) \\ &\notag = \int_{\mcM^2} d\mu_x d\mu_y \, h(x) f(y) \mathrm E (u_1, x) \mathrm E (u_2, x) \mathrm E (u_3, x) \\ &\notag \left[\frac23 \nabla_\mu^{(y)} \Delta_\mathrm{A}^\mathscr{V} (x,y)\nabla_\nu^{(y)} \mathrm E (u_4,y) + \frac23 \nabla_\nu^{(y)} \Delta_\mathrm{A}^\mathscr{V}(x,y)\nabla_\mu^{(y)} \mathrm E (u_4,y) \right. \\&\notag \left. -\frac13 g_{\mu \nu}(y) \nabla_\sigma^{(y)} \Delta_\mathrm{A}^\mathscr{V}(x,y) \nabla^{\sigma, (y)} \mathrm E(u_4,y) \right. \\ &\notag\left. \frac13 \left(G_{\mu \nu}(y) - g_{\mu \nu}(y) \frac{R(y)}{3} \right) \Delta_\mathrm{A}^\mathscr{V}(x,y) \mathrm E(u_4, y) \right. \\&\notag \left. +\frac23 g_{\mu \nu}(y) \mathrm E )(u_4,y) \Box_g^{(y)} \Delta_\mathrm{A}^\mathscr{V}(x,y) +\frac23 g_{\mu \nu}(y) \Delta_\mathrm{A}^\mathscr{V}(x,y) \Box_g^{(y)} \mathrm E(u_4,y) \right. \\&\left. -\frac13 \mathrm E (u_4,y) \nabla_\nu^{(y)} \nabla_\mu^{(y)} \Delta_\mathrm{A}^\mathscr{V}(x,y) -\frac13 \Delta_\mathrm{A}^\mathscr{V}(x,y) \nabla_\nu^{(y)} \nabla_\mu^{(y)} \mathrm E (u_4, y)\right], 
\end{flalign}
and 
\begin{flalign}
\label{Eq: Gamma F2}
\notag
    \Gamma[F_2](u_1, u_2) &= \int_{\mcM^2} d\mu_x d\mu_y \, h(x) f(y) \mathrm{E}(u_1, x) \mathrm{E}(u_2,x) D_{\mu \nu, 2}(x,y) \\& \notag = \int_{\mcM^2} d\mu_x d\mu_y \, h(x) f(y) \mathrm{E}(u_1, x) \mathrm{E}(u_2,x) \\ &\notag\left[\frac23 \nabla_\mu^{(y)} H_F(x,y) \nabla_\nu^{(y)} H_F(x,y) - \frac23 \nabla_\mu^{(y)} H(x,y) \nabla_\nu^{(y)} H(x,y) \right. \\&\notag \left. -\frac16 g_{\mu \nu}(y) \nabla_\sigma^{(y)} H_F(x,y) \nabla^{\sigma, (y)} H_F(x,y) + \frac16 \nabla_\sigma^{(y)} H(x,y) \nabla^{\sigma, (y)} H(x,y) \right. \\&\notag\left. \frac16 \left(G_{\mu \nu}(y) - g_{\mu \nu}(y) \frac{R(y)}{3}\right) (H_F^2(x,y) - H^2(x,y)) \right. \\&\notag \left. +\frac23 g_{\mu \nu}(y) H_F(x,y) \Box_g^{(y)} H_F(x,y) - \frac23 g_{\mu \nu}(y) H(x,y) \Box_g^{(y)} H(x,y) \right. \\&\left. -\frac13 H_F(x,y) \nabla_\mu^{(y)} \nabla_\nu^{(y)} H_F(x,y) +  \frac13 H(x,y) \nabla_\mu^{(y)} \nabla_\nu^{(y)} H(x,y) \right], 
\end{flalign}
where for the sake of brevity, we denoted by $\mathrm E (u_i, x) \doteq \tilde{\Delta}^\mathscr{V}_\mathrm{S}(u_i, x) \Omega^{-3}(x)$. Equations \eqref{Eq: Gamma F4} and \eqref{Eq: Gamma F2} encode the effect of a quartic bulk interaction on the asymptotic behaviour of the stress-energy tensor up to first order in the coupling. More precisely, the contribution associated with $F_4$ corresponds to the classical first-order correction since it survives in the classical limit $\hbar \to 0$, while $F_2$ represents a genuinely quantum contribution arising from the difference between the time-ordered and star products. Thus, already at first order in perturbation theory, the asymptotic
observable retains information both about the classical interacting dynamics in the bulk and about the corresponding quantum fluctuations.

\vskip .2cm

Beyond the explicit form of the projected observables, an important feature of the construction is the loss of locality under the bulk-to-boundary map. Although the original bulk observables are local, their images at null infinity are, in general, non-local. This behaviour is intrinsic to the definition of the bulk-to-boundary map, which propagates bulk information to the asymptotic boundary through the causal dynamics of the theory. As hinted in Remark \ref{Rem: two couplings}, the same strategy can, in principle, be applied order by order in perturbation theory and extended to different interaction potentials, \textit{e.g.}, including also quadratic mass term.

\begin{remark}
 We highlight that, similarly to Section \ref{Sec: Projection of Interacting Observables}, one could remove the cut-off, hence taking the adiabatic limit, considering only observables localised in a neighbourhood of a Cauchy surface. In this case one would have to exploit the time-slice axiom for interacting theories as discussed in \cite{Chilian:2008ye}. We refrain from implementing this procedure, since, contrary to the case of a massive real scalar field, we will be left with a formal power series in the underlying coupling constant, whose convergence cannot be established.
\end{remark}

\section*{Acknowledgements}
\addcontentsline{toc}{section}{Acknowledgements}

We are grateful to Nicola Pinamonti for enlightening discussions on preliminary versions of this manuscript. The work of BC has been supported by a fellowship of the INFN -- Sezione di Pavia, whose support is gratefully acknowledged also by CD. BC and CD both acknowledge the support of Gruppo Nazionale di Fisica Matematica, part of INdAM. This work has appeared in the M.Sc. thesis of AF submitted on 18/09/26 in partial fulfilment of the requirements to obtain a M.Sc. Degree in Theoretical Physics at the University of Pavia.

\vskip.2cm

\noindent\textbf{Data availability statement}. Data sharing is not applicable to this article as no new data were created or analysed in this study.

\vskip .2cm

\noindent\textbf{Conflict of interest statement.} The authors certify that they have no affiliations with or involvement in any
organization or entity with any financial interest or non-financial interest in the subject matter discussed in
this manuscript.

\vskip .2cm

\noindent\textbf{AI disclosure statement}. The authors acknowledge the use of ChatGPT (OpenAI) as an auxiliary tool during the preparation of this manuscript, in particular for linguistic editing, improving exposition, and discussing technical aspects of some arguments. All mathematical statements and proofs were independently checked and derived by the authors, who take full responsibility for the content of the paper.

\appendix

\end{document}